\documentclass[pdftex,twocolumn,epjc3]{svjour3}
\RequirePackage[T1]{fontenc}

\smartqed  

\RequirePackage{graphicx}
\RequirePackage{mathptmx}      
\RequirePackage{flushend}
\RequirePackage[numbers,sort&compress]{natbib}
\RequirePackage[colorlinks,citecolor=blue,urlcolor=blue,linkcolor=blue]{hyperref}

\journalname{Eur. Phys. J. C}

\usepackage{amsmath}
\usepackage{subcaption}
\usepackage{amssymb}

\begin{document}
\title{Triple Higgs Production at Muon Colliders in the Higgs Triplet Model}
%
\author{Bathiya Samarakoon \thanksref{e1,addr1,addr2}
\and 
Terrance M. Figy \thanksref{e2,addr1} %
}
%
\institute{
\label{addr1} Department of Mathematics, Statistics and Physics, Wichita State University, 1845 Fairmount St, Wichita, KS, 67260, USA
\and
\label{addr2} Department of Natural Sciences and Mathematics, McPherson College, 1600 E. Euclid,
McPherson, KS, 67460, USA
}

\thankstext{e1}{e-mail:samarakb@mcpherson.edu}
\thankstext{e2}{e-mail:terrance.figy@wichita.edu}
\date{Received: date / Revised version: date}
%
\maketitle

\begin{abstract}   
The production of three Higgs bosons provides a unique opportunity to probe both the trilinear and quartic Higgs couplings within the Standard Model and beyond.
In this study, we investigate triple Higgs production in the Higgs Triplet Model via photon fusion, $\gamma \gamma \to h^0 h^0 h^0$, allowing us to explore the contributions from singly and doubly charged Higgs bosons and their couplings to CP-even scalars.
We analyze the total cross sections and kinematic distributions, comparing our results with the corresponding SM predictions.
Our calculations utilize an automated one-loop calculator to evaluate the numerical amplitudes for the $2 \to 3$ process.
For the collider simulations, we developed a Monte Carlo phase-space integrator to predict outcomes for a future muon collider, covering a center-of-mass energy range from 2 TeV to 10 TeV.
\end{abstract} 

%

\section{Introduction}
\label{sec:intro}
In the Standard Model (SM), the scalar potential of the Higgs sector is given by
\begin{equation}
  V_{SM} \supset \frac{M_h^2}{2}h^2+\frac{1}{3!}\frac{3M_h^2}{v}h^3+\frac{1}{4!}\frac{3M_h^2}{v^2}h^4 .
\end{equation} 
Here, $h$ represents the physical Higgs field, $v$ represents the vacuum expectation value, and $M_{h}$ is the Higgs boson mass. In this framework, the trilinear Higgs coupling and quartic Higgs coupling are expressed as $ \lambda_{hhh} = \frac{3M_h^2}{v} $ and $ \lambda_{hhhh} = \frac{3M_h^2}{v^2}$~\cite{Englert:1964et, Higgs:1964pj, Guralnik:1964eu, Weinberg:1967tq, ATLAS:2018rvj, Salam:1968rm, Grober:2016yuo}. Any deviations from these SM predictions suggest the presence of physics beyond the SM \cite{ATLAS:2024ish}. Precise measurements of these couplings provide a valuable approach for probing new physics and unraveling the structure of the Higgs potential, which plays a crucial role in electroweak symmetry breaking~\cite{Higgs:1966ev, Higgs:1964ia, Abouabid:2024gms, Goldstone:1962es, Guralnik:1965uza, Kibble:1967sv, Plehn:2005nk, Baglio:2022wkx, Baglio:2020wgt}. Additionally, exploring the Higgs potential in detail is key to understanding the mechanisms underlying the electroweak phase transition and the matter-antimatter asymmetry in the universe~\cite{Weinberg:1974hy, Biermann:2024oyy, Dolan:1973qd, Coleman:1973jx, Kirzhnits:1976ts, Anderson:1991zb, Kuzmin:1985mm, Goncalves:2017iub, Karkout:2024ojx, Arnold:1992rz, Kajantie:1996mn}.

The process of producing three Higgs bosons offers a unique avenue to directly study the Higgs self-coupling. While double Higgs boson production provides insight into the trilinear Higgs coupling~\cite{Dawson:2022zbb, Lewis:2024yvj, deFlorian:2019app, Bahl:2022jnx, Plehn:1996wb}, triple Higgs boson production goes a step further by probing the quartic Higgs self-coupling. The quartic self-coupling is an essential component of the Higgs potential~\cite{Chiesa:2020awd}. Despite its importance, this aspect of the potential remains experimentally unexplored, highlighting the need for further investigations into the fundamental properties of the Higgs boson~\cite{Arhrib:2014nya}.  

For proton colliders, the dominant production mechanisms for double Higgs (\(hh\)) and triple Higgs (\(hhh\)) processes occur via gluon fusion. At a center-of-mass energy of \(\sqrt{s} = 14~\mathrm{TeV}\), the cross section for \(hh\) production is approximately 300 times larger than that for \(hhh\) production. At next-to-next-to-leading order (NNLO), the cross sections are given by $ \sigma(hh) = 36.69 \pm 3.0\%~{\rm fb}$ and $\sigma(hhh) = 0.103 \pm 15\%~{\rm fb}$~\cite{Shao:2013bz, deFlorian:2015moa, LHCHiggsCrossSectionWorkingGroup:2016ypw, deFlorian:2019app, Abouabid:2024gms}.

The production of three Higgs bosons at the Large Hadron Collider (LHC) is currently beyond the detection threshold due to its extremely low cross section. This, coupled with significant background noise and the LHC's limited integrated luminosity, makes it nearly impossible to observe a sufficient number of $hhh$ events for robust analysis~\cite{Asakawa:2008se, Papaefstathiou:2015paa, Binoth:2006ym, Plehn:2005nk}. Even at a center-of-mass energy of $\sqrt{s} = 14~\mathrm{TeV}$, the process remains highly suppressed. Future collider projects, Refs.~\cite{Acar:2016rde, Murayama:1996ec}, with increased energy and luminosity, such as the Future Circular Collider (FCC)~\cite{Benedikt:2022kan, FCC:2018evy}, would be necessary to achieve meaningful sensitivity to triple Higgs production~\cite{Telnov:2006rj, Arhrib:2009gg}.

In a muon collider~\cite{Delahaye:2019omf, Palmer:2014nza, Long:2020wfp, InternationalMuonCollider:2025sys} the collision of high-energy muon beams can result in the emission of two collinear photons. These photons may interact to produce multiple Higgs bosons, including triple Higgs boson production. This process is facilitated by a loop-induced mechanism, where virtual particles such as heavy fermions and gauge bosons mediate the effective coupling between the photons and Higgs bosons~\cite{vonWeizsacker:1934nji, Chiesa:2021qpr, Williams:1934ad}. The photon-photon interaction in this scenario can significantly enhance the effective cross-section for triple Higgs production, making it a rare yet important process to study~\cite{Costantini:2020stv, AlAli:2021let, Sonmez:2018mcq}. Moreover, the muon collider's clean experimental environment and high luminosity provide a unique opportunity to examine the Higgs boson self-couplings and explore the structure of the Higgs potential in greater detail \cite{Black:2022cth, MuonCollider:2022xlm, Capdevilla:2024ydp, Capdevilla:2020qel, Capdevilla:2024bwt, Han:2020pif, Capdevilla:2021rwo}.

In scenarios extending Beyond the Standard Model (BSM), the production of multiple Higgs bosons~\cite{Phan:2024vfy, Phan:2024vxm, Arhrib:2009gg, Figy:2011yu, Demirci:2016fri, Demirci:2019kop, Muhlleitner:2001kw, Fuks:2025gjv} such as three Higgs bosons, can be substantially enhanced due to the presence of new particles and interactions that modify the Higgs sector. BSM frameworks, including the Higgs Triplet Model (HTM)~\cite{Lee:1973iz, Schechter:1980gr, Cheng:1980qt, Mohapatra:1979ia, Konetschny:1977bn, Arbabifar:2012bd, Magg:1980ut}, and Two-Higgs Doublet Models, (2HDM) \cite{Deshpande:1977rw, Weinberg:1967tq, Glashow:1976nt, Ellis:1988er, Branco:2011iw, Demirci:2020zgt, ElKaffas:2007nii, WahabElKaffas:2007xd, Sonmez:2018mcq}, introduce additional scalar fields, such as charged Higgs bosons, which contribute to higher production rates compared to the SM predictions. Moreover, these models may predict the existence of Higgs-like particles absent in the SM, providing a window into the structure of an extended Higgs sector. For example, the HTM allows for the production of three charged Higgs bosons, enabling the study of new couplings derived from the extended scalar potential, thereby distinguishing these models from the SM. A detailed analysis of three Higgs production via photon fusion and VBF processes in the SM context is presented in Ref.~\cite{Chiesa:2021qpr}. Investigations into triple Higgs production \cite{Chiesa:2020awd} also reveal insights into the complex mechanisms of electroweak symmetry breaking~\cite{Englert:1964et, Higgs:1964pj, Goldstone:1962es, Kibble:1967sv, Weinberg:1967tq} in BSM theories, helping differentiate among various extensions. A comprehensive study on the production of three Higgs bosons within the framework of the Two Real Singlet Extension Model is presented in Ref.~\cite{Papaefstathiou:2020lyp}.

Although the total cross sections for triple Higgs production via $W^{\pm}W^{\pm}$ and $Z^0Z^0$
fusion are typically larger than those induced by photon fusion \cite{Chiesa:2021qpr, Shen:2015bna}, the
$\gamma\gamma$-initiated process provides a uniquely clean and sensitive probe of
the extended scalar sector of the HTM. In particular,
photon fusion is dominantly loop-induced and receives sizable contributions from
charged and doubly charged Higgs bosons, whose effects are directly governed by
the scalar self-couplings of the model.

In the HTM, the couplings of the doubly charged Higgs boson to electroweak gauge
bosons are proportional to the triplet vacuum expectation value $v_\Delta$ \cite{Rahili:2019ixf, Arhrib:2011uy}, as
defined later in Sec.~\ref{sec:theory}. For
small values of $v_\Delta$, these couplings are strongly
suppressed, leading to a reduced sensitivity of $W^{\pm}W^{\pm}$-fusion processes to the
doubly charged Higgs sector. By contrast, photon fusion processes are not
directly controlled by $v_\Delta$ and remain sensitive to charged and doubly
charged Higgs bosons through loop effects. As a result,
$\gamma\gamma$-initiated triple Higgs production offers a complementary and, in
certain regions of parameter space, more direct probe of the scalar
self-interactions and non-decoupling effects characteristic of the HTM.

In 2012, the ATLAS and CMS collaborations announced the discovery of the Higgs boson, with its mass measured at $125.6 \pm 0.3 \, {\rm GeV}$~\cite{CMS:2022dwd, CMS:2012qbp}. While this result aligns closely with the predictions of the SM, it is important to explore scenarios beyond the SM that could influence Higgs decay channels and self-couplings~\cite{BhupalDev:2013xol}. The SM assumes neutrinos are massless, as there are no right-handed neutrinos to generate a Dirac mass term. However, measurements from experiments like MAINZ,~\cite{Kraus:2004zw, Weinheimer:1999tn}, TROITSK~\cite{Lobashev:1999tp, Belesev1995ResultsOT}, and KATRIN~\cite{KATRIN:2021uub, Drexlin:2013lha, KATRIN:2019yun}, suggest the need for a new mass scale in the neutrino sector.

There are various models for generating neutrino mass, including the addition of right-handed neutrinos or mechanisms such as the inverse seesaw. Our work focuses on the Type II Seesaw mechanism. The core idea of the seesaw mechanism is to introduce additional fields that couple to the SM fermions, with a mass scale significantly higher than the electroweak scale. In the HTM (Type II seesaw model), left-handed lepton doublets are coupled to a triplet field, leading to neutrino masses proportional to the Yukawa coupling and the vacuum expectation value of the triplet~\cite{Schechter:1980gr, Primulando:2019evb}.

This paper is organized as follows: Sections \ref{sec:theory} and \ref{sec:param} provide an overview of the HTM and introduces the parameter scanning methodology. Section \ref{sec:methodology} details the calculations and the Monte Carlo methods used. In Section \ref{sec:results}, we present and analyze the results. Finally, Section \ref{sec:conclusions} concludes the paper, summarizing our findings.

\section{The Higgs Triplet Model}
\label{sec:theory}
To generate mass for neutrinos, the SM, which includes a Higgs doublet \(\Phi \sim (1,2,1/2)\), is extended by incorporating a \(SU(2)_L\) Higgs triplet field \(\Delta\). This extension allows for the creation of a Yukawa coupling between the SM leptons and the new scalar fields, which is expressed as follows \cite{Ashanujjaman:2025scr, Arhrib:2011uy}:

\begin{equation}\label{eq:leptonfield}
    \mathcal{L} \supset Y_{jk}L_j^T C i \sigma_2 \Delta L_k + h.c.
\end{equation}

Here, $L_j$ and $L_k$ denote the left-handed lepton doublets, $C$ is the charge conjugation operator, $\sigma_m$, ($m=1,2,3$), represents the Pauli matrices, and $Y_{jk}$ corresponds to the Yukawa coupling associated with the neutrino sector, which is an element of a complex symmetric matrix. The indices $j$ and $k$ refer to the lepton flavors $e$, $\mu$, and $\tau$~\cite{Arhrib:2011uy, Rahili:2019ixf, PhysRevD.22.2227, PhysRevD.25.2951, Babu:2016gpg, Du:2018eaw, Akeroyd:2012ms,Akeroyd:2012rg, Akeroyd:2011ir}.

The doublet field, which has a weak hypercharge of $Y_{\Phi} = 1$, and the triplet field $\Delta$, represented in a $2 \times 2$ matrix form with a weak hypercharge of $Y_{\Delta} = 2$, are defined as follows:

\begin{equation*}\label{eq:fields}
    \Phi = \begin{pmatrix}
        \phi^+  \\
        \phi^0 
    \end{pmatrix}, 
    \quad 
    \Delta = \begin{pmatrix}
        \frac{\delta^{+}}{\sqrt{2}} & \delta^{++} \\
\delta^0 & -\frac{\delta^+}{\sqrt{2}}   
    \end{pmatrix} . 
\end{equation*}

The triplet Higgs field $\Delta$ acquires a small vacuum expectation value (VEV), $v_\Delta$, which plays a crucial role in generating neutrino masses through the Type II Seesaw mechanism, while the Higgs doublet field acquires its VEV, denoted as $v_{\Phi}$. After electroweak symmetry breaking, the neutrino mass matrix is given by $m_\nu = \sqrt{2}Y_\nu v_\Delta$. The smallness of neutrino masses is naturally explained by the tiny value of \(v_\Delta\), which is typically on the order of \(\mathcal{O}(1)\) eV, ensuring consistency with experimental observations of neutrino mass scales~\cite{Babu:2016gpg, Akeroyd:2012ms}.

The scalar sector, including the kinetic terms and potential for $\Phi$ and $\Delta$, is given by:
\begin{equation}\label{eq:laga}
\mathcal{L}_{s} = (D_{\mu}\Phi)^{\dagger} (D^{\mu}\Phi) + \text{Tr}[(D_{\mu}\Delta)^{\dagger}(D^{\mu}\Delta)] - V(\Phi, \Delta),
\end{equation}

where the covariant derivatives, 

\begin{equation}\label{eq:coveDer1}
D_{\mu}\Phi = \partial_{\mu}\Phi + i\frac{g}{2}\sigma^mW^m_{\mu}\Phi+i\frac{g^{'}}{2}B_{\mu}\Phi,
\end{equation}
and,
\begin{equation}\label{eq:coveDer2}
D_{\mu}\Delta = \partial_{\mu}\Delta + ig[\frac{1}{2}\sigma^m W^m_{\mu},\Delta]+i\frac{g^{'}}{2}Y_{\Delta}B_{\mu}\Delta,
\end{equation}
are defined with the $SU(2)$ and $U(1)$ gauge field couplings, $g$ and $g'$, respectively. The general renormalizable potential term is introduced as~\cite{Akeroyd:2012ms, Haba:2016zbu}:

\begin{eqnarray} \label{eq:potentailterm}
V(\Phi,\Delta)&=&-\mu_{\Phi}^2\Phi^{\dagger}\Phi + \frac{\lambda}{4}(\Phi^{\dagger}\Phi)^2 + \mu^2_{\Delta}Tr(\Delta^{\dagger}\Delta) 
\\&+&[\mu(\Phi^T i\sigma^2 \Delta^{\dagger} \Phi)+ h.c]+\lambda_{1}(\Phi^{\dagger}\Phi) Tr(\Delta^{\dagger}\Delta)\nonumber\\&+& \lambda_{2} (Tr \Delta^{\dagger}\Delta)^{2} 
 + \lambda_{3} Tr(\Delta^{\dagger}\Delta)^{2} 
 + \lambda_{4} \Phi^{\dagger} \Delta \Delta^{\dagger} \Phi\nonumber. 
\end{eqnarray}

The scalar potential in the HTM involves several parameters with distinct roles and dimensions. The mass parameters include $\mu_{\Phi}^2>0$, which has dimensions of $[\text{mass}^2]$ and determines the  mass term for the SM Higgs doublet $\Phi$. A positive value for $\mu_{\Phi}^2$ guarantees spontaneous symmetry breaking \cite{Haba:2016zbu}, leading to the generation of the Higgs VEV, $v_{\Phi}$. Similarly, $\mu_{\Delta}^2$, also with dimensions of $[ \text{mass}^2 ]$, controls the mass term for the Higgs triplet $\Delta$. A positive value for $\mu_{\Delta}^2$ implies that the triplet does not acquire a large VEV, $v_{\Delta}=\mu^2v^2/(2\mu_{\Phi}^2)$. Additionally, the dimensionful parameter $\mu$, with dimensions of $[ \text{mass} ]$, couples the doublet $\Phi$ to the triplet $\Delta$ and plays a critical role in generating a small vacuum expectation value for $\Delta$ through the Type II Seesaw mechanism~\cite{Akeroyd:2010je}. The minimization of the Higgs potential in the HTM is essential for ensuring the correct pattern of electroweak symmetry breaking. By minimizing the potential, one obtains the following conditions that relate the parameters of the model to the VEVs: 

\begin{equation}\label{eq:mu_phi}
    \mu_{\Phi}^2=\frac{\lambda v^2_{\Phi}}{4}-\sqrt{2}\mu v_{\Delta}+\frac{(\lambda_1+\lambda_4)v_{\Delta}}{2}
\end{equation}
and
\begin{equation}\label{eq:mu_delta}
    \mu_{\Delta}^2=\frac{2\mu v^2_{\Phi}-\sqrt{2}(\lambda_1+\lambda_4)v_{\Phi}^2v_{\Delta}-2\sqrt{2}(\lambda_2+\lambda_3)v_{\Delta}^3}{2\sqrt{2}v_{\Delta}}.
\end{equation}

The dependence of $\mu_{\Phi}^2$ and $\mu_{\Delta}^2$ on the $\lambda_i$ couplings is also important because these Lagrangian parameters modify the couplings of the Higgs boson to gauge bosons and fermions, which can be investigated at collider experiments.

In HTM, the Higgs sector includes a doublet and a triplet, resulting in 10 real scalar degrees of freedom before EWSB. After EWSB, three Goldstone bosons are absorbed by the $W^{\pm}$ and $Z^0$ bosons, leaving seven physical Higgs states: two CP-even neutral Higgs bosons $h^0$ and $H^0$, one CP-odd neutral Higgs boson $A^0$, one singly charged Higgs $H^{\pm}$, and one doubly charged Higgs $H^{\pm\pm}$~\cite{Rahili:2019ixf}. 

The mixing between these states is determined by three mixing angles: \( \alpha \) for the CP-even Higgses, \( \beta \) for the CP-odd Higgses, and \( \beta^{'} \) for the charged Higgses. These mixings arise from the interactions between the Higgs doublet and triplet components, affecting the phenomenology and collider signatures of the Higgs sector~\cite{Primulando:2019evb}. The mixing between the light and heavy Higgs states is described by the following equations:
\begin{equation}\label{eq:hHmix-1}
    h^0=h\cos{\alpha} + \xi\sin{\alpha} 
\end{equation}
and
\begin{equation}\label{eq:hHmix-2}
    H^0=-h\sin{\alpha} +\xi\cos{\alpha} .
\end{equation}
Here, $ h = \text{Re}[\phi^0] $ and $ \xi = \text{Re}[\delta^0]$ are shifted by their VEVs. The triplet VEV ($v_{\Delta}$) is crucial in determining the strength of these mixings, influencing the production and decay channels of the extended Higgs sector in the Type-II Seesaw Model~\cite{Ashanujjaman:2021txz}.

Probing quartic and trilinear scalar couplings in the HTM provides an opportunity to examine the shape of the Higgs potential by reparameterizing $\mu_{\Delta}$ and $\mu_{\Phi}$ in terms of $\lambda$, $\lambda_{i}$ (where $i \in {1,2,3,4}$), and $\mu$. To establish a parameter space for collider simulations, we propose the following input set, noting that $v = \sqrt{v_{\Phi}^2 + 2v_{\Delta}^2} = 246 ~\mathrm{GeV}$:

\begin{equation}\label{eq:parammset}
\begin{aligned}
\mathcal{P}_{1}= \{ M_{h^0}, M_{H^0}, M_{A^0}, M_{H^{\pm}}, M_{H^{\pm\pm}}, v_{\Delta}, \cos{\alpha}\}.
\end{aligned}
\end{equation}

To establish a well-reasoned selection of parameters within the framework of the HTM, one must consider the $\rho$-parameter, the mass splitting between charged Higgs bosons, perturbativity, and vacuum stability. The $\rho$-parameter,  which deviates from unity at tree level, is given by  

\begin{equation}  
    \rho = \frac{1+2v^2_{\Delta}/v_{\Phi}^2}{1+4v^2_{\Delta}/v_{\Phi}}.  
\end{equation} \label{eq:rhoparameter}  

This establishes an upper limit on $v_{\Delta}$ of approximately 8 GeV to maintain that the experimental value of the $\rho$-parameter remains close to unity~\cite{Aoki:2012yt}.

The mass splitting between $ H^{\pm}$ and $H^{\pm\pm}$, constrained by electroweak precision data, implies that $|\Delta M| < 50$ GeV ~\cite{Das:2016bir}. The mass hierarchy of the charged Higgs bosons depends on the sign of $\Delta M$. 

To guarantee vacuum stability, ensuring that the scalar potential $V(\Phi, \Delta)$ remains bounded from below and does not enter an instability regime, the parameters in the scalar potential must satisfy the following conditions, which are valid only at leading order (LO) \cite{ Arhrib:2011uy, Babu:2016gpg}:
\begin{equation}
 \lambda \geq 0 ; ~\lambda_2+\lambda_3 \geq 0 ; ~\lambda_2+\frac{\lambda_3}{2} \geq 0 \ ,
 \label{eq:b1}
\end{equation}

\begin{equation}
    \lambda_1+\sqrt{\lambda(\lambda_2+\lambda_3)}\geq 0 ;~ \lambda_1+ 
\sqrt{\lambda(\lambda_2+\frac{\lambda_3}{2})} \geq 0 
\label{eq:b2}
\end{equation}
and,
\begin{equation}
    \lambda_1+\lambda_4+\sqrt{\lambda(\lambda_2+\lambda_3)}\geq 0 ; ~\lambda_1+\lambda_4+ 
\sqrt{\lambda(\lambda_2+\frac{\lambda_3}{2})} \geq 0 \ .
\label{eq:b3}
\end{equation}
To preserve theoretical consistency, the model parameters must respect
perturbative unitarity. A detailed discussion of perturbative unitarity in the
HTM can be found in Refs.~\cite{Arhrib:2011uy,Rahili:2019ixf, Haba:2016zbu, Ashanujjaman:2024lnr}. In
Ref.~\cite{Arhrib:2011uy}, the bounded-from-below (BFB) conditions and perturbative
unitarity constraints are explicitly derived and discussed in Section.~IV. In that
work, a scaling parameter $\kappa$ is introduced within the unitarity framework
and is allowed to take the values $\kappa = 16$ or $8$, as motivated by
$S$-matrix unitarity constraints from elastic scattering. In the present study,
we adopt the more conservative choice $\kappa = 16$ when imposing the unitarity
bounds.

\section{Framework for Parameter Space Investigation}\label{sec:param}

When studying SM-like Higgs production and decay processes in the HTM or other extended Higgs scalar models, selecting the input parameters efficiently and consistently is essential for ensuring theoretical and phenomenological accuracy. The masses of the Higgs scalars and the mixing angle $\alpha$ are physical parameters that inherently depend on the scalar potential parameters ($\lambda_i$, $\lambda$, $\mu$). To evaluate the scalar potential parameters, one can use the following expressions~\cite{Arhrib:2014nya, Samarakoon:2023crt, Arhrib:2011uy}:

\begin{equation}\label{eq:lam}
    \begin{aligned}
        \lambda = -\frac{2}{v_{\Phi}^2}(c_{\alpha}^2M^2_{h}+s^2_{\alpha}M^2_H) \ ,
    \end{aligned}
\end{equation}

\begin{equation}\label{eq:lam1}
    \begin{aligned}
        \lambda_1 = -\frac{2}{v_{\Phi}^2+4v^2_{\Delta}}M_A^2+ \frac{2}{v_{\Phi}^2+2v^2_{\Delta}}M_{H^{\pm}}^2\\  
        - \frac{\sin{2\alpha}}{2v_{\Delta}v_{\Phi}}(M_H^2-M_h^2) \ ,
    \end{aligned}
\end{equation}
 
\begin{equation}\label{eq:lam2}
    \begin{aligned}
        \lambda_2 = \frac{1}{2v_{\Delta}^2} \Big( c_{\alpha}^2M^2_{h}+s^2_{\alpha}M^2_H + \frac{v^2_{\Phi}}{v_{\Phi}^2+4v^2_{\Delta}}M_A^2 - \\ \frac{4v^2_{\Phi}}{v_{\Phi}^2+2v^2_{\Delta}}M_{H^{\pm}}^2 +2M^2_{H^{\pm\pm}}\Big) \ , 
    \end{aligned}
\end{equation}

\begin{equation}\label{eq:lam3}
    \begin{aligned}
        \lambda_3 = \frac{1}{v_{\Delta}^2} \Big(  -\frac{v^2_{\Phi}}{v_{\Phi}^2+4v^2_{\Delta}}M_A^2 + \frac{2v^2_{\Phi}}{v_{\Phi}^2+2v^2_{\Delta}}M^2_{H^{\pm}} -M^2_{H^{\pm\pm}}\Big) \ ,
    \end{aligned}
\end{equation} 

\begin{equation}\label{eq:lam4}
    \begin{aligned}
        \lambda_4 =   \frac{4}{v_{\Phi}^2+4v^2_{\Delta}}M_A^2 - \frac{4}{v_{\Phi}^2+2v^2_{\Delta}}M^2_{H^{\pm}} \ ,
    \end{aligned} 
\end{equation} 
and
\begin{equation}\label{eq:mu}
    \begin{aligned}
        \mu = \frac{\sqrt{2}v_{\Phi}}{v_{\Phi}^2+4v^2_{\Delta}}M^2_A \ .
    \end{aligned}
\end{equation} 

The masses of the Higgs bosons are directly related to observable quantities measurable in collider experiments. Utilizing $\mathcal{P}_1$ space, Eq.~(\ref{eq:parammset}), secures that simulations remain consistent with experimental constraints. The choice of $\alpha$ is critical in determining the degree of mixing between the doublet-like and triplet-like states, ensuring consistency with current experimental data and precise Higgs coupling measurements~\cite{Du:2018eaw}. However, during simulations, calculating $\lambda_i$, $\lambda$, $\mu$ from $\mathcal{P}_1$ can be computationally challenging, especially when considering the mass splittings between the charged Higgs bosons and the neutral Higgs bosons. Therefore, this highlights the need to develop a reliable method for deducing the parameter functions from the mass values in $\mathcal{P}_1$. However, this is demanding due to the high sensitivity of $\lambda_2$ and $\lambda_3$, particularly when ensuring compliance with vacuum stability and perturbativity constraints.

If one calculates $\lambda_2$ and $\lambda_3$ using the parameters in $\mathcal{P}_1$ and Eqs.~(\ref{eq:lam2})-(\ref{eq:lam3}), and then introduces a minute adjustment, ($\sim 10^{-4}~\mathrm{GeV}$), to the scalar masses, $\delta M_{A^0, H^{0}, H^{\pm}, H^{\pm\pm}}$, the values of $\lambda_2$ and $\lambda_3$ can increase significantly. If these parameters become excessively large, the contributions from $s$-channel scalar exchange to Drell-Yan processes would dominate,~\cite{Ducu:2024xxf,Arbabifar:2012bd,Arhrib:2014nya}, potentially leading to incorrect tree-level results. As discussed in Refs.~\cite{Ducu:2024xxf,Arhrib:2014nya}, this pronounced sensitivity represents a theoretical issue rather than merely a numerical challenge. This behavior, which can be described as highly reactive, is not always a direct consequence of variations in $\delta M_{A^0, H^{0}, H^{\pm}, H^{\pm\pm}}/v_{\Delta}$. Instead, it can arise purely from inconsistencies in the mass parameters, particularly when evaluated at $v_{\Delta} = 1\,\mathrm{GeV}$. These inconsistencies emphasize the necessity for precise parameter tuning to maintain theoretical consistency and prevent inaccurate results in cross sections and kinematic distributions.

To develop an effective approach for parameter scanning, we begin by ensuring that the square root terms in Eqs.~(\ref{eq:b2}) and (\ref{eq:b3}) remain real. This requirement imposes the conditions that both $\lambda_2 + \lambda_3$ and $\lambda_2 + \lambda_3/2$ must be positive~\cite{Samarakoon:2023crt}. These scenarios form the basis for deriving the resultant expressions in Eqs.~(\ref{eq:MAmax}) and~(\ref{eq:MHpmax}), which are then used to deduce the masses of the singly charged Higgs boson and the pseudoscalar. 

\begin{equation}\label{eq:MAmax}
    M_{A^0}^{\text{max}} = \frac{(v^2_{\Phi}+4v_{\Delta}^2)^{1/2}}{v_{\Phi}}(\cos{\alpha}^2M_{H^0}^2+\sin{\alpha}^2M_{h^0}^2)^{1/2} \ . 
\end{equation}
\begin{equation}\label{eq:MHpmax}
   M_{H^\pm}^{\text{max}} = \frac{(v^2_{\Phi}+2v_{\Delta}^2)^{1/2}}{\sqrt{2}v_{\Phi}}(M^2_{H^{\pm\pm}}+\cos{\alpha}^2M_{H^0}^2+\sin{\alpha}^2M_{h^0}^2)^{1/2}. 
 \end{equation}

The approach relies on input parameters including $M_{H^{\pm\pm}}$, $M_{H^0}$, $M_{h^0}$, $\cos{\alpha}$, and $v_{\Delta}$. By carefully selecting these inputs, the method ensures consistency between theoretical predictions and experimental constraints, providing a sufficient and robust framework for exploring the parameter space of the model. 
 
When selecting numerical inputs for the mass of the singly charged Higgs boson, based on the masses of the heavy Higgs boson and the doubly charged Higgs boson as per Eq.~(\ref{eq:MHpmax}), one can determine it as: \begin{equation}\label{eq:MHpinput} M_{H^\pm} = M_{H^\pm}^{\text{max}} \pm \delta M_{H^{\pm}}. 
\end{equation}

\begin{figure}[htb]
    \centering
    \includegraphics[width=1.0\linewidth]{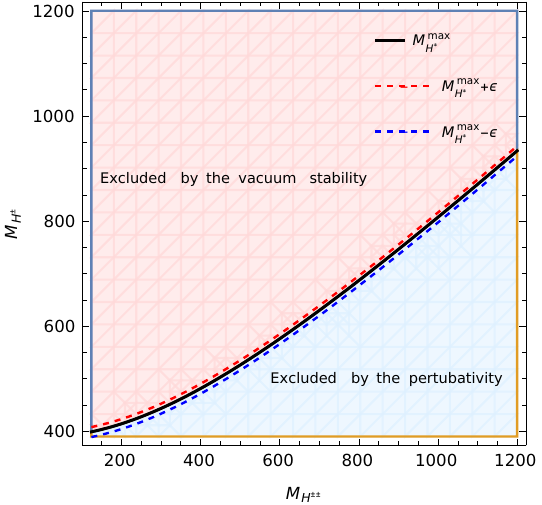} 
    \caption{The mass of \( H^{\pm} \) as a function of \( M_{H^{\pm\pm}} \), with \( M_{H^0} = 600~\mathrm{GeV} \), \( v_{\Delta} = 1 \) GeV, and \( \cos{\alpha} = 0.9999 \). For any value chosen above the red dashed lines, the potential term becomes unstable, and for values below the blue dashed lines, the parameter functions of the potential violate perturbativity.
}
    \label{figure:pertandsatb}
\end{figure}

The Eqs. (\ref{eq:MAmax}) and (\ref{eq:MHpmax}) highlight that the singly charged Higgs boson and the pseudoscalar boson have maximum values they can attain within the parameter space $\mathcal{P}_2 = \mathcal{P}_1 \setminus \{M_{H^{\pm}}, M_{A^0}\}$, which serves as a sufficient condition for stability. If one chooses $M_{H^{\pm}}^{\text{input}}$ as an input value such that $M_{H^{\pm}}^{\text{input}} \lessgtr M_{H^{\pm}}^{\text{max}}$, it could lead to either a violation of vacuum stability or a breakdown of perturbativity, particularly for sufficiently large variations $\epsilon = \delta M_{H^{\pm}}$. This emphasizes the need for careful parameter selection to maintain theoretical consistency, see the Fig.~\ref{figure:pertandsatb}. For sufficiently large values of $\epsilon$, $M^{\text{max}}_{H^{\pm}} + \epsilon$ can result in a violation of vacuum stability. Similarly, for certain values of $\epsilon$, $M^{\text{max}}_{H^{\pm}} - \epsilon$ may lead to a breakdown of perturbativity.

Now, let us express $\lambda_{2,3}$ as functions of the parameters in the space $\mathcal{P}_2 \cup \{M_{H^{\pm}}^{\text{max}} - \epsilon, M_{A^0}^{\text{max}}\}$. The corresponding results, denoted as $\tilde{\lambda}_2$ and $\tilde{\lambda}_3$, are given by  
\begin{equation}  
    \tilde{\lambda}_2 = \lambda_2+N\epsilon M^{\text{max}}_{H^{\pm}} ~, 
\end{equation}\label{eq:lam2til} 
and  
\begin{equation}  
    \tilde{\lambda}_3 =  \lambda_3-N\epsilon M^{\text{max}}_{H^{\pm}} ~,  
\end{equation}\label{eq:lam3til}  

with 
\begin{equation}  
    N=\frac{4v_{\Phi}^2}{v_{\Delta}^2(v_{\Phi}^2+2v_{\Delta}^2)} ~.
\end{equation}

This indicates that, for certain nonzero values of $|\epsilon|$, the inequalities $|\lambda_{2,3}| \leq 8\pi$ still hold \cite{Rahili:2019ixf}. This implies that for fixed values of the parameters in $\mathcal{P}_2$, there exists an $\epsilon > 0$ such that $|\lambda_3(M_{H^{\pm}}^{\text{max}} - \epsilon)| \geq 8\pi$ and $|\lambda_2(M_{H^{\pm}}^{\text{max}} + \epsilon)| \geq 8\pi$. Moreover, there exists a $\epsilon_P \in (0, \epsilon)$ such that $|\lambda_3(M_{H^{\pm}}^{\text{max}} - \epsilon_P)| \leq 8\pi$\label{eq:ineqpert3} and $|\lambda_2(M_{H^{\pm}}^{\text{max}} + \epsilon_P)| \leq 8\pi$\label{eq:inequpert2}. When $\epsilon_P < \epsilon$, $\lambda_2$ and $\lambda_3$ are nonzero and satisfy the conditions for perturbativity. However, when $\epsilon = \epsilon_P = 0$, $\lambda_2$ and $\lambda_3$ approach extremely small values. For nonzero values of $\lambda_2$ and $\lambda_3$, these parameters have a considerable impact on the couplings $\lambda^{HTM}_{H^0H^{+}H^{-}}$ and $\lambda^{HTM}_{H^0H^{++}H^{--}}$. Although their contributions to $\lambda^{HTM}_{h^0H^{+}H^{-}}$ and $\lambda^{HTM}_{h^0H^{++}H^{--}}$ couplings are negligible in the limit $\alpha \to 0$, they still play a significant role in the scattering amplitude for the processes $\gamma \gamma \rightarrow h^0 h^0, h^0h^0h^0$. By solving the last two inequalities in this section for $\epsilon_{P}$, we find that:

\begin{equation}
    \epsilon_P = 2\sqrt{2}\pi v \frac{v^2_{\Delta}}{v_{\Phi}}\frac{1}{\sqrt{M^2_{H^{\pm\pm}}+\cos^2{\alpha}M_{H0}^2+\sin^2{\alpha}M_{h^0}^2}}.
\end{equation}\label{eq:espP}
Hence, for all $\epsilon \in \mathbb{R}^+\setminus [0, \epsilon_P]$, the perturbativity breaks down in the $\mathcal{P}_1$ parameter space.

To verify compatibility with the last inequality in the conditions shown in Eq.~(\ref{eq:b3}), We consider the modifications introduced by Eq.~(\ref{eq:MHpmax}). In the parameter space $\mathcal{P}_2 \cup \{{M_{H^{\pm}}^{\text{max}}} + \epsilon_V, M_{A^0}^{\text{max}}\}$, assume that there exists a value $\epsilon_V > 0$ such that the condition $\tilde{\lambda}_2 + \frac{\tilde{\lambda}_3}{2} > 0$ is satisfied. 

However,  
\begin{equation}  
    \tilde{\lambda}_2 + \frac{\tilde{\lambda}_3}{2} = - \frac{v_{\Phi}^2}{2v^2_{\Delta}v^2}\epsilon_V \Big( \epsilon_V + 2M^{\text{max}}_{H^{\pm}} \Big),
\end{equation}  
which implies that $\tilde{\lambda}_2 + \frac{\tilde{\lambda}_3}{2}$ remains positive only if $-2M^{\text{max}}_{H^{\pm}} < \epsilon_V < 0$.  

This results in a contradiction because $\epsilon_V$ does not satisfy the positivity condition. Therefore, we can conclude that for every sufficiently large $\epsilon_V > 0$, the input $M^{\text{max}}_{H^{\pm}} + \epsilon_V$ violates the vacuum stability of the Higgs potential.

In the scenario where \(\alpha = \pi/4\), the above conditions may not hold, as for non-zero \(\alpha\), \(\lambda_1 + \lambda_4\) can become negative, potentially could not satisfy
$
\lambda_1 + \lambda_4 > -\sqrt{\lambda(\lambda_2 + \lambda_3/2)}.
$
This violates the bounded-from-below condition presented in Eq.~(\ref{eq:b3}).  

Therefore, without explicitly solving the inequality to derive the scalar mass limits, we deduced sufficient conditions for \(\lambda_1 + \lambda_4 > 0\), which provides an upper bound on the heavy Higgs mass for \(\alpha \in (\tan^{-1}(2v_{\Delta}^2/v_{\Phi}^2), \sim0.143\pi)\). This inequality consequently leads to a lower bound on the pseudo-scalar mass, expressed as:  
\begin{equation}
    M^{\text{min}}_{A^0} = \frac{1}{2}\sqrt{\frac{v_{\Phi}^2+4 v_{\Delta}^2}{v_{\Delta}v_{\Phi}}}\sqrt{(\Delta M_s^2)\sin{2\alpha}}.
\end{equation}\label{eq:lover_massA0}
Where $\Delta M_s^2 = M^2_{H^0}-M_{h^0}^2$. Now, comparing this with Eq.~(\ref{eq:MAmax}), we find that \(M_{A^0}^{\text{min}} \leq M_{A^0}^{\text{max}}\), which provides:

\begin{equation}\label{eq:MHmax}
    M_{H^0}^{\text{max}} = \Big( 1+ \frac{2v_{\Delta}}{v_{\Phi}} \tan{\alpha}\Big)^{1/2} \Big( 1- \frac{2v_{\Delta}}{v_{\Phi}} \frac{1}{\tan{\alpha}}\Big)^{-1/2}M_{h^0}.
\end{equation}

This relationship holds for $\tan{\alpha} > \frac{2v_{\Delta}}{v_{\Phi}}$, and the constraint arising from the $\rho$ parameter bounds $v_{\Delta} < 8 \, \mathrm{GeV}$ \cite{Akeroyd:2012ms}.

\section{Methodology}
\label{sec:methodology}

\begin{figure*}[htb]
\centering
\includegraphics[width=0.75\linewidth]{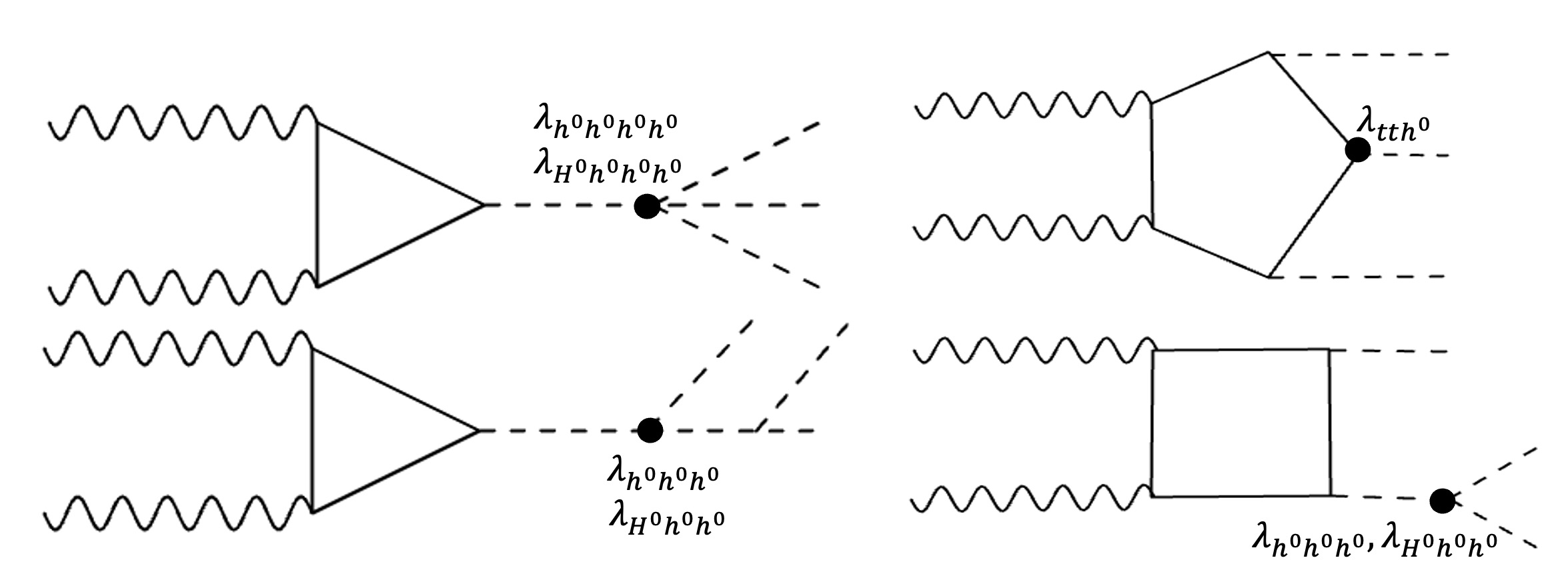} 
    \caption{These Feynman diagrams illustrate the loop-induced production of three Higgs bosons in both the SM and the HTM. The wavy lines represent collinear photons, the dashed lines indicate scalar bosons, and the solid lines depict fermions and $W^{\pm}$-bosons. In the SM, these processes are facilitated by SM fermions and $ W^{\pm}$-bosons. The HTM, in addition to the SM contributions, also includes contributions from charged Higgs bosons, as shown by the solid lines.
}
    \label{fig:feynmandiaghhh}
\end{figure*}

In both the SM and the HTM, the one-loop process
$\gamma\gamma \to h^0 h^0 h^0$ is mediated by triangle, box, and pentagon loop
topologies, together with bubble-type diagrams arising from quartic scalar
couplings. In the HTM, these diagrams are sensitive to contributions from fermions, $W^{\pm}$ bosons, and charged Higgs bosons within the loops, as illustrated in Fig.~\ref{fig:feynmandiaghhh}. Due to the large number of diagrams, here we display only generic Feynman diagrams for simplicity. If all the contributions mediated by $H^{\pm}$ and $H^{\pm\pm}$ are excluded, the remaining diagrams correspond to the SM case.

To perform these calculations, we begin by using \texttt{Mathematica} and \texttt{FeynRules}~\cite{Christensen:2008py, Alloul:2013bka} to generate the Universal FeynRules Output (UFO) model file for the HTM. The \texttt{FeynRules} model file employed in this work was previously utilized in Ref.~\cite{Samarakoon:2023crt}. The computation of triple Higgs production in the SM via $\gamma \gamma$ fusion is performed using \texttt{FormCalc}~\cite{Hahn:1998yk, Denner:1991kt, Hahn:2016ktb, Hahn:2004fe}, and \texttt{FeynArts}~\cite{Hahn:2000kx}, with minor modifications. However, the analytic expression for the scattering amplitude in the HTM is considerably complex, posing computational challenges. One major difficulty arises in implementing Breit-Wigner propagator replacements~\cite{Plehn:1996wb}, to regulate pole singularities, which significantly increases computational demands. Additionally, evaluating the partonic-level cross section, $\hat{\sigma}(\gamma \gamma \rightarrow h^0h^0h^0)^{HTM}$, in the HTM using \texttt{FormCalc} requires an extensive amount of computation time, making the process highly demanding.

The study of three Higgs production in the HTM involves significant complexity, requiring the algebraic reduction of over 5,400 Feynman diagrams at the one-loop level. Additionally, there is currently no existing Monte Carlo code capable of calculating the kinematic distributions for $\mu^+\mu^- \rightarrow \gamma \gamma \rightarrow h^0h^0h^0$ within the HTM framework. These challenges can be successfully addressed through the following approach.

The UFO file generated from the \texttt{FeynRules} model is exported to \texttt{GoSam-2.0}, which automates the calculation of one-loop amplitudes for processes involving multiple particles. The analytic expression for the scattering amplitude of $\gamma \gamma \rightarrow h^0h^0h^0$ is computed using \texttt{GoSam-2.0}, which integrates tools such as \texttt{QGRAF}~\cite{Nogueira:1991ex}, \texttt{FORM}~\cite{Vermaseren:2000nd}, \texttt{spinney}~\cite{Cullen:2010jv} and and \texttt{haggies}~\cite{Reiter:2009ts}. The amplitude involves loop integrals, which are reduced through the use of the \texttt{Ninja}~\cite{Mastrolia:2010nb}, \texttt{Golem95}~\cite{Binoth:2008uq, Cullen:2011kv, Guillet:2013msa}, and \texttt{Samurai}~\cite{Mastrolia:2010nb}, libraries. \texttt{Ninja} carried out the integral reduction through Laurent expansion, as detailed in Ref.~\cite{vanDeurzen:2013saa, Peraro:2014cba}. For further details, the author recommends consulting the \texttt{Gosam-2.0} manual in Ref.~\cite{GoSam:2014iqq}. After generating the large analytical expression, a numerical amplitude for this 2-to-3 process is produced using the provided inputs. To compute total cross-sections and kinematic distributions, we developed a Monte Carlo phase-space integration method \cite{Samarakoon:2025uyw} based on the \texttt{RAMBO} algorithm~\cite{Kleiss:1985gy}, which utilizes isotropic phase-space sampling~\cite{Platzer:2013esa, Kleiss:1985gy, Campbell:2017hsr}. 

The production of three Higgs bosons in photon--photon collisions arises as a
subprocess of $\mu^+\mu^-$ interactions at a muon collider. The corresponding
total cross section can be obtained from
\begin{equation}\label{fullXsec}
\sigma(s, h^0 h^0 h^0)
= \int_{\tau_{0}}^{1} d\tau\,
\frac{d\mathcal{L}_{\gamma\gamma}}{d\tau}\,
\hat{\sigma}\!\left(\hat{s} = \tau s,\; h^0 h^0 h^0 \right),
\end{equation}
where the photon luminosity is defined as
\begin{equation}\label{LuminosityFunc}
\frac{d\mathcal{L}_{\gamma\gamma}}{d\tau}
= \int_{\tau}^{1} \frac{dx}{x}\;
f_{\gamma/\mu}(x)\,
f_{\gamma/\mu}\!\left(\tfrac{\tau}{x}\right),
\end{equation}
with $\tau_{0}=(3 m_{H})^2/s$. $\sqrt{\hat{s}}$, $\sqrt{s}$, and $m_{H}$ denote the center-of-mass energies of the $\gamma\gamma$, $\mu^+\mu^-$ systems, and final-state Higgs boson mass, respectively. 

The distribution of photon energies radiated by a charged lepton of initial
energy $\mathcal{E}$, carrying a momentum fraction $x$, follows the
Weizsäcker–Williams spectrum~\cite{1934ZPhy...88..612W, PhysRev.45.729, Landau:1934zj}
and is well described by the Effective Photon Approximation (EPA)%
~\cite{Garosi:2023bvq, Han:2020uid, Han:2021kes, Delahaye:2019omf, Peskin:1995ev}.  
At leading logarithmic accuracy, the EPA spectrum is given by
\begin{equation}\label{eq:EPA}
f_{\gamma/\ell}(x)
\simeq \frac{\alpha_e}{2\pi}\,
P_{\gamma/\ell}(x)\,
\ln\!\left(\frac{\mathcal{E}^2}{m_\ell^2}\right),
\end{equation}
where the relevant splitting function is
\begin{equation}\label{eq:split1}
P_{\gamma/\ell}(x)
= \frac{1 + (1-x)^2}{x},
\qquad \ell \rightarrow \gamma.
\end{equation}
  
For consistency with existing SM results%
~\cite{Chiesa:2021qpr} and to facilitate comparison, we retain only the leading
logarithmic term of the photon splitting function.  
We note that non-logarithmic corrections can be sizable at a muon collider%
~\cite{Budnev:1975poe, Frixione:1993yw}, but they are not included in the
present work.

To validate our implementation, we computed the parton-level cross section
$\hat{\sigma}(hhh)$ for the process $\gamma\gamma \to h^0 h^0 h^0$ within the
SM and compared our results with those obtained using
\texttt{FormCalc}.  
For the SM calculation, we employed the \texttt{smdiag} model option in
\texttt{GoSam-2.0}.  
Excellent agreement was found between both codes.

\section{Results and Discussion}
\label{sec:results}

\begin{figure}[htb]
    \centering
    \includegraphics[width=1.01\linewidth]{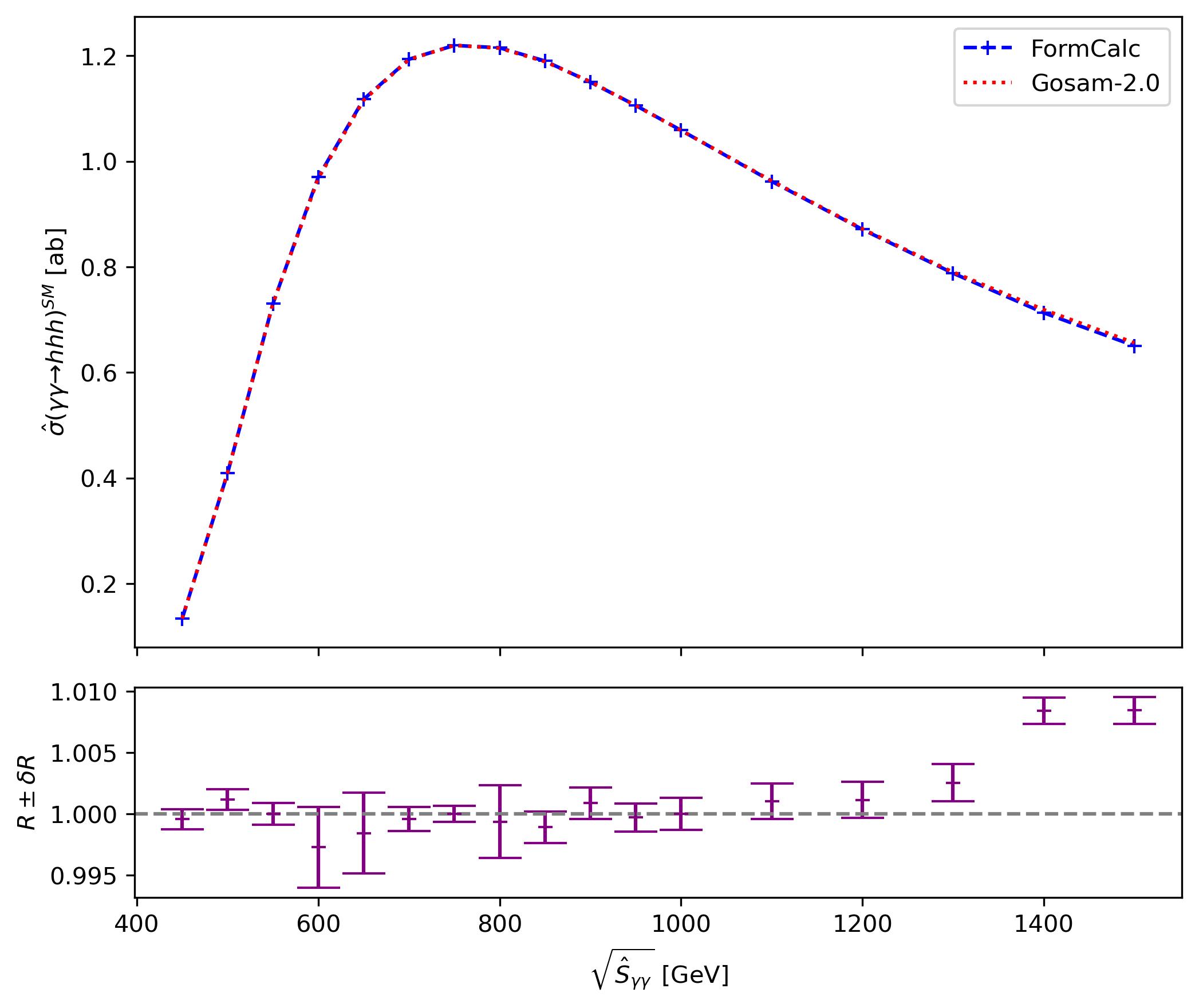} 
    \caption{Partonic cross-sections as a function of the center-of-energy of \(\gamma\gamma\), $\sqrt{\hat{s}_{\gamma \gamma}}$. This shows the results obtained from \texttt{FormCalc} and \texttt{Gosam-2.0.} The lower panel presents $R=\hat{\sigma}_{\texttt{FormCalc}}/\hat{\sigma}_{\texttt{Gosam-2.0}}$ with errors.
}
    \label{figure:parton_sm}
\end{figure}

Our goal is to produce predictions for the production of three light Higgs bosons in $\mu^+ \mu^-$ colliders via a $\gamma \gamma$ loop-induced process. In this section, we present the numerical results for triple Higgs production ($h^0h^0h^0$) via photon fusion in the HTM, focusing on cross-sections and kinematic distributions, and comparing them to SM predictions. In the previous section, we calculated the partonic cross-sections of $\gamma \gamma \rightarrow h^0h^0h^0$ within the range of $500 \, \mathrm{GeV} \leq \sqrt{\hat{s}_{\gamma \gamma}} \leq 1.5 \, \mathrm{TeV}$ in the SM as part of the code validation process. The partonic cross-section $\hat{\sigma}(\gamma \gamma \rightarrow h^0h^0h^0)$ reaches its maximum value at approximately $750 \, \mathrm{GeV}$. The results obtained, as shown in Fig.~\ref{figure:parton_sm}, from \texttt{FormCalc} and \texttt{GoSam-2.0} are in agreement with each other and also align with the calculations in Ref.~\cite{Chiesa:2021qpr}.
At higher center-of-mass energies ($\hat{s}_{\gamma\gamma} \gtrsim 1.3~\text{TeV}$), a small deviation of our \texttt{GoSam-2.0} results from the \texttt{FormCalc} predictions is observed, with the ratio
\begin{equation}
R = \frac{\sigma_{\text{GoSam}}}{\sigma_{\text{FormCalc}}}
\end{equation}
slightly exceeding unity by up to $\sim 0.5\%$. This deviation can be attributed primarily to numerical uncertainties inherent in the Monte Carlo integration over the large multi-particle phase space, as well as minor differences in the evaluation of one-loop amplitudes between the two codes. \texttt{GoSam-2.0} and \texttt{FormCalc} employ different tensor reduction and numerical strategies, which can produce small discrepancies in regions where the cross section varies rapidly or the loop integrals are numerically sensitive. 

Importantly, the observed deviation is within the estimated Monte Carlo error bars (see the lower panel of Fig.~\ref{figure:parton_sm}) and does not affect the overall behavior of the cross section. We have verified that increasing the number of integration points slightly reduces this deviation, confirming that it arises from numerical rather than physical effects. Therefore, the excellent agreement at lower and intermediate energies validates the correctness of our implementation.

\begin{figure}[htb]
    \centering
    \includegraphics[width=1.01\linewidth]{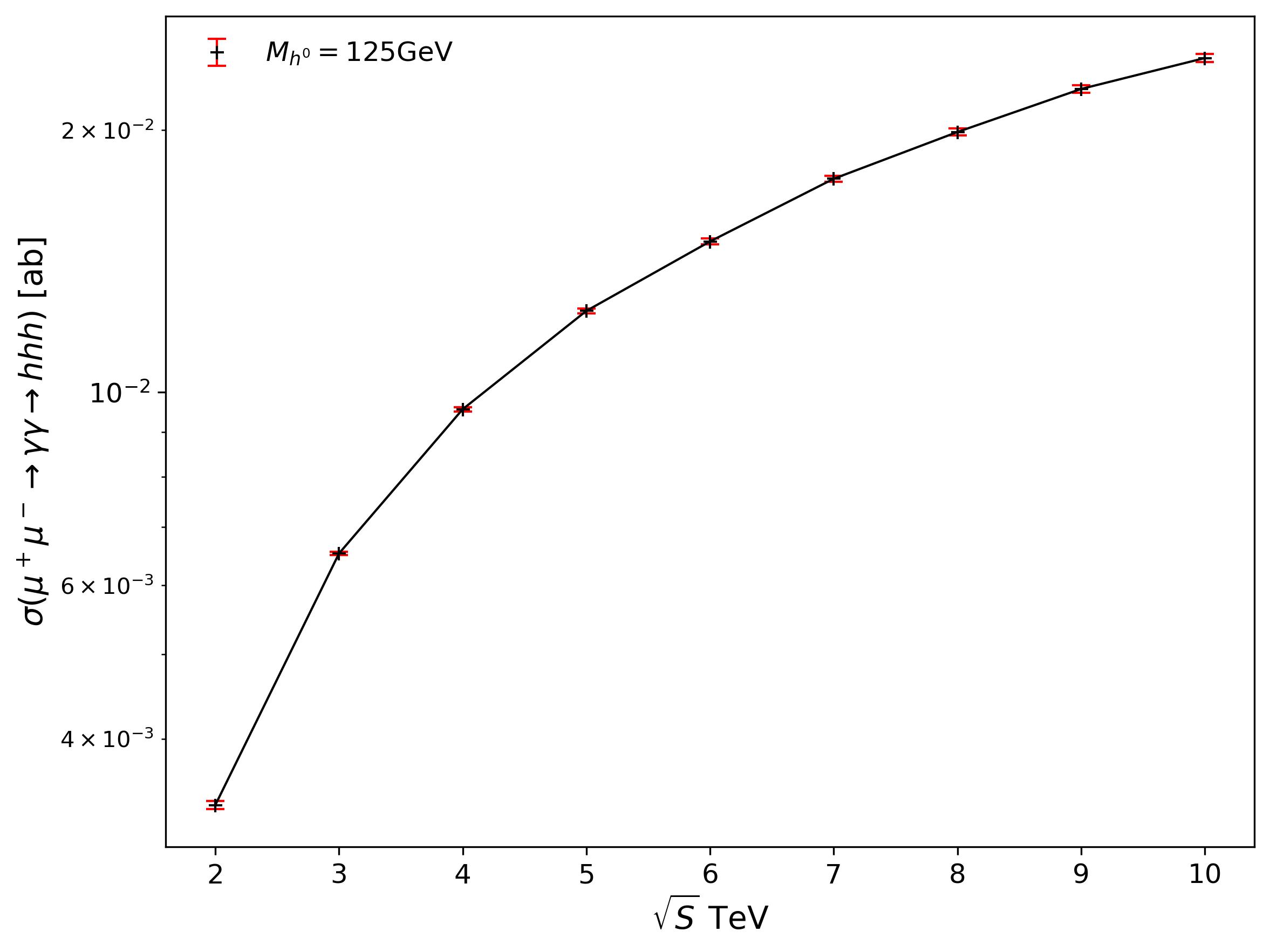} 
    \caption{The cross-section for $h^0h^0h^0$ production in a muon collider for the SM via photon fusion as a function of the center-of-mass energy of the muon collider. The results are obtained using the leading-order EPA.
}
    \label{figure:sm_crosshhh}
\end{figure}

\begin{figure}[htb]
    \centering
    \begin{subfigure}[b]{0.45\textwidth}
        \centering
        \includegraphics[width=\linewidth]{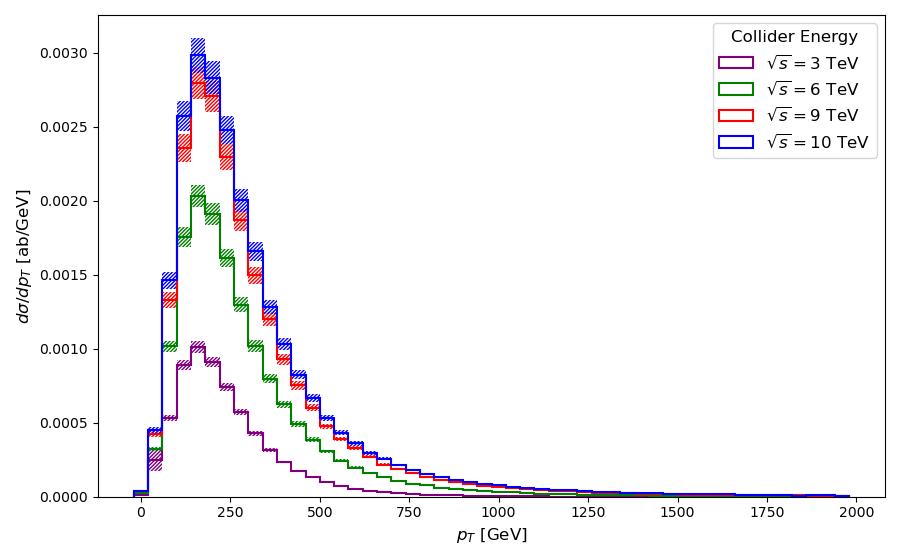}
        \caption{Higgs transverse-momentum distribution.}
        \label{fig:subfig1-1}
    \end{subfigure}
    \hfill
    \begin{subfigure}[b]{0.45\textwidth}
        \centering
        \includegraphics[width=\linewidth]{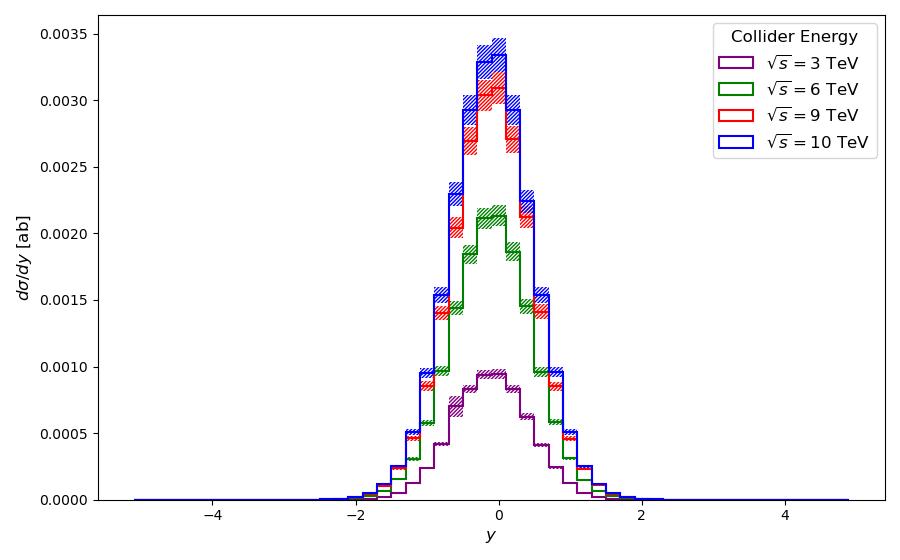}
        \caption{Higgs rapidity distribution.}
        \label{fig:subfig2-1}
    \end{subfigure}

    \caption{Higgs transverse-momentum of the hardest Higgs (a) and rapidity (b) distributions for three-Higgs production via photon-induced processes in $\mu^+ \mu^-$ collisions within the SM.}
    \label{fig:histo_sm_3tev}
\end{figure}

In the second step, we calculated the total cross-section for $\mu^+ \mu^- \rightarrow \gamma \gamma \rightarrow h^0h^0h^0$ and found that the results generated by \texttt{Gosam-2.0} do not agree with Ref.~\cite{Chiesa:2021qpr}. In Fig.~\ref{figure:sm_crosshhh}, the full cross sections for SM predictions are shown as a function of center of mass energy of the muon collider with the Monte Carlo errors. The cross-section in the SM ranges from $3.35 \times 10^{-3}$ to $2.42 \times 10^{-2}$~ab, as shown in Fig.~\ref{figure:sm_crosshhh}. With integrated luminosities between $0.4$ and $10$ ab$^{-1}$, the expected number of events is less than one. This implies that SM three-Higgs production via photon fusion is not observable at the center-of-mass energies considered here. In Fig.~\ref{fig:histo_sm_3tev}, we present the transverse momentum and rapidity distributions for the SM case, selecting events with transverse momentum $P_T > 0$ and rapidity $|y| < 5$. The distribution shows peaks at different collider energies, around 130 GeV.

\begin{table*}[htb]    
\centering
\caption{The singly charged Higgs and pseudoscalar masses were obtained using Eq.~(\ref{eq:MHpinput}) and Eq.~(\ref{eq:MAmax}), respectively, for both Scenario I and Scenario II. For the second scenario, the mass of the heavy Higgs was obtained from Eq.~(\ref{eq:MHmax}). Here we have dropped the \text{max} and \text{min} labels for simplicity. For Scenario I, we have chosen $\epsilon_P = 4.248 \times 10^{-3}~\mathrm{GeV}$, whereas for Scenario II, it is $3.2822 \times 10^{-3}~\mathrm{GeV}$ when evaluating the singly charged Higgs mass using Eq.~(\ref{eq:MHpmax}) and Eq.~(\ref{eq:MHpinput}).}
    \label{tab:scenariosIandII}
    \begin{tabular}{ccccccc}
        Scenario &$\alpha$& $M_{h^0}$ & $M_{H^0}$ & $M_{A^0}$ & $M_{H^{\pm\pm}}$ & $M_{H^{\pm}}$ \\ 
        \hline 
        I &0& $125$ & 376.500 & 376.512 & 290.5 & 336.262 \\ 
        II &$\pi/4$& 125 & 126.0204 & 125.515 & 240.0 & 191.511\\
    \end{tabular}
    
\end{table*}


\begin{table}
\caption{The calculated Lagrangian parameters are based on the inputs from Table \ref{tab:scenariosIandII}. The sign of the $\lambda_4$ parameter is a discriminator for the hierarchy between doubly and singly charged Higgs bosons.}
\label{tab:my_label}

\begin{tabular}{ccccccc}
        Scenario &$\lambda$& $\lambda_1$ & $\lambda_2$ & $\lambda_3$ & $\lambda_4$ & $\mu$ \\ 
        \hline 
        I &0.51& 2.7 & 5.61 & -5.61 & 1.89 & 3.31 \\ 
        II &0.52& 1.382 & 2.514 & -2.514 & -1.382 & 0.368 \\
\end{tabular}

\end{table}

In the HTM, at the one-loop level, the process $\gamma \gamma \rightarrow h^0h^0h^0$ is mediated by $H^{\pm}$ and $H^{\pm\pm}$, unlike in the SM. This process relies on the couplings $H^0h^0h^0$ and $H^0h^0h^0h^0$. The Feynman diagrams in Fig.~\ref{fig:feynmandiaghhh} show that the triangle diagrams are particularly sensitive to these couplings. Based on the two scenarios described in Sec.~\ref{sec:param}, we have established the input sets shown in Table \ref{tab:scenariosIandII}. In Scenario I, when the mixing angle between $h^0$ and $H^0$ is minimized, the amplitude is consequently reduced. The final states of the process will be SM-like Higgs bosons as $\alpha \rightarrow 0$, which is characteristic of the HTM. One main reason for the suppression of the amplitude is that the $H^0h^0h^0$ and $H^0h^0h^0h^0$ couplings are approaching their weak-coupling decoupling limits~\cite{Haber:2013mia}. This scenario is not favorable for detecting the heavy Higgs boson and examining its contributions. This issue can be addressed in Scenario I by setting the intermediate state on-shell and ensuring that the threshold $M_{H^0} \geq 3M_{h^0}$ is met. Unfortunately, these contributions are not maximally enhanced, resulting in a moderate increase in the cross sections due to the relatively weak couplings mentioned earlier. Nevertheless, this enhancement improves the likelihood of observing the heavy Higgs boson and examining its contributions, which is essential for gaining a more comprehensive understanding of the HTM.

However, we cannot overlook the contributions of $H^0$ and its couplings to neutral scalars. Therefore, we have established a second scenario (Scenario II) where $\alpha = \pi/4$. This scenario has some limitations, as it cannot reach large mass scales for the heavy Higgs. Our parameter scanning method shows that $M_{H^0}$ can reach up to approximately 150 GeV for $\cos{\alpha} \in [0, 0.9)$ at $v_{\Delta} = 8~\mathrm{GeV}$, as shown in panel (b) of Fig.~\ref{fig:panel3}. In the second scenario, we keep $v_{\Delta} = 1 \mathrm{GeV}$ and allow the heavy Higgs to be slightly heavier than the SM-like Higgs, as discussed in Ref.~\cite{Akeroyd:2012ms}. Although the decay $H^0 \rightarrow h^0h^0$ is kinematically forbidden, the $H^0h^0h^0$ and $H^0h^0h^0h^0$ couplings are significantly larger than those in Scenario I. The significant increase in the $H^0h^0h^0$ coupling in Scenario II compared to Scenario I, by a factor of $\sim 743$ (Table~\ref{tab:coup_comp}), allows $H^0$ to still appear as a virtual (off-shell~\cite{Goncalves:2017iub}) propagator in loop diagrams contributing to $\gamma \gamma \to h^0h^0h^0$. These couplings are much stronger due to the specific parameter settings, which enhance the interactions involving the heavy Higgs. This dramatic increase suggests that Scenario II could provide a more favorable environment for studying the effects and contributions of the heavy Higgs boson, despite its limitations in reaching larger mass scales. To analyze the scalar couplings in these calculations, we used the expressions presented in~\ref{app:appendixA}, which were evaluated in the cross-section calculations for each scenario.

Table~\ref{tab:coup_comp} provides a comprehensive comparison of the trilinear and quartic neutral scalar couplings between the SM and the two scenarios in the HTM with $\alpha = 0$ and $\alpha = \pi/4$. This comparison is crucial for understanding the variations in coupling strengths under different parameter settings. The table highlights how the $h^0h^0h^0$ and $h^0h^0h^0h^0$ couplings in Scenario I are closer to the SM values, while those in Scenario II show more substantial deviations. 

In Scenario I, the couplings $\lambda^{HTM}_{h^0h^0h^0h^0}$ and $\lambda^{HTM}_{h^0h^0h^0}$ are very close to the SM values ($0.98 \times \lambda_{h^0h^0h^0h^0}^{SM}$ and $0.98 \times \lambda_{h^0h^0h^0}^{SM}$, respectively), indicating minimal mixing between the doublet and triplet Higgs fields. In contrast, Scenario II shows significant deviations from the SM, with $\lambda^{HTM}_{h^0h^0h^0h^0}$ reduced to $\frac{1}{4} \times \lambda_{h^0h^0h^0h^0}^{SM}$ and $\lambda^{HTM}_{h^0h^0h^0}$ reduced to $0.353 \times \lambda_{h^0h^0h^0}^{SM}$.

In the triangle, pentagon, and bubble-type Feynman diagrams, which are sensitive to vertices that produce the squared couplings in the cross sections, the couplings $(\lambda_{h^0H^{\pm}H^{\mp}}^{HTM})^2$ and $(\lambda_{h^0H^{\pm\pm}H^{\mp\mp}}^{HTM})^2$ are approximately $22 \times (\lambda_{h^0h^0h^0}^{SM})^2$ and $12 \times (\lambda_{h^0h^0h^0}^{SM})^2$, respectively, in Scenario I. In contrast, these couplings are approximately $0.4 \times (\lambda_{h^0h^0h^0}^{SM})^2$ and $1.639 \times (\lambda_{h^0h^0h^0}^{SM})^2$ in Scenario II. In the box type diagrams, the doubly charged Higgs provides more contributions in the second scenario since $\lambda_{h^0H^{\pm\pm}H^{\mp\mp}}^{HTM}(\alpha = \pi/4) \approx 2^{1/4} \times \lambda_{h^0H^{\pm\pm}H^{\mp\mp}}^{HTM}(\alpha = 0)$ and $\lambda_{h^0H^{\pm}H^{\mp}}^{HTM}(\alpha = \pi/4) \approx \frac{1}{11} \times \lambda_{h^0H^{\pm}H^{\mp}}^{HTM}(\alpha = 0)$.

\begin{table*}
\centering
\caption{Comparison of trilinear and quartic neutral scalar couplings between the SM and the two scenarios in the HTM. The first three rows compare the scalar couplings between Scenarios I and II, while the last two rows show the comparison of these couplings in the scenarios with the SM couplings separately.}

    \label{tab:coup_comp}
    \begin{tabular}{ccc}
        Coupling Expression & Scenario I ($\alpha = 0$) & Scenario II ($\alpha = \pi/4$) \\
        \hline
        $\lambda_{hhhh}^{HTM}(\alpha)$ & $\approx 4 \times \lambda_{hhhh}^{HTM} (\alpha = \pi/4)$ & $\approx \frac{1}{4} \times \lambda_{hhhh}^{HTM} (\alpha = 0)$ \\
        $\lambda_{hhhH}^{HTM}(\alpha)$ & $\approx 0$ & $\gg \lambda_{hhhH}^{HTM} (\alpha = 0)$ \\
        $\lambda_{hhH}^{HTM}(\alpha)$ & $\approx 0$ & $743 \times \lambda_{hhH}^{HTM}(\alpha=0)$\\
        \hline\hline
        $\lambda_{hhhh}^{HTM}(\alpha)$ & $0.98 \times \lambda_{hhhh}^{SM}$ & $\frac{1}{4} \times \lambda_{hhhh}^{SM}$ \\
        $\lambda_{hhh}^{HTM}(\alpha)$ & $0.98 \times \lambda_{hhh}^{SM}$ & $0.353 \times \lambda_{hhh}^{SM}$ \\
    \end{tabular}
 
\end{table*}

In addition to trilinear and quartic scalar couplings, Feynman diagrams with box, triangle, and pentagon topologies are particularly sensitive to the Higgs couplings with fermions and $W^\pm$ bosons. Evaluating their contributions to the cross section is straightforward, as the coupling between the neutral Higgs boson and fermions, $h^0 f\bar{f}$, in the HTM is modified relative to the SM and is given by:

\begin{equation}\label{eq:ratiofermionsh}
\cos{\alpha} \left(1+\frac{2v_{\Delta}^2}{v_{\Phi}^2}\right)^{1/2}.
\end{equation}

Notably, in two scenarios, the coupling strength follows the relation:

\begin{equation}\label{eq:couhffsci}
\lambda^{HTM}_{h^0f\bar{f}}(\alpha = 0) \approx \sqrt{2} \times \lambda^{HTM}_{h^0f\bar{f}}(\alpha = \pi/4).
\end{equation}

Similarly, the Higgs coupling to $W^{\pm}W^{\mp}$ bosons in the HTM is given by:

\begin{equation}\label{eq:hwwcoupling}
\lambda_{h^0W^{\pm}W^{\mp}}^{HTM} = \frac{2M_W^2}{v^2} (\cos{\alpha} \, v_{\Phi} + 2\sin{\alpha} \, v_{\Delta}).
\end{equation}

which shows that:

\begin{equation}\label{eq:hwwsci}
\lambda^{HTM}_{h^0W^{\pm}W^{\mp}}(\alpha = 0) = \sqrt{2} \times \lambda^{HTM}_{h^0W^{\pm}W^{\mp}}(\alpha = \pi/4).
\end{equation}

From these expressions, it is evident that in the first scenario ($\alpha = 0$), both the Higgs-fermion and Higgs-gauge boson couplings are stronger compared to the second scenario ($\alpha = \pi/4$) and the SM.

\begin{figure}[htb]
    \centering
    \includegraphics[width=1.010\linewidth]{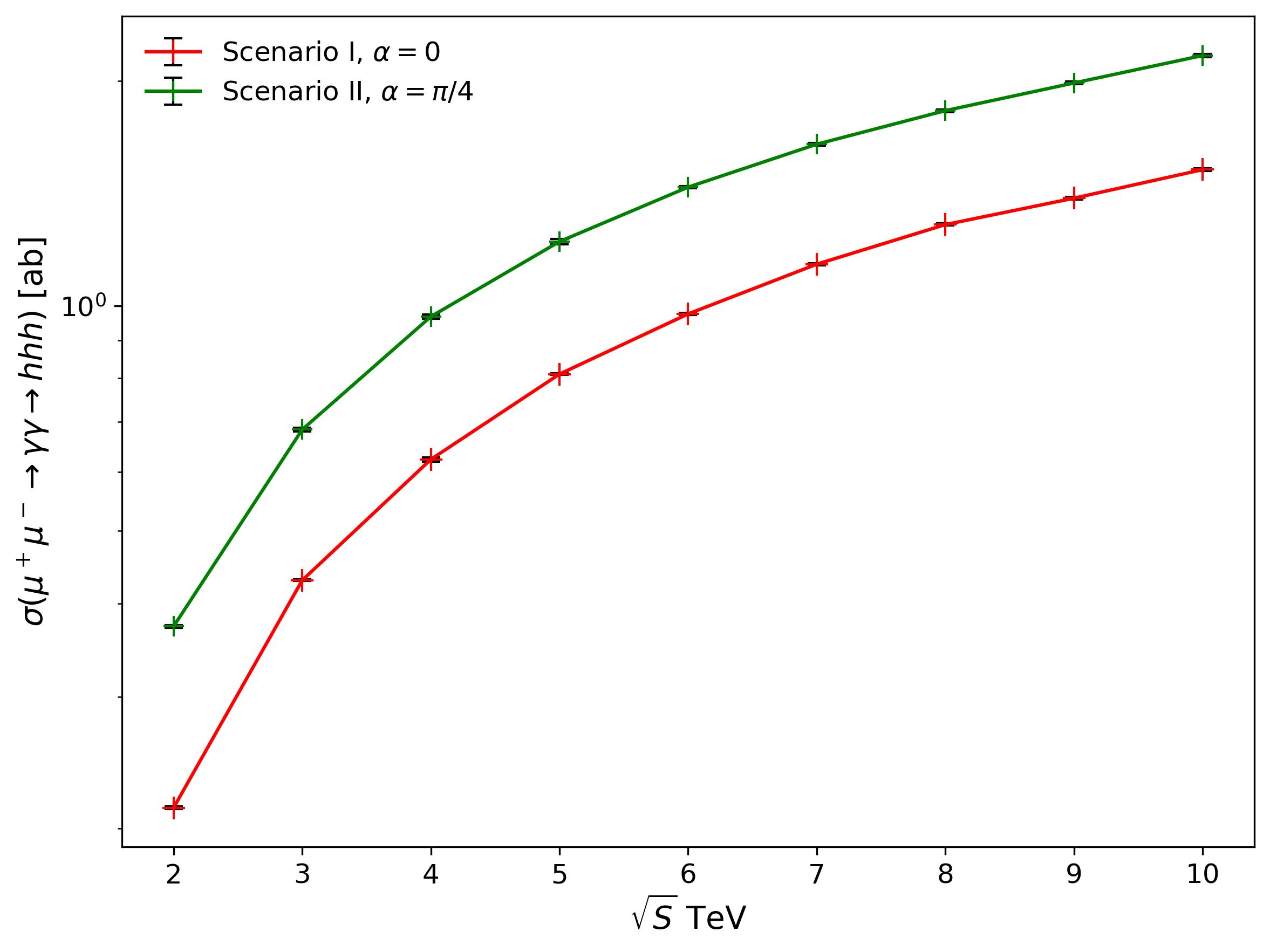} 
    \caption{The cross-section for $h^0h^0h^0$ production in a muon collider for the HTM via photon fusion as a function of the center-of-mass energy of the muon collider. The results are obtained using the leading-order EPA.
}
    \label{figure:htm_crosshhh}
\end{figure}

\begin{figure}[htb]
    \centering
    \includegraphics[width=1.01\linewidth]{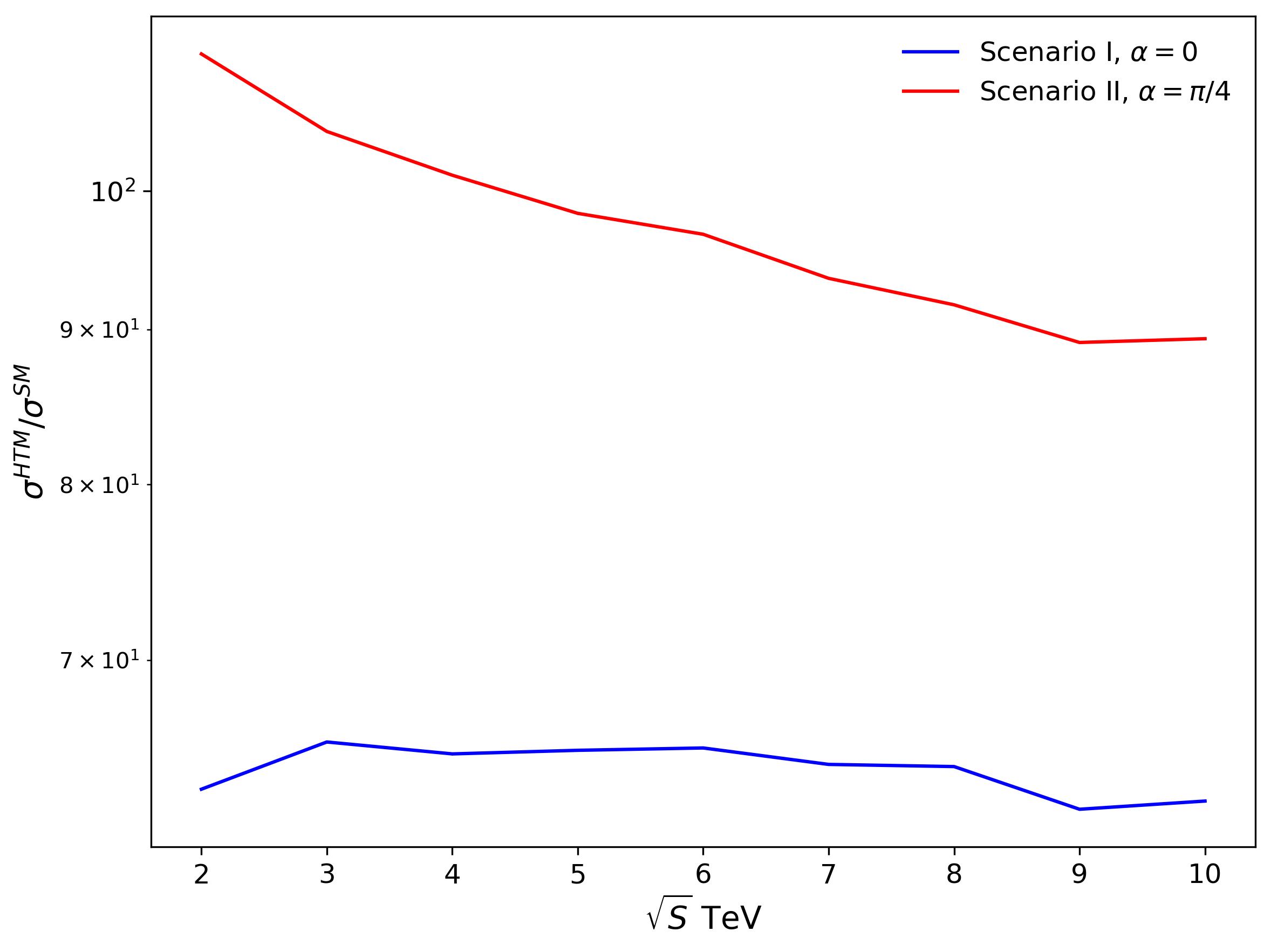} 
    \caption{The ratio of $\sigma^{HTM}/\sigma^{SM}$ as a function of center-of-mass energy shows a significant enhancement in the cross sections for both scenarios compared to the SM.
}
    \label{figure:sm_htm_comparison}
\end{figure}

The cross sections for the process $\mu^+ \mu^- \rightarrow \gamma \gamma \rightarrow h^0 h^0 h^0$ at a muon collider with $\sqrt{s} = 2-10$ TeV are shown in Fig.~\ref{figure:htm_crosshhh} for each scenario. Notably, the cross sections in Scenario II exhibit a significant enhancement compared to Scenario I.  

In Fig.~\ref{figure:sm_htm_comparison}, the cross sections for both scenarios are compared to the SM predictions. For Scenario I, the ratio of the HTM cross section to the SM cross section is in the range $50 < \sigma^{HTM}/\sigma^{SM} < 60$, whereas for Scenario II, this ratio increases to $90 < \sigma^{HTM}/\sigma^{SM} < 110$. This suggests that the coupling $\lambda_{H^0 h^0 h^0}$ plays a crucial role in enhancing the production cross section.  

\begin{figure}[htb]
    \centering
    \begin{subfigure}[t]{0.45\textwidth}
        \centering
        \includegraphics[width=\linewidth]{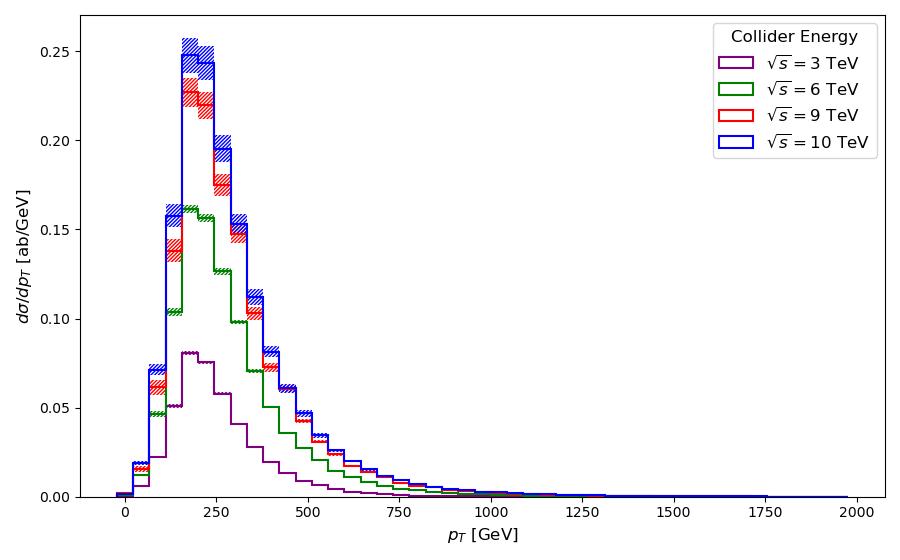}
        \caption{Higgs transverse-momentum distribution.}
        \label{fig:subfig1-2}
    \end{subfigure}

    \begin{subfigure}[t]{0.45\textwidth}
        \centering
        \includegraphics[width=\linewidth]{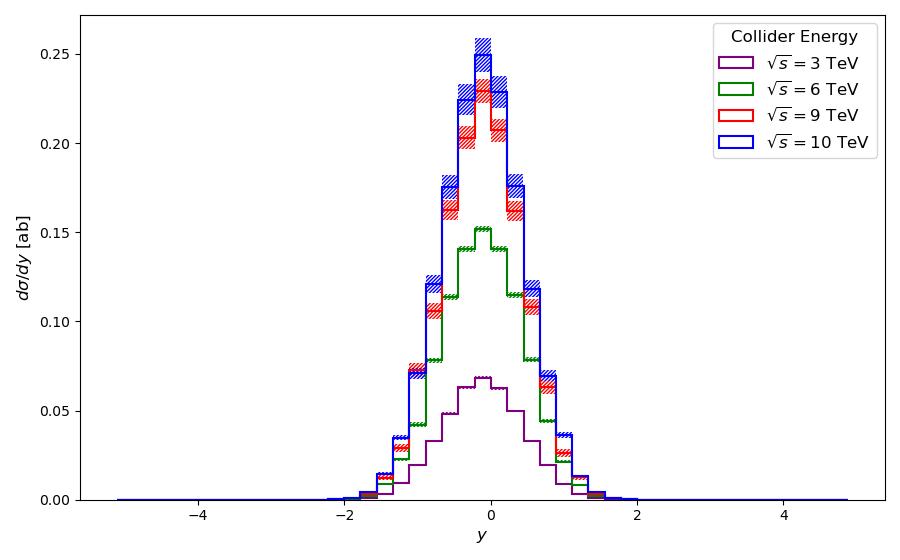}
        \caption{Higgs rapidity distribution.}
        \label{fig:subfig2-2}
    \end{subfigure}

    \caption{Higgs transverse-momentum of the hardest Higgs (a) and rapidity (b) distributions for three-Higgs production via photon-induced processes in $ \mu^+ \mu^- $ collisions within the HTM under scenario I.}
    \label{fig:histo_htm_scI}
\end{figure}

\begin{figure}[htb]
    \centering
    \begin{subfigure}[t]{0.45\textwidth}
        \centering
        \includegraphics[width=\linewidth]{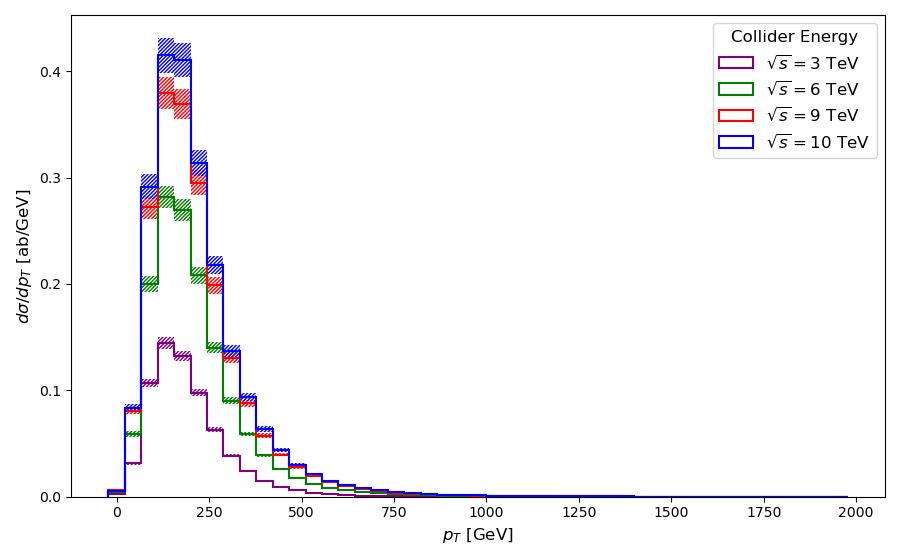}
        \caption{Higgs transverse-momentum distribution.}
        \label{fig:subfig1-3}
    \end{subfigure}

    \begin{subfigure}[t]{0.45\textwidth}
        \centering
        \includegraphics[width=\linewidth]{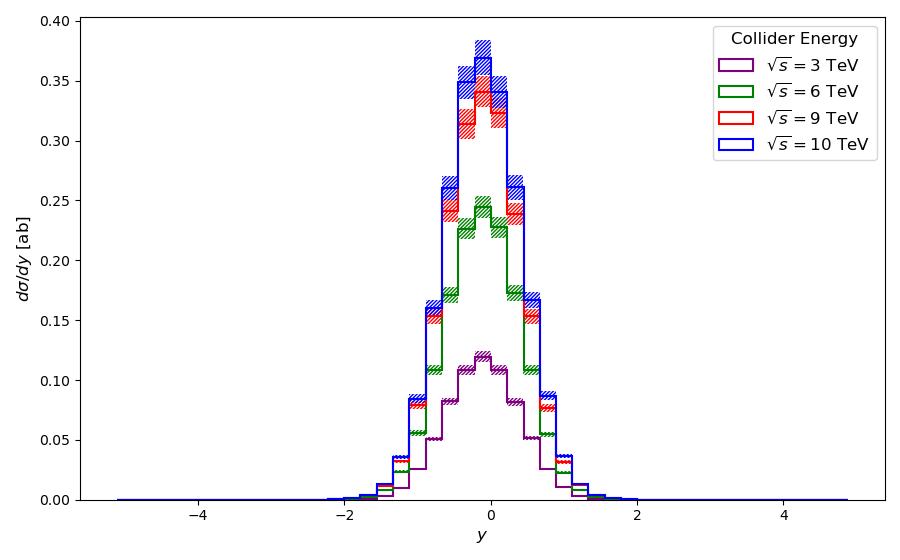}
        \caption{Higgs rapidity distribution.}
        \label{fig:subfig2-3}
    \end{subfigure}

    \caption{Higgs transverse-momentum of the hardest Higgs (a) and rapidity (b) distributions for three-Higgs production via photon-induced processes in $ \mu^+ \mu^- $ collisions within the HTM under scenario II}
    \label{fig:histo_htm_scII}
\end{figure}

In the HTM, Figs.~\ref{fig:histo_htm_scI} and~\ref{fig:histo_htm_scII} present the transverse momentum $P_T$ distributions of the Higgs boson with the highest $P_T$ ($P_T > 0$ and $|y| < 5$) for both scenarios I and II, at center-of-mass energies $\sqrt{s} = 3, 6, 9,$ and $10$ TeV. In Scenario II, the peak of the transverse momentum distribution is higher compared to both Scenario I and the SM. Meanwhile, the rapidity distributions exhibit a similar shape across different center-of-mass energies.  

Additionally, the charged Higgs mass hierarchies differ between the two scenarios. In Scenario~I, the hierarchy is $M_{H^\pm} > M_{H^{\pm\pm}}$, with the neutral Higgs masses approximately satisfying $M_{H^0} \approx M_{A^0}$. In contrast, Scenario~II exhibits a reversed hierarchy, where $ M_{H^\pm} < M_{H^{\pm\pm}}$ and $M_{H^0} > M_{A^0}$.

\begin{table}
\centering 
\caption{Event counts ($N$) for three Higgs production in $\mu^+ \mu^-$ colliders. The event count obtained by $N=\sigma\mathcal{L}$~\cite{Plehn:2009nd}.}
\label{table:events}

\begin{tabular}{cccc}
$\sqrt{s}~(\mathrm{TeV})$ & Luminosity $[\text{ab}^{-1}$] & $N_{\text{Scenario I}}$ & $N_{\text{Scenario II}}$ \\
\hline
2 & 0.4 & 0 & 0 \\
3 & 1 & 0 & 1 \\
4 & 2 & 2 & 2 \\
5 & 3 & 3 & 4 \\
6 & 4 & 4 & 6 \\
7 & 5 & 6 & 9 \\
8 & 7 & 9 & 13 \\
9 & 9 & 13 & 18 \\
10 & 10 & 16 & 22 \\
\end{tabular}

\end{table}

The observability of these scenarios at different muon collider energies is summarized in Table~\ref{table:events}. In Scenario I, the event rate is too low for detection at $\sqrt{s} = 2$ TeV and $\sqrt{s} = 3$ TeV, making it unobservable at these energies. However, in Scenario II, the process remains unobservable only at $\sqrt{s} = 2$ TeV, while higher energies yield a sufficient number of events for detection.

\section{Conclusions}
\label{sec:conclusions}
The study of multi-Higgs final states, such as $h^0h^0h^0$ production at $\mu^+\mu^-$ colliders, provides a means to probe the Higgs potential within the HTM and extract information about its underlying parameters and interactions, offering insights into physics beyond the SM. In the HTM, double and triple Higgs production involves trilinear and quartic charged Higgs couplings, which are absent in the SM and many BSM scenarios. Their contributions can be explored through the process $\gamma \gamma \rightarrow h^0h^0h^0$, and to investigate this, we have calculated the total cross sections and kinematic distributions for $\mu^-\mu^- \rightarrow \mu^+\mu^- h^0h^0h^0$, mediated by the one-loop-induced $\gamma \gamma \rightarrow h^0h^0h^0$ process using the EPA. 

Prior to performing calculations, we established our own parameter scanning process and Monte Carlo code using \texttt{Fortran} and \texttt{Gosam-2.0} for $\alpha \in [-\pi/2, \pi/2]$, where the input parameters for the Monte Carlo simulation are the masses of the scalar Higgs bosons. Our parameter scanning method determines the upper and lower bounds of $M_{H^{\pm}}$, $M_{A^0}$, and $M_{H^0}$ based on $M_{H^{\pm\pm}}$, $v_{\Delta}$ and $\alpha$. 

To modify \texttt{Gosam-2.0} for Monte Carlo simulations, we utilized the \texttt{RAMBO} algorithm and incorporated the EPA as a \texttt{Fortran} subroutine. Our calculations were performed by establishing two benchmark parameter sets (Scenario I and Scenario II), with a particular focus on the mixing angle $\alpha$ between the heavy Higgs and the SM-like Higgs. In Scenario I, we obtained cross sections and transverse momentum distributions, including rapidity dependence, for small values of $\alpha$, corresponding to the decoupling limit of $H^0h^0h^0$ and $H^0h^0h^0h^0$ interactions.

In Scenario II, with $\alpha = \pi/4$ and $M_{H^0} \sim M_{h^0}$, most trilinear scalar couplings are smaller than in Scenario I, except for the $h^0h^0h^0H^0$ and $h^0h^0H^0$ couplings, which remain sizable. According to Eqs.~(\ref{eq:ratiofermionsh}) and (\ref{eq:couhffsci}), the Higgs boson in this scenario is more fermiophobic than in Scenario I, with suppressed couplings to fermions relative to both Scenario I and the SM.

Despite this behavior, the dominance of heavy Higgs couplings to the SM-like Higgs in Scenario II leads to a significant enhancement in the production cross section, yielding $\sigma^{HTM}/\sigma^{SM} > 90$, whereas in Scenario I, we find $\sigma^{HTM}/\sigma^{SM} > 48$. For an integrated luminosity given by $\mathcal{L} \sim (\sqrt{s}/10~\mathrm{TeV})^2\times10~\mathrm{ab}^{-1}$, ranging from $2$ to $10~\mathrm{ab}^{-1}$, all scenarios in the HTM predict more than one observable event in muon collider experiments, whereas the SM prediction remains undetectable within this center-of-mass energy range.


\begin{acknowledgements}
We would like to express our sincere gratitude to Thomas Hahn at the Max-Planck-Institut für Physik for his invaluable assistance in debugging the \texttt{FormCalc} code. We also extend our appreciation to the BeoShock High-Performance Computing Service at Wichita State University (WSU) for providing free access to their computing cluster, which enabled us to perform all the calculations. The total CPU time for the computations exceeded 500 hours. We would like to extend our gratitude to Daniel Grady at WSU for his invaluable discussions, and to the Physics Division of the Department of Mathematics, Statistics, and Physics at WSU for their travel support. Finally, we thank the Academic Affairs at WSU for providing funding for the HPC Graduate Assistantship held by Bathiya Samarakoon.
\end{acknowledgements}

\appendix

\section{SCALAR HIGGS COUPLINGS IN TERMS OF $\lambda$, $\lambda_i$s and VEVs}
\label{app:appendixA}
The following expressions represent the scalar couplings evaluated in the cross-section calculations for Scenario I.

\begin{equation}\label{eq:threehiggshhhh}
\lambda_{h^0h^0h^0h^0}^{HTM}  = 3\frac{\lambda}{2}
\end{equation}

\begin{equation}\label{eq:threehiggsHHHH}
\lambda_{h^0h^0h^0H^0}^{HTM} \approx 0
\end{equation}

\begin{equation}\label{eq:threehiggshhh}
\lambda_{h^0h^0h^0}^{HTM} = 3\frac{\lambda v_{\Phi}}{2}
\end{equation}

\begin{equation}\label{eq:threehiggsHhh}
\lambda_{h^0h^0H^0}^{HTM} \approx \sqrt{2}\mu
-(\lambda_1+\lambda_4)v_{\Delta}
\end{equation}

\begin{equation}\label{eq:threehiggshHH}
\lambda_{h^0H^{\pm}H^{\pm}}^{HTM} \approx -(\lambda_1+\lambda_4/2)v_{\Phi}-2\sqrt{2}\mu \frac{v_{\Delta}}{v_{\Phi}}
\end{equation}

\begin{equation}\label{eq:threehiggsHHHpp}
\lambda_{h^0H^{\pm\pm}H^{\pm\pm}}^{HTM} \approx \lambda_1 v_{\Phi}
\end{equation}

\begin{equation}\label{eq:threehiggshhHplusHplus}
\lambda_{h^0h^0H^{\pm}H^{\pm}}^{HTM} \approx -\left( \lambda_1 + \frac{\lambda_4}{2} \right)
\end{equation}

\begin{equation}\label{eq:threehiggshhHppHpp}
\lambda_{h^0h^0H^{\pm\pm}H^{\pm\pm}}^{HTM} \approx  \lambda_1
\end{equation}

The following expressions represent the scalar couplings evaluated in the cross-section calculations for Scenario II.

\begin{equation}\label{eq:threehiggshhhh}
\lambda_{h^0h^0h^0h^0}^{HTM}  = 3\frac{\lambda}{8}+\frac{3}{2}(\lambda_1+\lambda_2+\lambda_3+\lambda_4)
\end{equation}

\begin{equation}\label{eq:threehiggsHHHH}
\lambda_{h^0h^0h^0H^0}^{HTM} \approx \frac{3}{8}\lambda-\frac{3}{2}(\lambda_2+\lambda_3)
\end{equation}

\begin{align}\label{eq:threehiggshhh}
\lambda_{h^0h^0h^0}^{HTM} = -\frac{1}{8}\Big\{12\mu-3\sqrt{2}(\lambda+2\lambda_1+2\lambda_4)\Big\}v_{\Phi}\nonumber\\+\frac{3}{2\sqrt{2}}\Big\{ \lambda_1+\lambda_4+2(\lambda_2+\lambda_3)\Big\}v_{\Delta}
\end{align}

\begin{align}\label{eq:threehiggsHhh}
\lambda_{h^0h^0H^0}^{HTM} = \frac{1}{8}\Big\{-4\mu+\sqrt{2}(3\lambda-2\lambda_1-2\lambda_4)\Big\}v_{\Phi}\nonumber\\+\frac{1}{2\sqrt{2}}\Big\{ \lambda_1+\lambda_4-6(\lambda_2+\lambda_3)\Big\}v_{\Delta}
\end{align}

\begin{align}
    \label{eq:threehiggshhh}
\lambda_{h^0H^{\pm}H^{\pm}}^{HTM} = -\frac{1}{\sqrt{2}}\left( \lambda_1 + \frac{\lambda_4}{2} \right) v_{\Phi}\nonumber\\+ \Big\{ \frac{-2(\lambda_2+\lambda_3)+\lambda_4}{\sqrt{2}}-\frac{\mu}{v_{\Phi}} \Big\}v_{\Delta}
\end{align}

\begin{equation}\label{eq:threehiggshHH}
\lambda_{h^0H^{\pm\pm}H^{\pm\pm}}^{HTM} = -
\frac{\lambda_1}{\sqrt{2}}v_{\Phi}-2\sqrt{2}\lambda_2v_{\Delta}
\end{equation}

\begin{equation}\label{eq:threehiggshhHpHp}
\lambda_{h^0h^0H^{\pm}H^{\pm}}^{HTM} = 
-\frac{1}{4} \left( 2\lambda_1+4(\lambda_2+\lambda_3)-\lambda_4 \right)+\frac{\lambda_4 v_{\Delta}}{v_{\Phi}}
\end{equation}

\begin{figure*}[tb]
    \centering
    \begin{subfigure}{0.45\textwidth}
        \includegraphics[width=\linewidth]{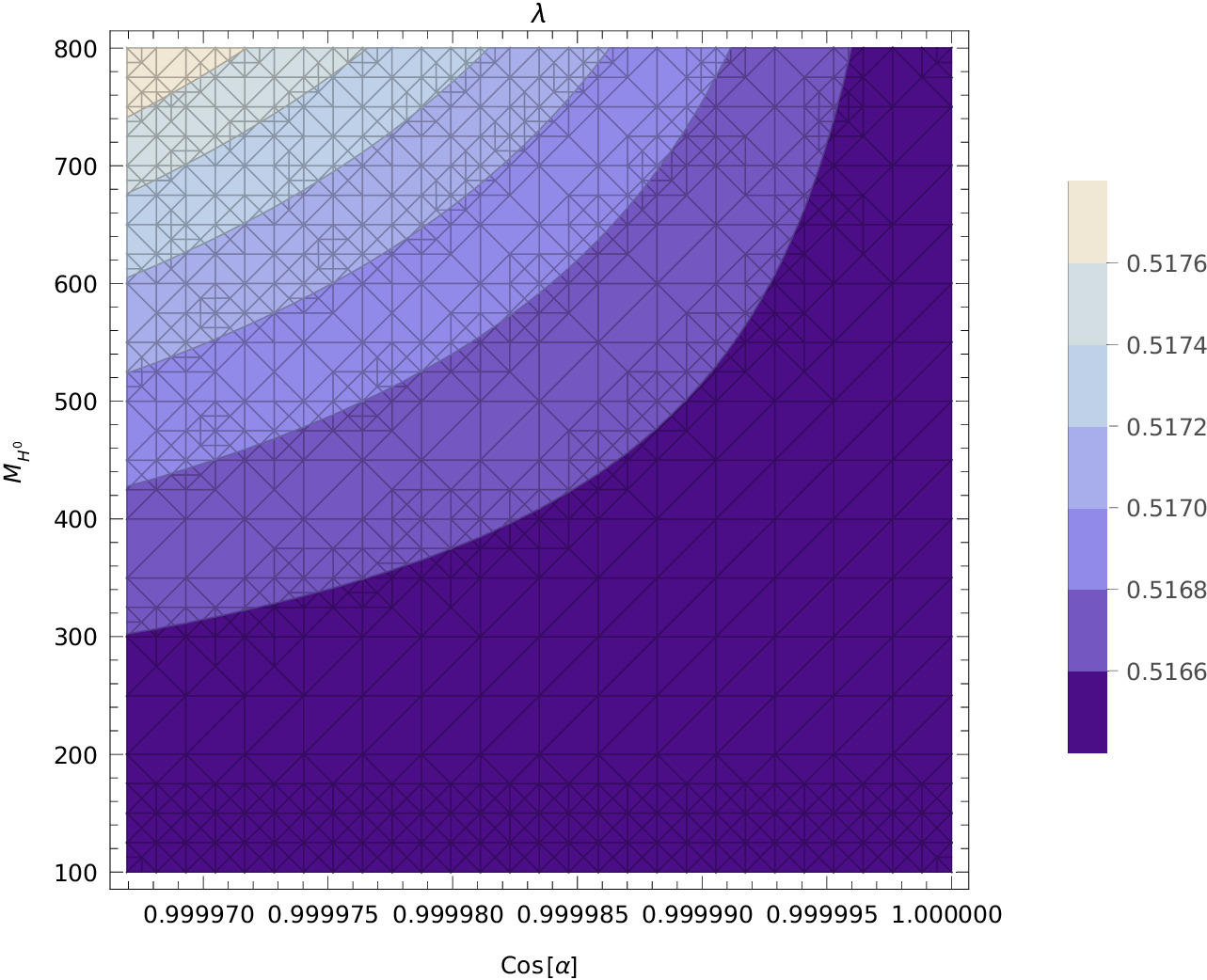}
        \caption{}
        \label{fig:subfig1-4}
    \end{subfigure}%
    \hspace{0.05\textwidth}
    \begin{subfigure}{0.45\textwidth}
        \includegraphics[width=\linewidth]{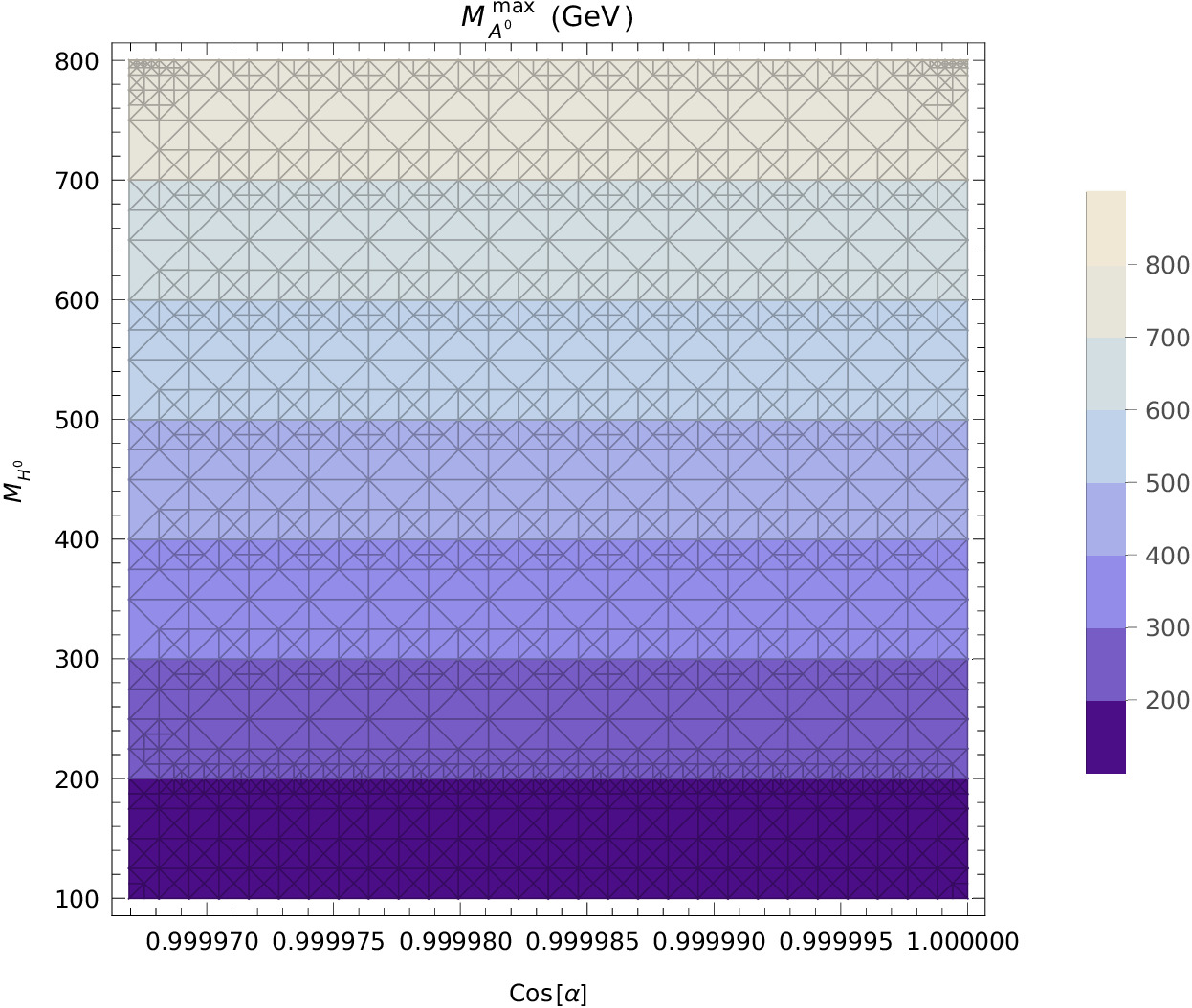}
        \caption{}
        \label{fig:subfig2-4}
    \end{subfigure}

    \begin{subfigure}{0.45\textwidth}
        \includegraphics[width=\linewidth]{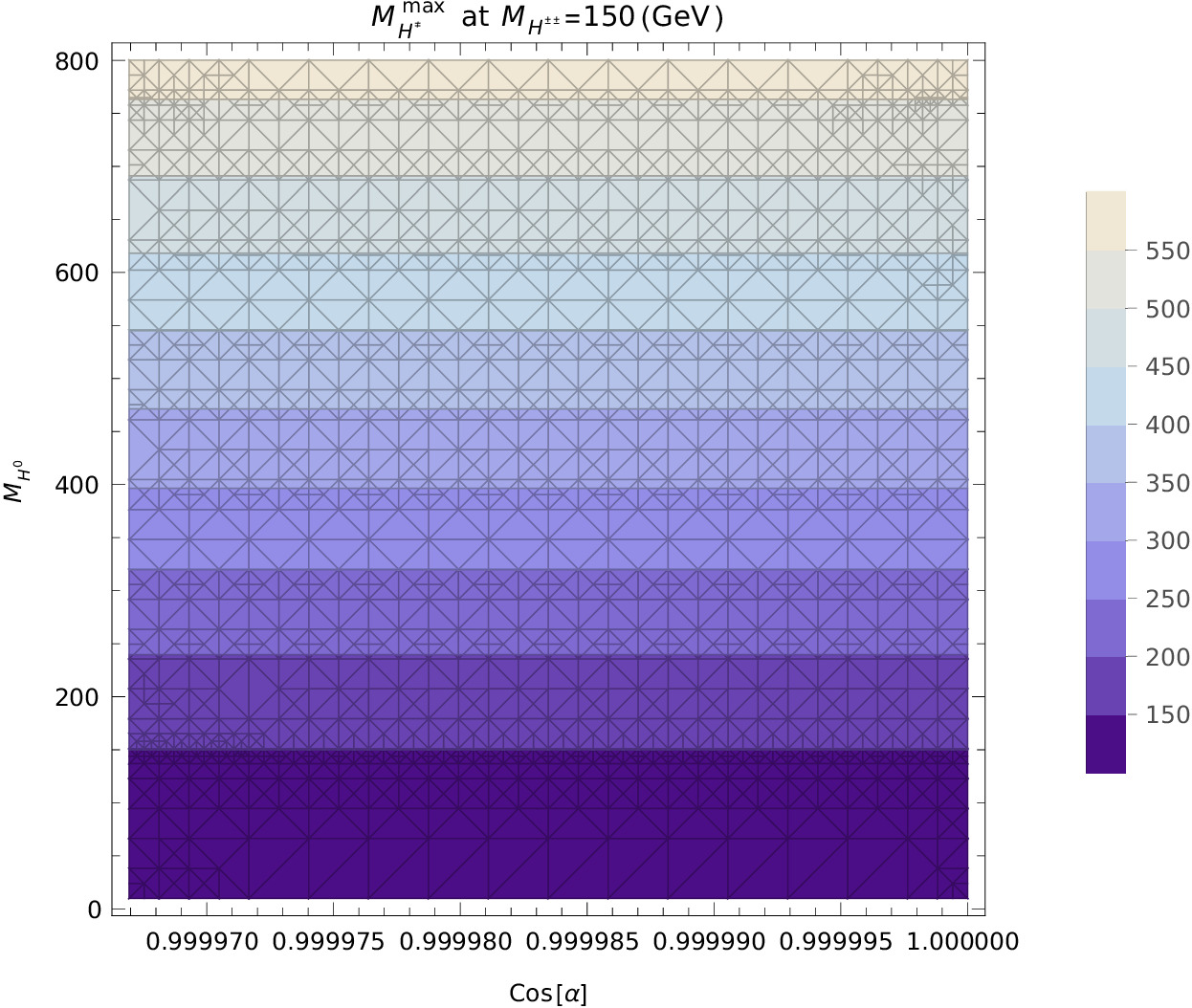}
        \caption{}
        \label{fig:subfig3-4}
    \end{subfigure}%
    \hspace{0.05\textwidth}
    \begin{subfigure}{0.45\textwidth}
        \includegraphics[width=\linewidth]{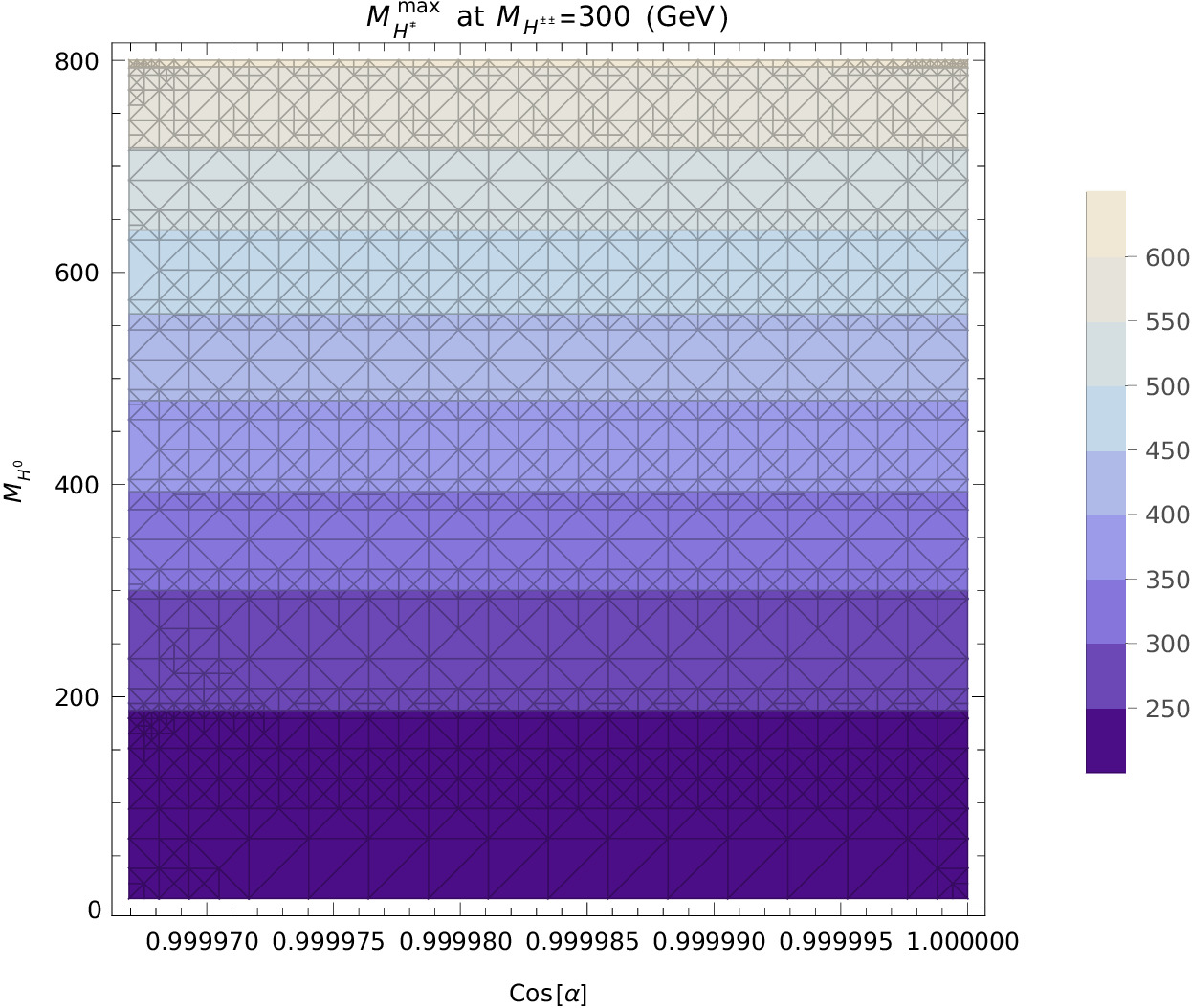}
        \caption{}
        \label{fig:subfig4-4}
    \end{subfigure}

    \begin{subfigure}{0.45\textwidth}
        \includegraphics[width=\linewidth]{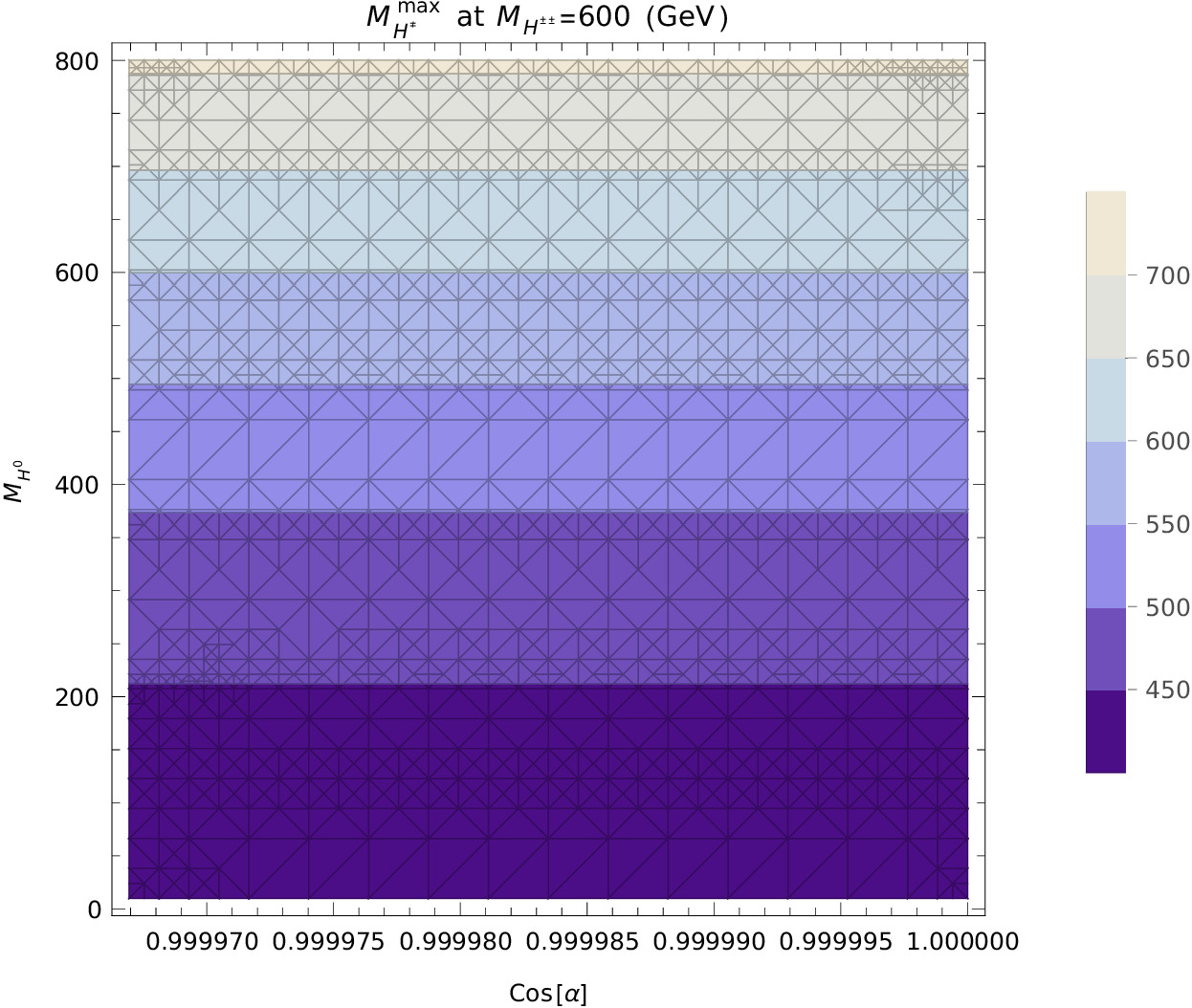}
        \caption{}
        \label{fig:subfig5-4}
    \end{subfigure}%
    \hspace{0.05\textwidth}
    \begin{subfigure}{0.45\textwidth}
        \includegraphics[width=\linewidth]{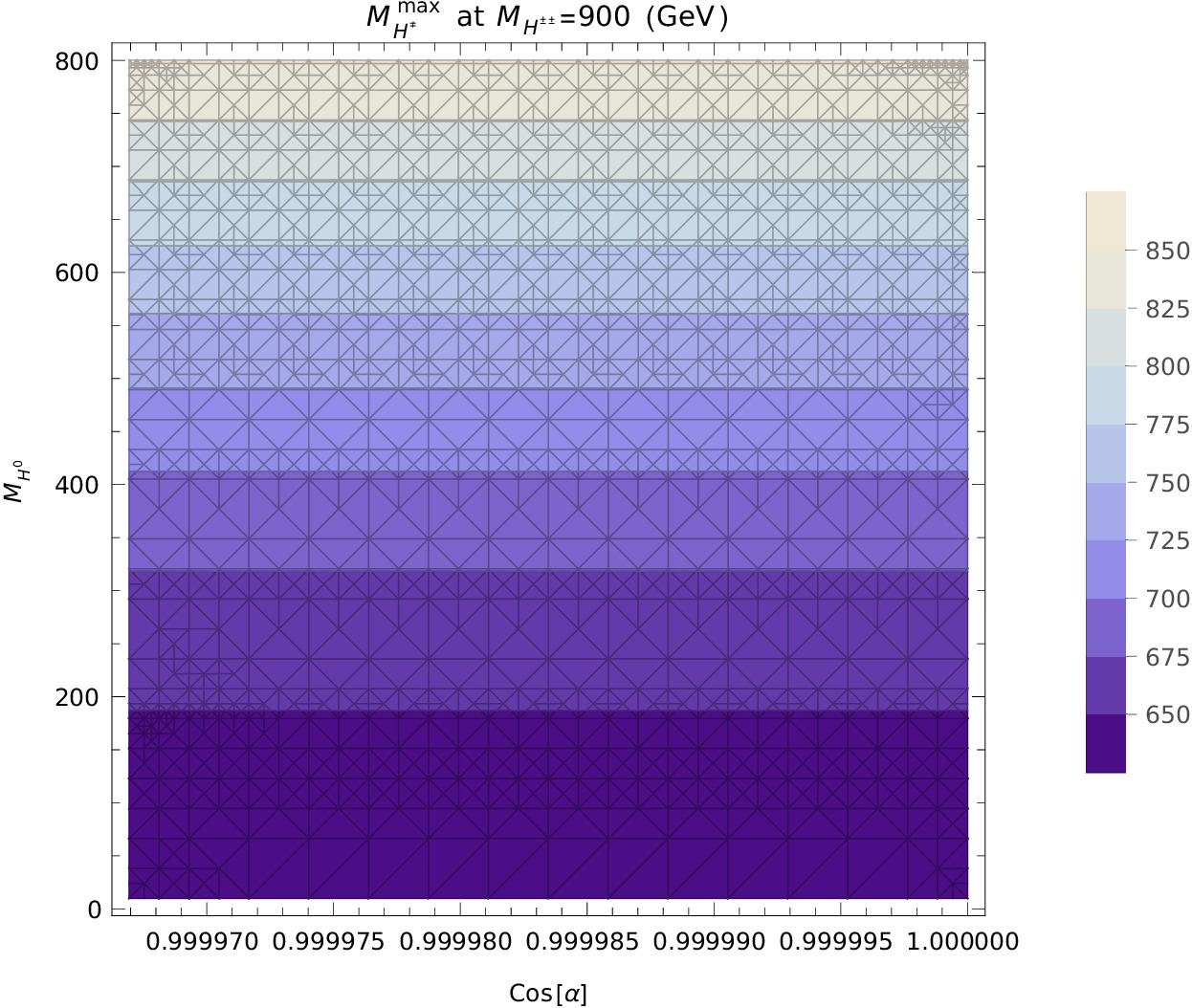}
        \caption{}
        \label{fig:subfig6-4}
    \end{subfigure}

    \caption{This figure presents the parameter \(\lambda\) (panel (a)), \(\mathrm{max}(M_{A^0})\) (panel (b)), and \(M_{H^{\pm}}\) (panels (c), (d), (e), and (f)) as functions of \(M_{H^0}\) and \(\cos{\alpha}\rightarrow 1\), with \(M_{h^0} = 125\,\mathrm{GeV}\). This is more relevant to Scenario~I.}
    \label{fig:panel1}
\end{figure*}

\begin{figure*}[tb]
    \centering
    \begin{subfigure}{0.45\textwidth}
        \includegraphics[width=\linewidth]{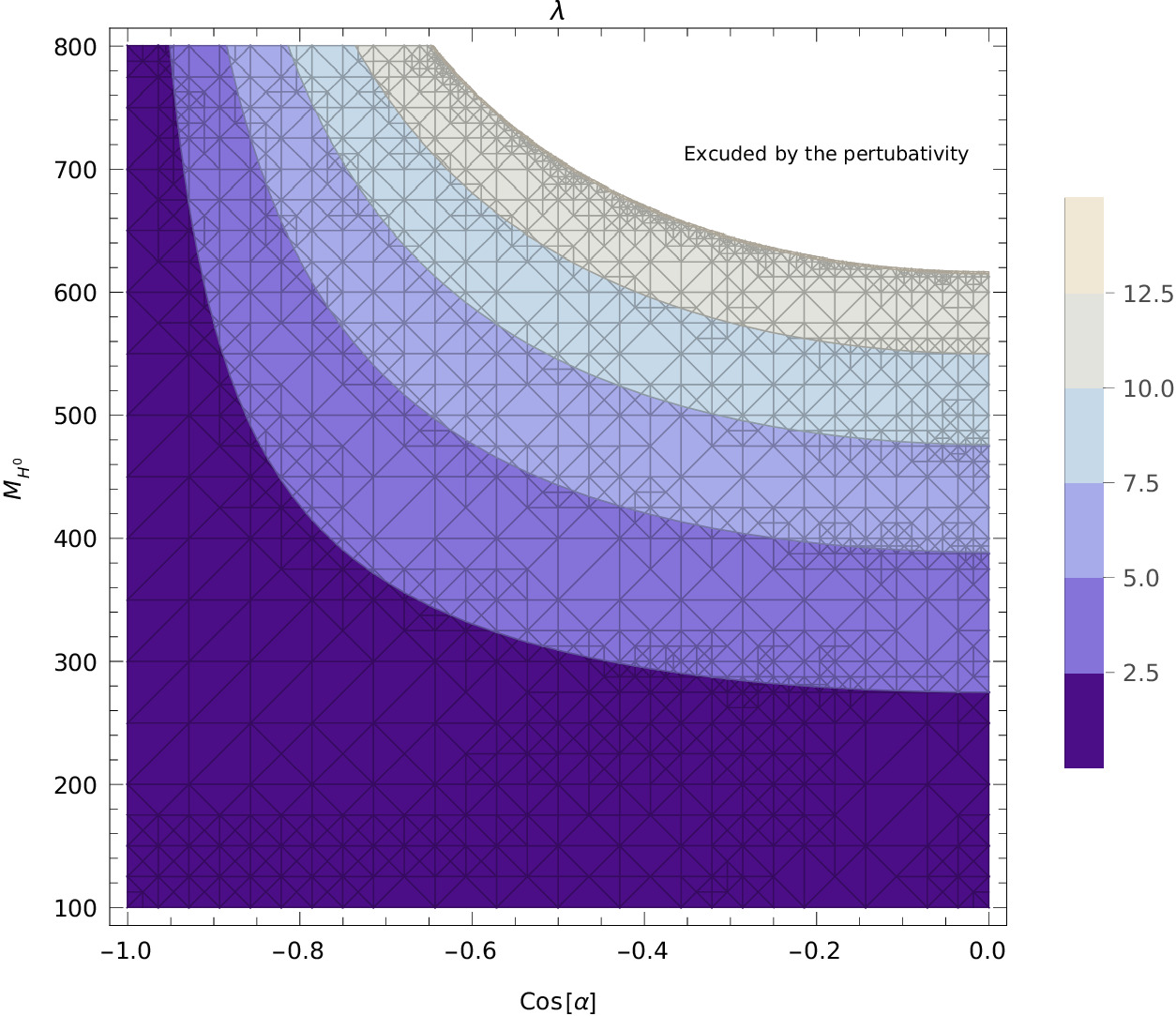}
        \caption{}
        \label{fig:subfig1-6}
    \end{subfigure}%
    \hspace{0.05\textwidth}
    \begin{subfigure}{0.45\textwidth}
        \includegraphics[width=\linewidth]{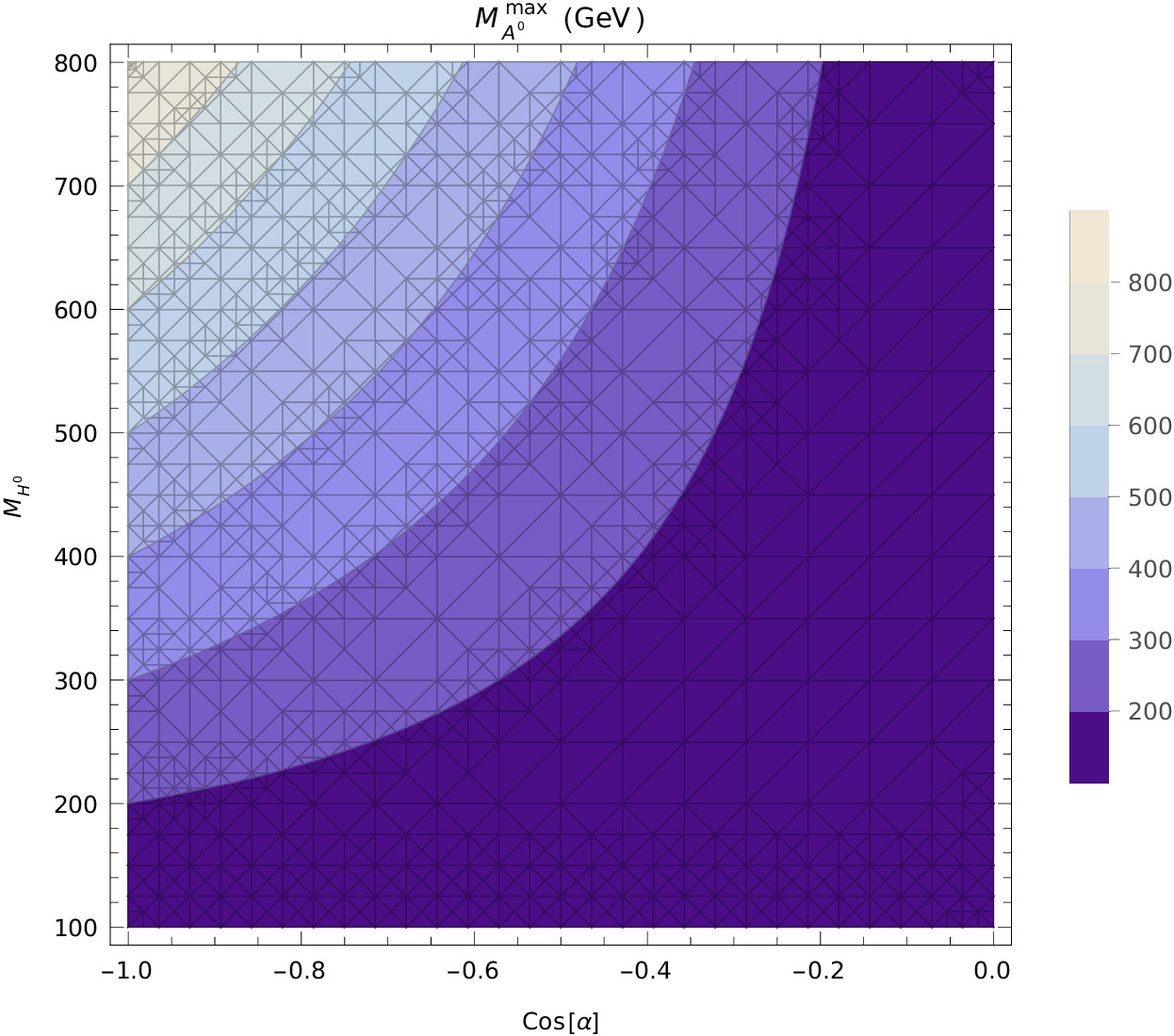}
        \caption{}
        \label{fig:subfig2-6}
    \end{subfigure}

    \begin{subfigure}{0.45\textwidth}
        \includegraphics[width=\linewidth]{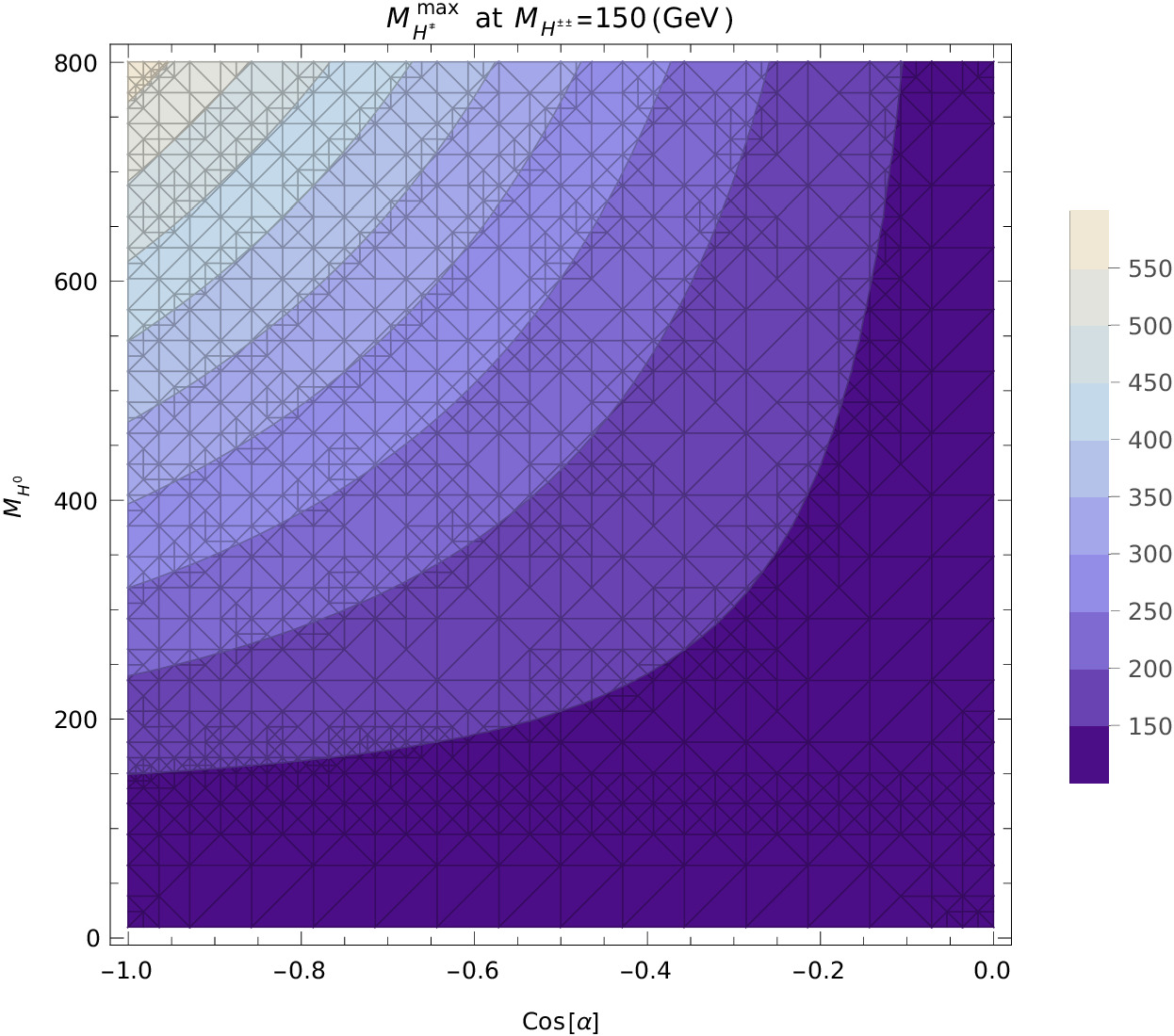}
        \caption{}
        \label{fig:subfig3}
    \end{subfigure}%
    \hspace{0.05\textwidth}
    \begin{subfigure}{0.45\textwidth}
        \includegraphics[width=\linewidth]{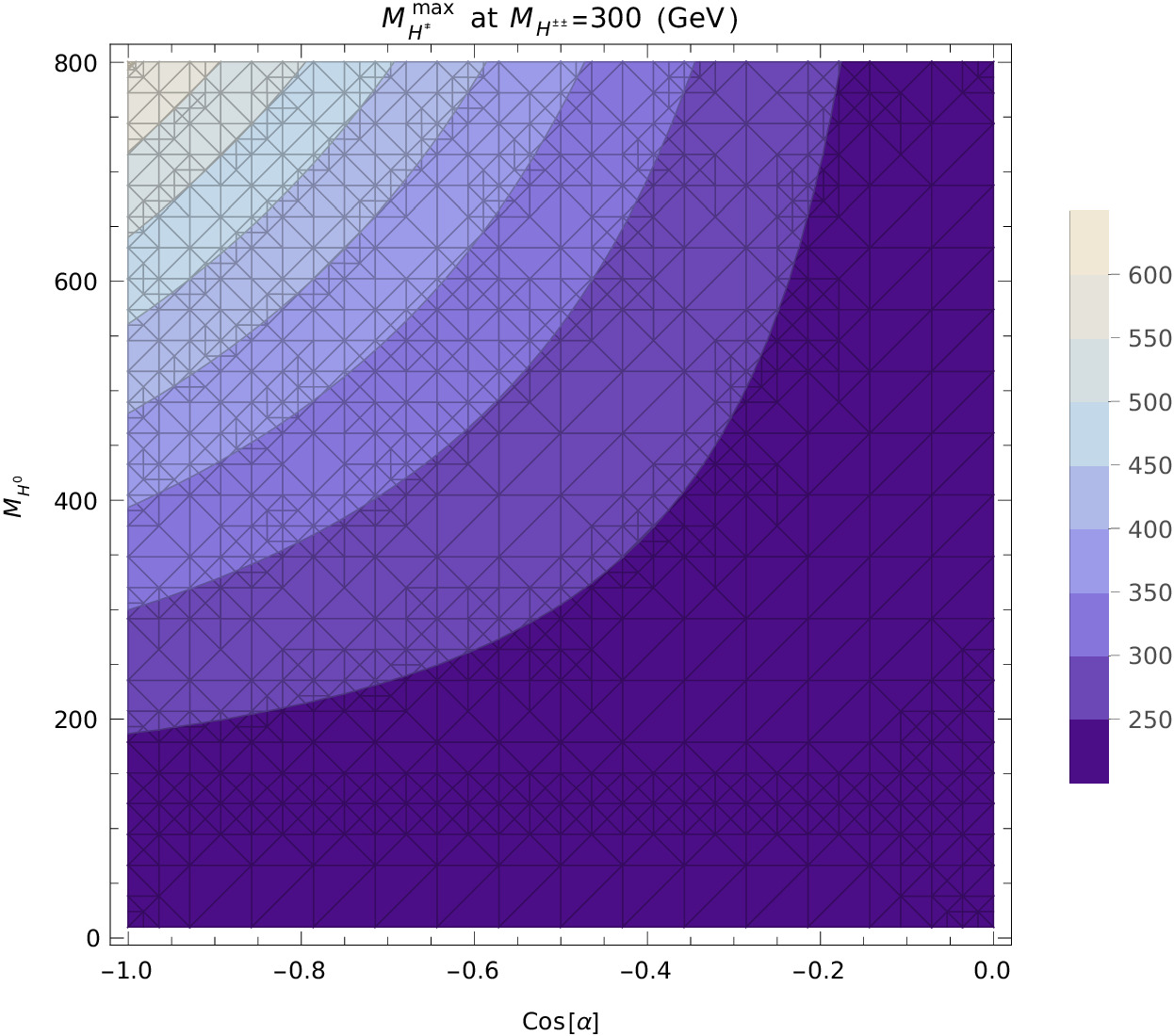}
        \caption{}
        \label{fig:subfig4-6}
    \end{subfigure}

    \begin{subfigure}{0.45\textwidth}
        \includegraphics[width=\linewidth]{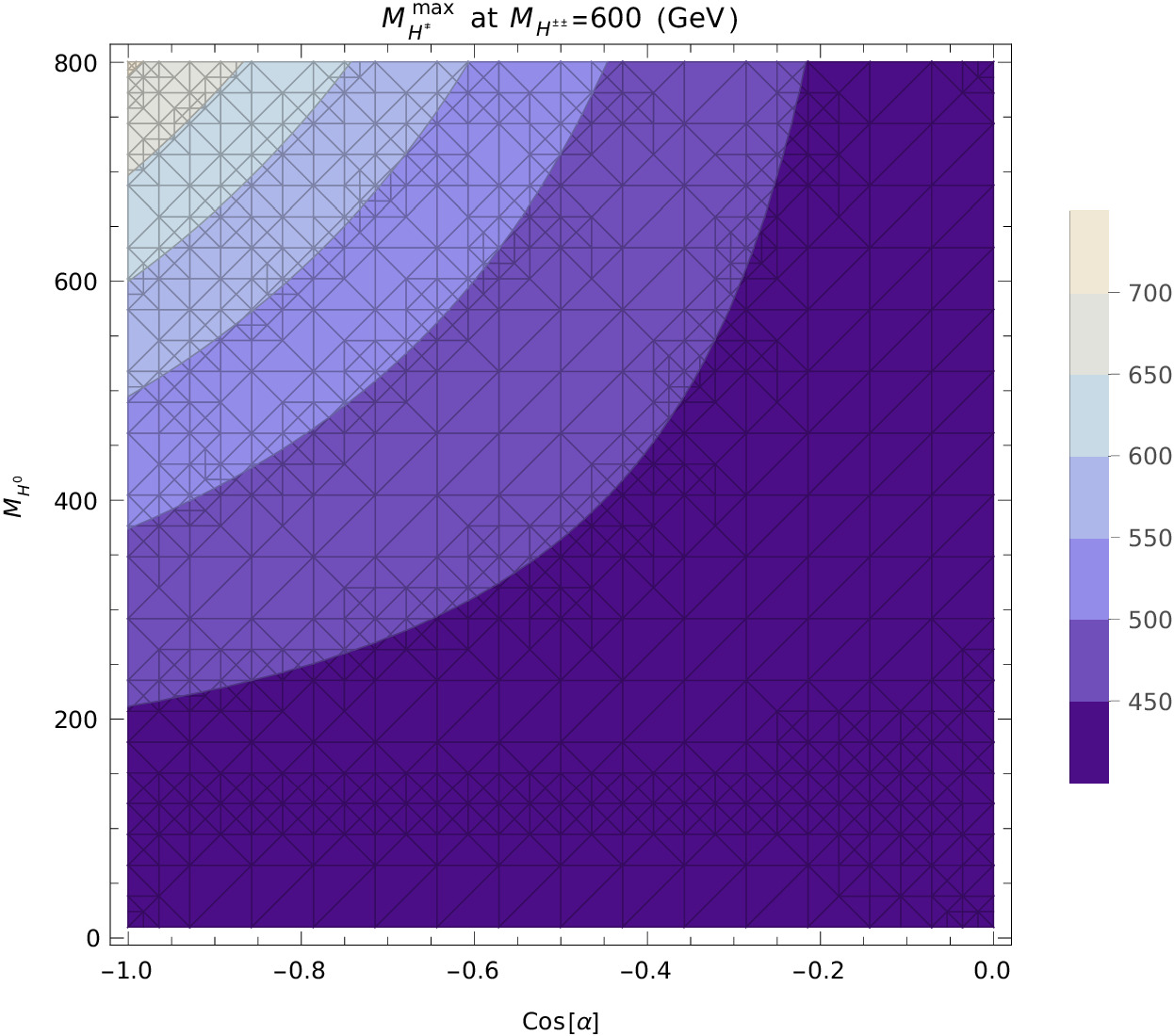}
        \caption{}
        \label{fig:subfig5-6}
    \end{subfigure}%
    \hspace{0.05\textwidth}
    \begin{subfigure}{0.45\textwidth}
        \includegraphics[width=\linewidth]{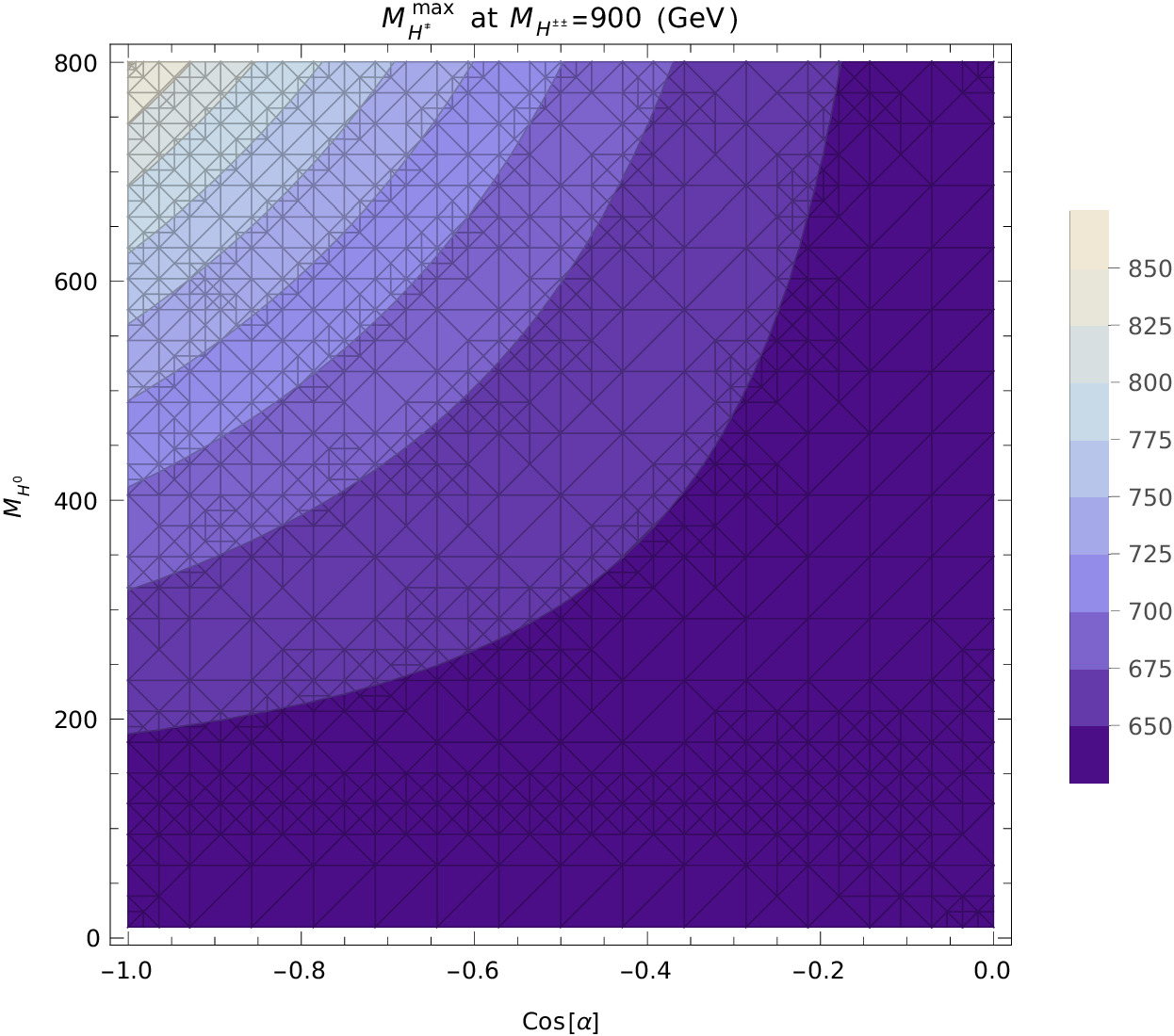}
        \caption{}
        \label{fig:subfig6-6}
    \end{subfigure}

    \caption{This figure presents the parameter \(\lambda\) (panel (a)), \(\mathrm{max}(M_{A^0})\) (panel (b)), and \(M_{H^{\pm}}\) (panels (c), (d), (e), and (f)) as functions of \(M_{H^0}\) and \(\cos{\alpha}\in [-1,0)\), with \(M_{h^0} = 125\,\mathrm{GeV}\). This is more relevant to Scenario~II.}
    \label{fig:panel2}
\end{figure*}

\begin{figure*}[tb]
    \centering
    \begin{subfigure}{0.45\textwidth}
        \includegraphics[width=\linewidth]{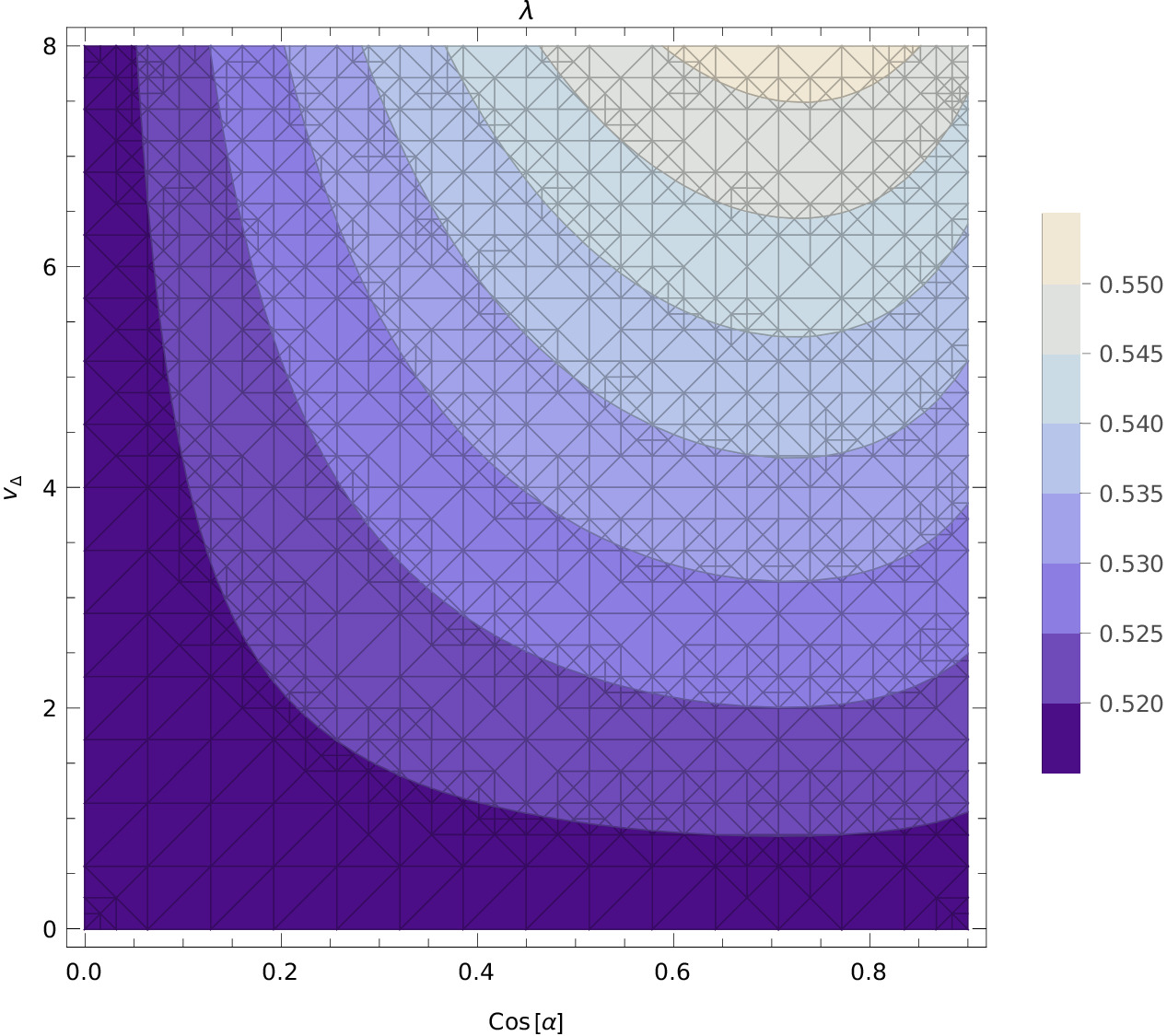}
        \caption{}
        \label{fig:subfig1-7}
    \end{subfigure}%
    \hspace{0.05\textwidth}
    \begin{subfigure}{0.45\textwidth}
        \includegraphics[width=\linewidth]{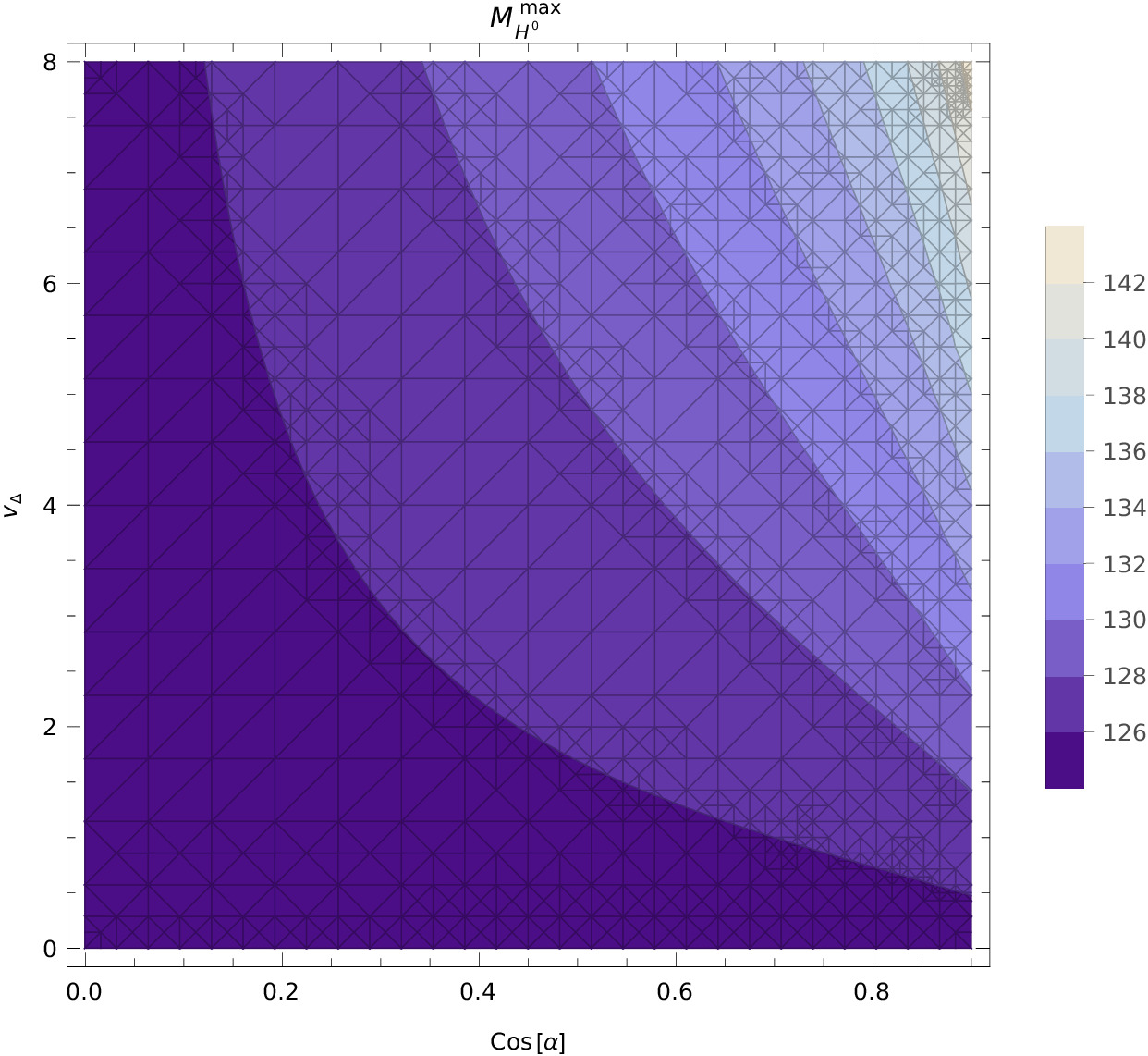}
        \caption{}
        \label{fig:subfig2-7}
    \end{subfigure}

    \begin{subfigure}{0.45\textwidth}
        \includegraphics[width=\linewidth]{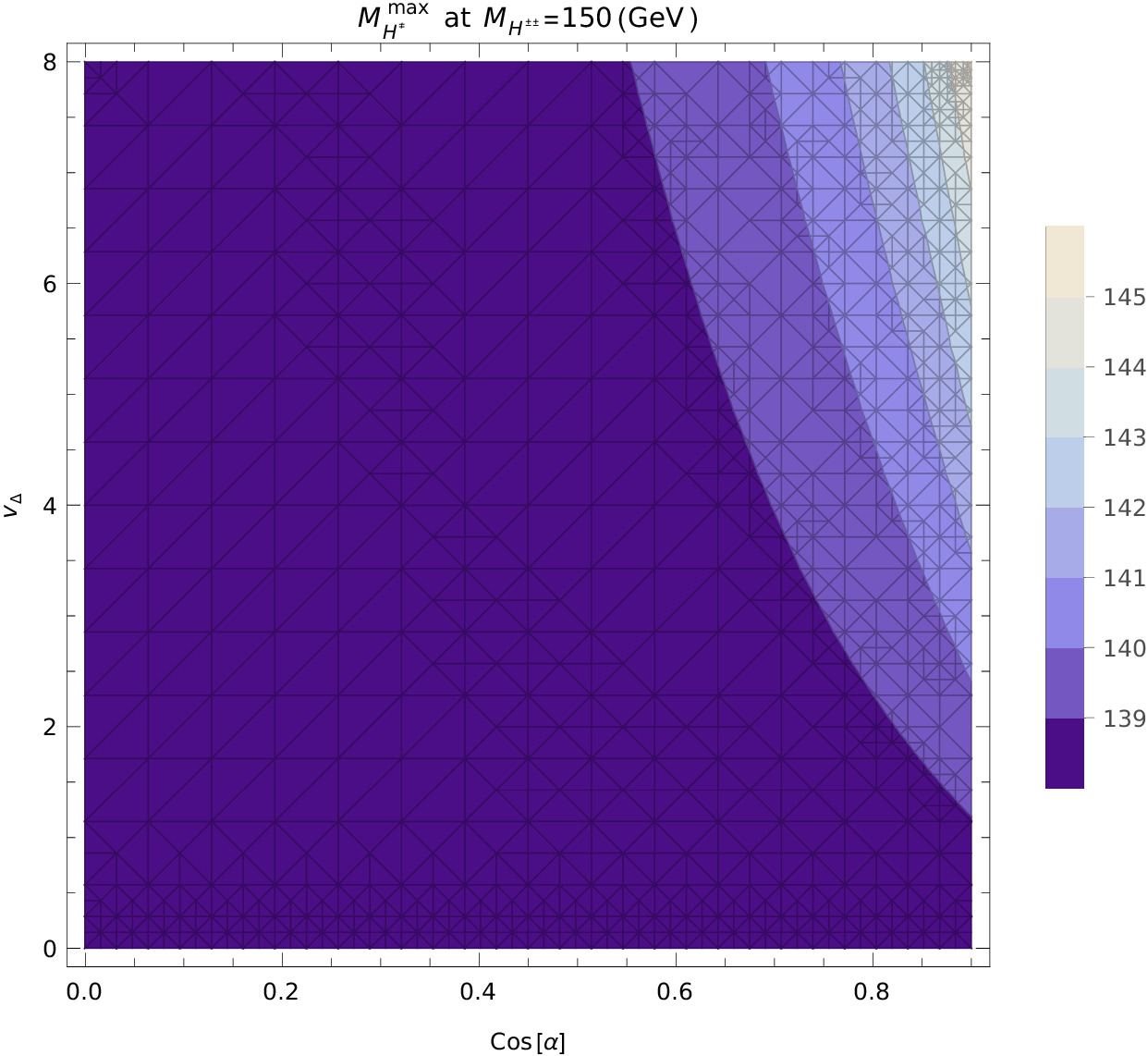}
        \caption{}
        \label{fig:subfig3-7}
    \end{subfigure}%
    \hspace{0.05\textwidth}
    \begin{subfigure}{0.45\textwidth}
        \includegraphics[width=\linewidth]{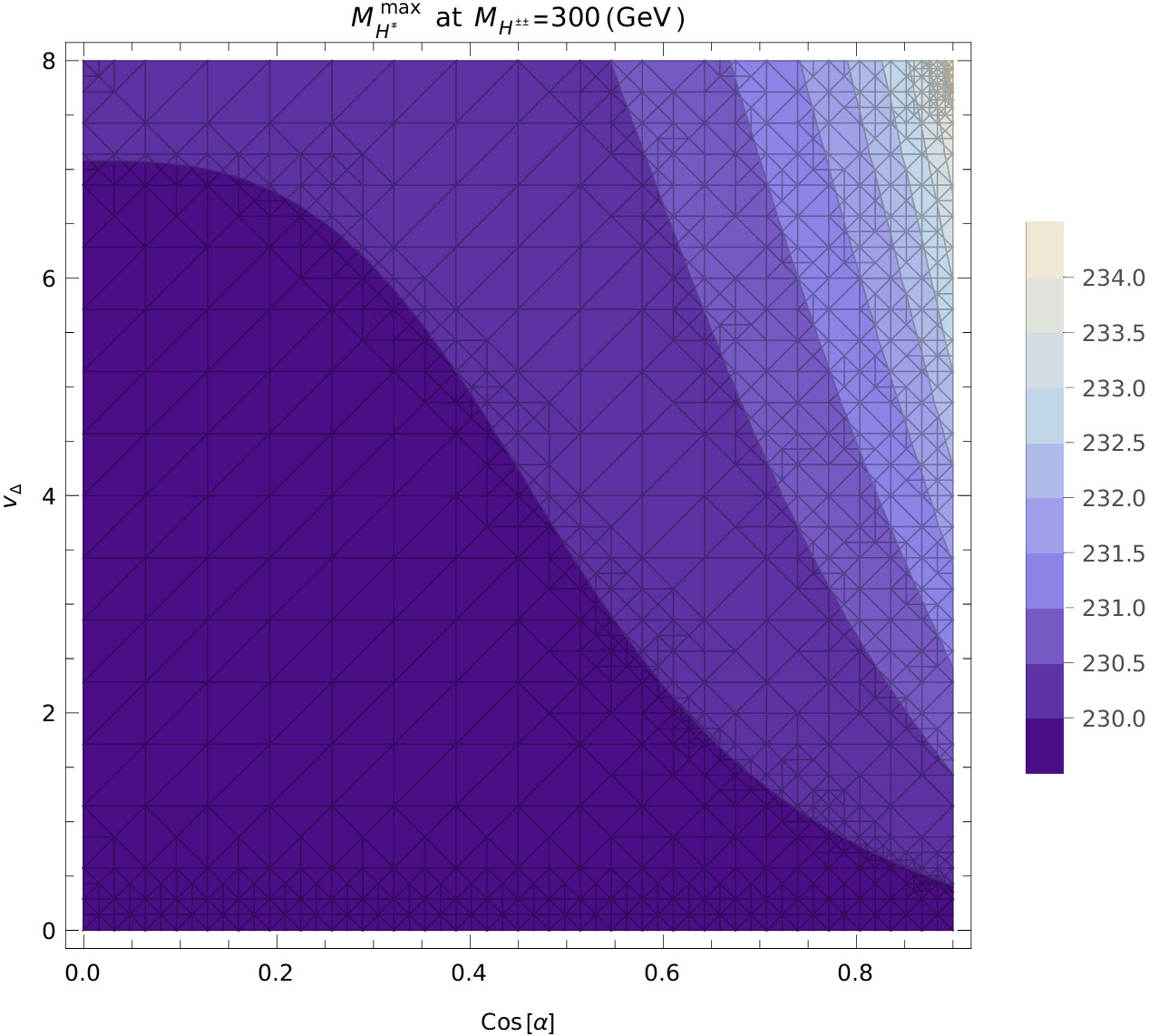}
        \caption{}
        \label{fig:subfig4-7}
    \end{subfigure}

    \begin{subfigure}{0.45\textwidth}
        \includegraphics[width=\linewidth]{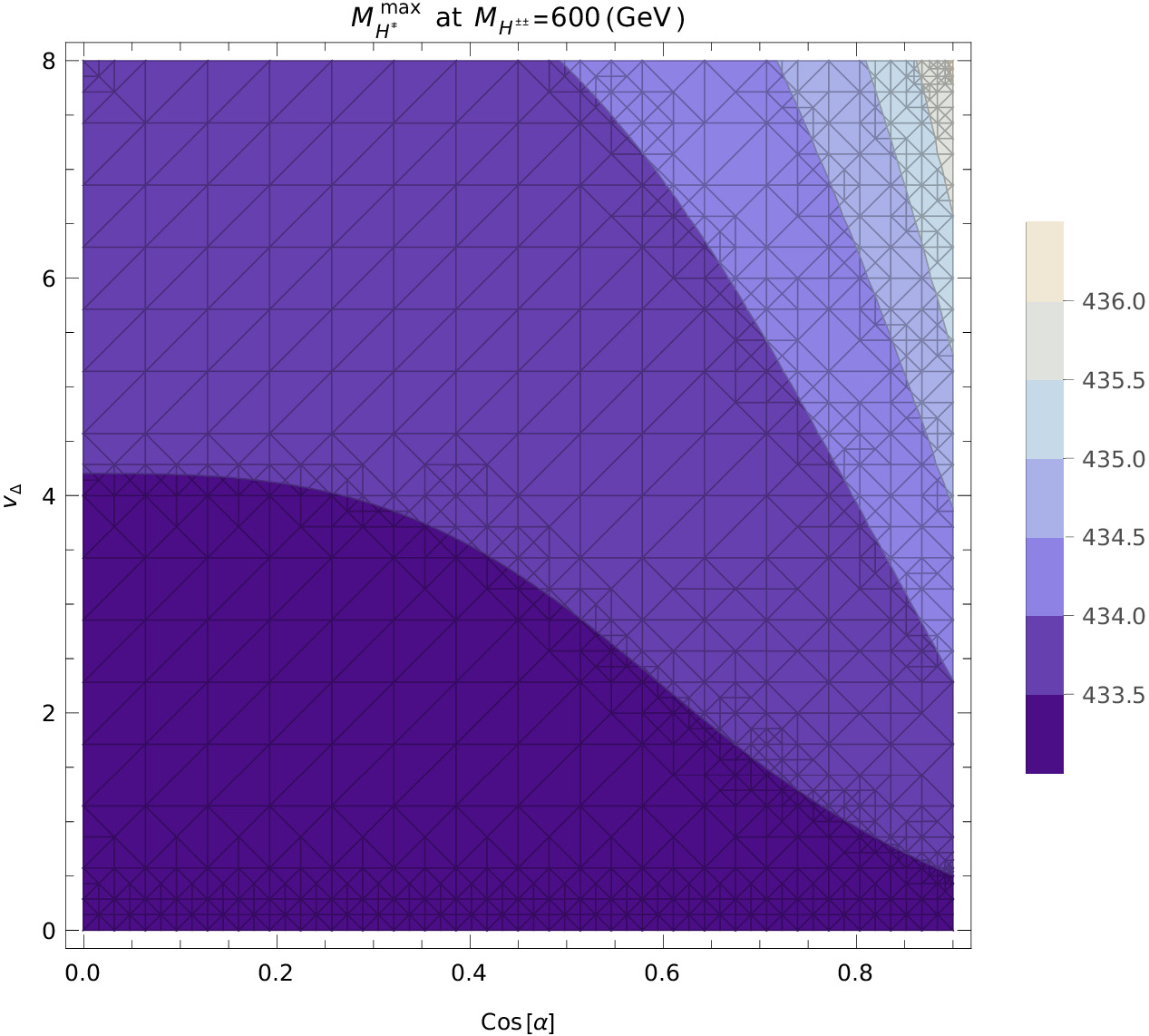}
        \caption{}
        \label{fig:subfig5-7}
    \end{subfigure}%
    \hspace{0.05\textwidth}
    \begin{subfigure}{0.45\textwidth}
        \includegraphics[width=\linewidth]{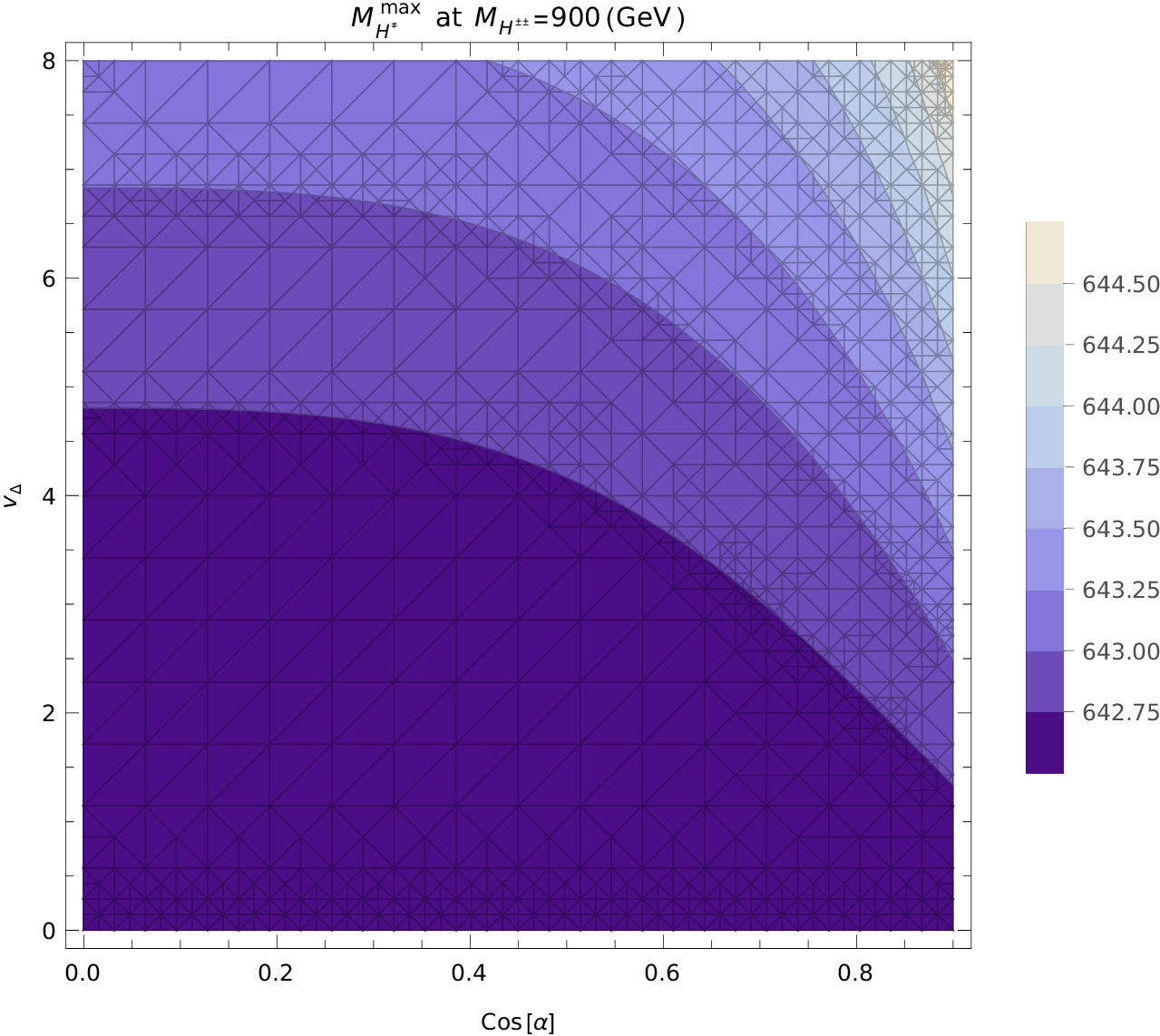}
        \caption{}
        \label{fig:subfig6-7}
    \end{subfigure}

    \caption{This figure presents the parameter \(\lambda\) (panel (a)), \(\mathrm{max}(M_{H^0})\) (panel (b)), and \(M_{H^{\pm}}\) (panels (c), (d), (e), and (f)) as functions of \(v_{\Delta}\) and $\cos{\alpha}\in [0,0.9)$ under scenario II, with \(M_{h^0} = 125\,\mathrm{GeV}\).}
    \label{fig:panel3}
\end{figure*}

\begin{equation}\label{eq:threehiggshhHppHpp}
\lambda_{h^0h^0H^{\pm\pm}H^{\pm\pm}}^{HTM}\approx \lambda_2+\frac{\lambda_1}{2}
\end{equation}
\vspace{0.02mm}
\section{Parameter Scanning and Model Variables: Plots and Analysis}
The following plots show the values of $\lambda$, $M_{H^{\pm}}^{\text{max}}$, $M_{A^0}^{\text{max}}$ in Fig.~\ref{fig:panel1} and Fig.~\ref{fig:panel2}, and $M_{H^0}^{\text{max}}$ in Fig.~\ref{fig:panel3}, depending on the parameter scanning method discussed in Sec.~\ref{sec:param}. The corresponding $\cos{\alpha}$ ranges are $(0.9,1]$, $[-1,0)$, and $[0,0.9]$, respectively.

\clearpage
\bibliographystyle{spphys}
\bibliography{references}

\begin{thebibliography}{100}
\providecommand{\url}[1]{{#1}}
\providecommand{\urlprefix}{URL }
\expandafter\ifx\csname urlstyle\endcsname\relax
  \providecommand{\doi}[1]{DOI \discretionary{}{}{}#1}\else
  \providecommand{\doi}{DOI \discretionary{}{}{}\begingroup \urlstyle{rm}\Url}\fi

\bibitem{Englert:1964et}
F.~Englert, R.~Brout, Phys. Rev. Lett. \textbf{13}, 321 (1964).
\newblock \doi{10.1103/PhysRevLett.13.321}

\bibitem{Higgs:1964pj}
P.W. Higgs, Phys. Rev. Lett. \textbf{13}, 508 (1964).
\newblock \doi{10.1103/PhysRevLett.13.508}

\bibitem{Guralnik:1964eu}
G.S. Guralnik, C.R. Hagen, T.W.B. Kibble, Phys. Rev. Lett. \textbf{13}, 585 (1964).
\newblock \doi{10.1103/PhysRevLett.13.585}

\bibitem{Weinberg:1967tq}
S.~Weinberg, Phys. Rev. Lett. \textbf{19}, 1264 (1967).
\newblock \doi{10.1103/PhysRevLett.19.1264}

\bibitem{ATLAS:2018rvj}
{ATLAS Collaboration},   (2018).
\newblock ATL-PHYS-PUB-2018-053

\bibitem{Salam:1968rm}
A.~Salam, Conf. Proc. C \textbf{680519}, 367 (1968).
\newblock \doi{10.1142/9789812795915_0034}

\bibitem{Grober:2016yuo}
R.~Gr\"ober, EPJ Web Conf. \textbf{164}, 05002 (2017).
\newblock \doi{10.1051/epjconf/201716405002}

\bibitem{ATLAS:2024ish}
G.~Aad, et~al., Phys. Rev. Lett. \textbf{133}(10), 101801 (2024).
\newblock \doi{10.1103/PhysRevLett.133.101801}

\bibitem{Higgs:1966ev}
P.W. Higgs, Phys. Rev. \textbf{145}, 1156 (1966).
\newblock \doi{10.1103/PhysRev.145.1156}

\bibitem{Higgs:1964ia}
P.W. Higgs, Phys. Lett. \textbf{12}, 132 (1964).
\newblock \doi{10.1016/0031-9163(64)91136-9}

\bibitem{Abouabid:2024gms}
H.~Abouabid, et~al., Eur. Phys. J. C \textbf{84}, 1183 (2024).
\newblock \doi{10.1140/epjc/s10052-024-13376-3}

\bibitem{Goldstone:1962es}
J.~Goldstone, A.~Salam, S.~Weinberg, Phys. Rev. \textbf{127}, 965 (1962).
\newblock \doi{10.1103/PhysRev.127.965}

\bibitem{Guralnik:1965uza}
G.S. Guralnik, C.R. Hagen, Nuovo Cim. \textbf{43}(1), 1 (1965).
\newblock \doi{10.1007/BF02753182}

\bibitem{Kibble:1967sv}
T.W.B. Kibble, Phys. Rev. \textbf{155}, 1554 (1967).
\newblock \doi{10.1103/PhysRev.155.1554}

\bibitem{Plehn:2005nk}
T.~Plehn, M.~Rauch, Phys. Rev. D \textbf{72}, 053008 (2005).
\newblock \doi{10.1103/PhysRevD.72.053008}

\bibitem{Baglio:2022wkx}
J.~Baglio, F.~Campanario, S.~Glaus, M.M. M\"uhlleitner, J.~Ronca, M.~Spira, PoS \textbf{PANIC2021}, 393 (2022).
\newblock \doi{10.22323/1.380.0393}

\bibitem{Baglio:2020wgt}
J.~Baglio, F.~Campanario, S.~Glaus, M.~M\"uhlleitner, J.~Ronca, M.~Spira, Phys. Rev. D \textbf{103}(5), 056002 (2021).
\newblock \doi{10.1103/PhysRevD.103.056002}

\bibitem{Weinberg:1974hy}
S.~Weinberg, Phys. Rev. D \textbf{9}, 3357 (1974).
\newblock \doi{10.1103/PhysRevD.9.3357}

\bibitem{Biermann:2024oyy}
L.~Biermann, C.~Borschensky, C.~Englert, M.~M\"uhlleitner, W.~Naskar, Phys. Rev. D \textbf{110}(9), 095012 (2024).
\newblock \doi{10.1103/PhysRevD.110.095012}

\bibitem{Dolan:1973qd}
L.~Dolan, R.~Jackiw, Phys. Rev. D \textbf{9}, 3320 (1974).
\newblock \doi{10.1103/PhysRevD.9.3320}

\bibitem{Coleman:1973jx}
S.R. Coleman, E.J. Weinberg, Phys. Rev. D \textbf{7}, 1888 (1973).
\newblock \doi{10.1103/PhysRevD.7.1888}

\bibitem{Kirzhnits:1976ts}
D.A. Kirzhnits, A.D. Linde, Annals Phys. \textbf{101}, 195 (1976).
\newblock \doi{10.1016/0003-4916(76)90279-7}

\bibitem{Anderson:1991zb}
G.W. Anderson, L.J. Hall, Phys. Rev. D \textbf{45}, 2685 (1992).
\newblock \doi{10.1103/PhysRevD.45.2685}

\bibitem{Kuzmin:1985mm}
V.A. Kuzmin, V.A. Rubakov, M.E. Shaposhnikov, Phys. Lett. B \textbf{155}, 36 (1985).
\newblock \doi{10.1016/0370-2693(85)91028-7}

\bibitem{Goncalves:2017iub}
D.~Goncalves, T.~Han, S.~Mukhopadhyay, Phys. Rev. Lett. \textbf{120}(11), 111801 (2018).
\newblock \doi{10.1103/PhysRevLett.120.111801}.
\newblock [Erratum: Phys.Rev.Lett. 121, 079902 (2018)]

\bibitem{Karkout:2024ojx}
O.~Karkout, A.~Papaefstathiou, M.~Postma, G.~Tetlalmatzi-Xolocotzi, J.~van~de Vis, T.~du~Pree, JHEP \textbf{11}, 077 (2024).
\newblock \doi{10.1007/JHEP11(2024)077}

\bibitem{Arnold:1992rz}
P.B. Arnold, O.~Espinosa, Phys. Rev. D \textbf{47}, 3546 (1993).
\newblock \doi{10.1103/PhysRevD.47.3546}.
\newblock [Erratum: Phys.Rev.D 50, 6662 (1994)]

\bibitem{Kajantie:1996mn}
K.~Kajantie, M.~Laine, K.~Rummukainen, M.E. Shaposhnikov, Phys. Rev. Lett. \textbf{77}, 2887 (1996).
\newblock \doi{10.1103/PhysRevLett.77.2887}

\bibitem{Dawson:2022zbb}
S.~Dawson, et~al., in \emph{{Snowmass 2021}} (2022).
\newblock ArXiv:2209.07510 (Contribution to Snowmass 2021)

\bibitem{Lewis:2024yvj}
I.M. Lewis, J.L. Scott, M.A.S. Alcaraz, M.~Sullivan, Phys. Rev. D \textbf{112}(9), 095024 (2025).
\newblock \doi{10.1103/n5rt-jvg3}

\bibitem{deFlorian:2019app}
D.~de~Florian, I.~Fabre, J.~Mazzitelli, JHEP \textbf{03}, 155 (2020).
\newblock \doi{10.1007/JHEP03(2020)155}

\bibitem{Bahl:2022jnx}
H.~Bahl, J.~Braathen, G.~Weiglein, Phys. Rev. Lett. \textbf{129}(23), 231802 (2022).
\newblock \doi{10.1103/PhysRevLett.129.231802}

\bibitem{Plehn:1996wb}
T.~Plehn, M.~Spira, P.M. Zerwas, Nucl. Phys. B \textbf{479}, 46 (1996).
\newblock \doi{10.1016/0550-3213(96)00418-X}.
\newblock [Erratum: Nucl.Phys.B 531, 655--655 (1998)]

\bibitem{Chiesa:2020awd}
M.~Chiesa, F.~Maltoni, L.~Mantani, B.~Mele, F.~Piccinini, X.~Zhao, JHEP \textbf{09}, 098 (2020).
\newblock \doi{10.1007/JHEP09(2020)098}

\bibitem{Arhrib:2014nya}
A.~Arhrib, R.~Benbrik, G.~Moultaka, L.~Rahili,  (2014).
\newblock ArXiv:1411.5645 [hep-ph]

\bibitem{Shao:2013bz}
D.Y. Shao, C.S. Li, H.T. Li, J.~Wang, JHEP \textbf{07}, 169 (2013).
\newblock \doi{10.1007/JHEP07(2013)169}

\bibitem{deFlorian:2015moa}
D.~de~Florian, J.~Mazzitelli, JHEP \textbf{09}, 053 (2015).
\newblock \doi{10.1007/JHEP09(2015)053}

\bibitem{LHCHiggsCrossSectionWorkingGroup:2016ypw}
D.~de~Florian, et~al.,  (2016).
\newblock \doi{10.23731/CYRM-2017-002}

\bibitem{Asakawa:2008se}
E.~Asakawa, D.~Harada, S.~Kanemura, Y.~Okada, K.~Tsumura, Phys. Lett. B \textbf{672}, 354 (2009).
\newblock \doi{10.1016/j.physletb.2009.01.050}

\bibitem{Papaefstathiou:2015paa}
A.~Papaefstathiou, K.~Sakurai, JHEP \textbf{02}, 006 (2016).
\newblock \doi{10.1007/JHEP02(2016)006}

\bibitem{Binoth:2006ym}
T.~Binoth, S.~Karg, N.~Kauer, R.~Ruckl, Phys. Rev. D \textbf{74}, 113008 (2006).
\newblock \doi{10.1103/PhysRevD.74.113008}

\bibitem{Acar:2016rde}
Y.C. Acar, A.N. Akay, S.~Beser, A.C. Canbay, H.~Karadeniz, U.~Kaya, B.B. Oner, S.~Sultansoy, Nucl. Instrum. Meth. A \textbf{871}, 47 (2017).
\newblock \doi{10.1016/j.nima.2017.07.041}

\bibitem{Murayama:1996ec}
H.~Murayama, M.E. Peskin, Ann. Rev. Nucl. Part. Sci. \textbf{46}, 533 (1996).
\newblock \doi{10.1146/annurev.nucl.46.1.533}

\bibitem{Benedikt:2022kan}
M.~Benedikt, et~al.,  (2022).
\newblock ArXiv:2203.07804 [physics.acc-ph] (FERMILAB-CONF-22-182-AD, Contribution to Snowmass 2021)

\bibitem{FCC:2018evy}
A.~Abada, et~al., Eur. Phys. J. ST \textbf{228}(2), 261 (2019).
\newblock \doi{10.1140/epjst/e2019-900045-4}

\bibitem{Telnov:2006rj}
V.I. Telnov, Acta Phys. Polon. B \textbf{37}, 633 (2006)

\bibitem{Arhrib:2009gg}
A.~Arhrib, R.~Benbrik, C.H. Chen, R.~Santos, Phys. Rev. D \textbf{80}, 015010 (2009).
\newblock \doi{10.1103/PhysRevD.80.015010}

\bibitem{Delahaye:2019omf}
J.P. Delahaye, M.~Diemoz, K.~Long, B.~Mansouli{\'e}, N.~Pastrone, L.~Rivkin, D.~Schulte, A.~Skrinsky, A.~Wulzer, Eur. Phys. J. C  (2019).
\newblock ArXiv:1901.06150 [physics.acc-ph]

\bibitem{Palmer:2014nza}
R.B. Palmer, Rev. Accel. Sci. Tech. \textbf{7}, 137 (2014).
\newblock \doi{10.1142/S1793626814300072}

\bibitem{Long:2020wfp}
K.~Long, D.~Lucchesi, M.~Palmer, N.~Pastrone, D.~Schulte, V.~Shiltsev, Nature Phys. \textbf{17}(3), 289 (2021).
\newblock \doi{10.1038/s41567-020-01130-x}

\bibitem{InternationalMuonCollider:2025sys}
C.~Accettura, et~al., Eur. Phys. J. C  (2025).
\newblock ArXiv:2504.21417 [physics.acc-ph]

\bibitem{vonWeizsacker:1934nji}
C.F. von Weizsacker, Z. Phys. \textbf{88}, 612 (1934).
\newblock \doi{10.1007/BF01333110}

\bibitem{Chiesa:2021qpr}
M.~Chiesa, B.~Mele, F.~Piccinini, Eur. Phys. J. C \textbf{84}(5), 543 (2024).
\newblock \doi{10.1140/epjc/s10052-024-12882-8}

\bibitem{Williams:1934ad}
E.J. Williams, Phys. Rev. \textbf{45}, 729 (1934).
\newblock \doi{10.1103/PhysRev.45.729}

\bibitem{Costantini:2020stv}
A.~Costantini, F.~De~Lillo, F.~Maltoni, L.~Mantani, O.~Mattelaer, R.~Ruiz, X.~Zhao, JHEP \textbf{09}, 080 (2020).
\newblock \doi{10.1007/JHEP09(2020)080}

\bibitem{AlAli:2021let}
H.~Al~Ali, et~al., Rept. Prog. Phys. \textbf{85}(8), 084201 (2022).
\newblock \doi{10.1088/1361-6633/ac6678}

\bibitem{Sonmez:2018mcq}
N.~S{\"o}nmez, Turk. J. Phys. \textbf{42}(6), 675 (2018).
\newblock \doi{10.3906/fiz-1807-6}

\bibitem{Black:2022cth}
K.M. Black, et~al., JINST \textbf{19}(02), T02015 (2024).
\newblock \doi{10.1088/1748-0221/19/02/T02015}

\bibitem{MuonCollider:2022xlm}
{Muon Collider Collaboration},  (2022).
\newblock ArXiv:2203.07261 (Contribution to Snowmass 2021)

\bibitem{Capdevilla:2024ydp}
R.~Capdevilla, F.~Garosi, D.~Marzocca, B.~Stechauner, Eur. Phys. J. C  (2024).
\newblock ArXiv:2410.21383 [hep-ph]

\bibitem{Capdevilla:2020qel}
R.~Capdevilla, D.~Curtin, Y.~Kahn, G.~Krnjaic, Phys. Rev. D \textbf{103}(7), 075028 (2021).
\newblock \doi{10.1103/PhysRevD.103.075028}

\bibitem{Capdevilla:2024bwt}
R.~Capdevilla, F.~Meloni, J.~Zurita, Eur. Phys. J. C  (2024).
\newblock ArXiv:2405.08858 [hep-ph]

\bibitem{Han:2020pif}
T.~Han, D.~Liu, I.~Low, X.~Wang, Phys. Rev. D \textbf{103}(1), 013002 (2021).
\newblock \doi{10.1103/PhysRevD.103.013002}

\bibitem{Capdevilla:2021rwo}
R.~Capdevilla, D.~Curtin, Y.~Kahn, G.~Krnjaic, Phys. Rev. D \textbf{105}(1), 015028 (2022).
\newblock \doi{10.1103/PhysRevD.105.015028}

\bibitem{Phan:2024vfy}
K.H. Phan, D.T. Tran, T.H. Nguyen, Eur. Phys. J. C  (2024).
\newblock ArXiv:2409.00662 [hep-ph]

\bibitem{Phan:2024vxm}
K.H. Phan, D.T. Tran, T.H. Nguyen, Eur. Phys. J. C  (2024).
\newblock ArXiv:2410.06827 [hep-ph]

\bibitem{Figy:2011yu}
T.~Figy, R.~Zwicky, JHEP \textbf{10}, 145 (2011).
\newblock \doi{10.1007/JHEP10(2011)145}

\bibitem{Demirci:2016fri}
M.~Demirci, A.I. Ahmadov, Phys. Rev. D \textbf{94}(7), 075025 (2016).
\newblock \doi{10.1103/PhysRevD.94.075025}

\bibitem{Demirci:2019kop}
M.~Demirci, Turk. J. Phys. \textbf{43}(5), 442 (2019).
\newblock \doi{10.3906/fiz-1903-15}

\bibitem{Muhlleitner:2001kw}
M.M. Muhlleitner, M.~Kramer, M.~Spira, P.M. Zerwas, Phys. Lett. B \textbf{508}, 311 (2001).
\newblock \doi{10.1016/S0370-2693(01)00321-5}

\bibitem{Fuks:2025gjv}
B.~Fuks, A.~Papaefstathiou, G.~Tetlalmatzi-Xolocotzi, Eur. Phys. J. C \textbf{85}(11), 1309 (2025).
\newblock \doi{10.1140/epjc/s10052-025-15051-7}

\bibitem{Lee:1973iz}
T.D. Lee, Phys. Rev. D \textbf{8}, 1226 (1973).
\newblock \doi{10.1103/PhysRevD.8.1226}

\bibitem{Schechter:1980gr}
J.~Schechter, J.W.F. Valle, Phys. Rev. D \textbf{22}, 2227 (1980).
\newblock \doi{10.1103/PhysRevD.22.2227}

\bibitem{Cheng:1980qt}
T.P. Cheng, L.F. Li, Phys. Rev. D \textbf{22}, 2860 (1980).
\newblock \doi{10.1103/PhysRevD.22.2860}

\bibitem{Mohapatra:1979ia}
R.N. Mohapatra, G.~Senjanovic, Phys. Rev. Lett. \textbf{44}, 912 (1980).
\newblock \doi{10.1103/PhysRevLett.44.912}

\bibitem{Konetschny:1977bn}
W.~Konetschny, W.~Kummer, Phys. Lett. B \textbf{70}, 433 (1977).
\newblock \doi{10.1016/0370-2693(77)90407-5}

\bibitem{Arbabifar:2012bd}
F.~Arbabifar, S.~Bahrami, M.~Frank, Phys. Rev. D \textbf{87}(1), 015020 (2013).
\newblock \doi{10.1103/PhysRevD.87.015020}

\bibitem{Magg:1980ut}
M.~Magg, C.~Wetterich, Phys. Lett. B \textbf{94}, 61 (1980).
\newblock \doi{10.1016/0370-2693(80)90825-4}

\bibitem{Deshpande:1977rw}
N.G. Deshpande, E.~Ma, Phys. Rev. D \textbf{18}, 2574 (1978).
\newblock \doi{10.1103/PhysRevD.18.2574}

\bibitem{Glashow:1976nt}
S.L. Glashow, S.~Weinberg, Phys. Rev. D \textbf{15}, 1958 (1977).
\newblock \doi{10.1103/PhysRevD.15.1958}

\bibitem{Ellis:1988er}
J.R. Ellis, J.F. Gunion, H.E. Haber, L.~Roszkowski, F.~Zwirner, Phys. Rev. D \textbf{39}, 844 (1989).
\newblock \doi{10.1103/PhysRevD.39.844}

\bibitem{Branco:2011iw}
G.C. Branco, P.M. Ferreira, L.~Lavoura, M.N. Rebelo, M.~Sher, J.P. Silva, Phys. Rept. \textbf{516}, 1 (2012).
\newblock \doi{10.1016/j.physrep.2012.02.002}

\bibitem{Demirci:2020zgt}
M.~Demirci, Nucl. Phys. B \textbf{961}, 115235 (2020).
\newblock \doi{10.1016/j.nuclphysb.2020.115235}

\bibitem{ElKaffas:2007nii}
A.W. El~Kaffas, O.M. Ogreid, P.~Osland, eConf \textbf{C0705302}, HIG08 (2007)

\bibitem{WahabElKaffas:2007xd}
A.~Wahab El~Kaffas, P.~Osland, O.M. Ogreid, Phys. Rev. D \textbf{76}, 095001 (2007).
\newblock \doi{10.1103/PhysRevD.76.095001}

\bibitem{Papaefstathiou:2020lyp}
A.~Papaefstathiou, T.~Robens, G.~Tetlalmatzi-Xolocotzi, JHEP \textbf{05}, 193 (2021).
\newblock \doi{10.1007/JHEP05(2021)193}

\bibitem{Shen:2015bna}
J.F. Shen, Z.X. Li, EPL \textbf{111}(3), 31001 (2015).
\newblock \doi{10.1209/0295-5075/111/31001}

\bibitem{Rahili:2019ixf}
L.~Rahili, A.~Arhrib, R.~Benbrik, Eur. Phys. J. C \textbf{79}(11), 940 (2019).
\newblock \doi{10.1140/epjc/s10052-019-7471-3}

\bibitem{Arhrib:2011uy}
A.~Arhrib, R.~Benbrik, M.~Chabab, G.~Moultaka, M.C. Peyranere, L.~Rahili, J.~Ramadan, Phys. Rev. D \textbf{84}, 095005 (2011).
\newblock \doi{10.1103/PhysRevD.84.095005}

\bibitem{CMS:2022dwd}
A.~Tumasyan, et~al., Nature \textbf{607}(7917), 60 (2022).
\newblock \doi{10.1038/s41586-022-04892-x}.
\newblock [Erratum: Nature 623, (2023)]

\bibitem{CMS:2012qbp}
S.~Chatrchyan, et~al., Phys. Lett. B \textbf{716}, 30 (2012).
\newblock \doi{10.1016/j.physletb.2012.08.021}

\bibitem{BhupalDev:2013xol}
P.S. Bhupal~Dev, D.K. Ghosh, N.~Okada, I.~Saha, JHEP \textbf{03}, 150 (2013).
\newblock \doi{10.1007/JHEP03(2013)150}.
\newblock [Erratum: JHEP 05, 049 (2013)]

\bibitem{Kraus:2004zw}
C.~Kraus, et~al., Eur. Phys. J. C \textbf{40}, 447 (2005).
\newblock \doi{10.1140/epjc/s2005-02139-7}

\bibitem{Weinheimer:1999tn}
C.~Weinheimer, B.~Degenddag, A.~Bleile, J.~Bonn, L.~Bornschein, O.~Kazachenko, A.~Kovalik, E.W. Otten, Phys. Lett. B \textbf{460}, 219 (1999).
\newblock \doi{10.1016/S0370-2693(99)00780-7}.
\newblock [Erratum: Phys.Lett.B 464, 352--352 (1999)]

\bibitem{Lobashev:1999tp}
V.M. Lobashev, et~al., Phys. Lett. B \textbf{460}, 227 (1999).
\newblock \doi{10.1016/S0370-2693(99)00781-9}

\bibitem{Belesev1995ResultsOT}
A.I. Belesev, A.I. Bleule, E.V. Geraskin, A.~Golubev, N.~Golubev, O.V. Kazachenko, E.P. Kiev, Y.~Kuznetsov, V.M. Lobashev, B.M. Ovchinnikov, V.I. Parfenov, I.V. Sekachev, A.P. Solodukhin, N.A. Titov, I.E. Yarykin, Y.I. Zakharov, S.N. Balashov, P.E. Spivak, Physics Letters B \textbf{350}, 263 (1995).
\newblock \urlprefix\url{https://api.semanticscholar.org/CorpusID:120963143}

\bibitem{KATRIN:2021uub}
M.~Aker, et~al., Nature Phys. \textbf{18}(2), 160 (2022).
\newblock \doi{10.1038/s41567-021-01463-1}

\bibitem{Drexlin:2013lha}
G.~Drexlin, V.~Hannen, S.~Mertens, C.~Weinheimer, Adv. High Energy Phys. \textbf{2013}, 293986 (2013).
\newblock \doi{10.1155/2013/293986}

\bibitem{KATRIN:2019yun}
M.~Aker, et~al., Phys. Rev. Lett. \textbf{123}(22), 221802 (2019).
\newblock \doi{10.1103/PhysRevLett.123.221802}

\bibitem{Primulando:2019evb}
R.~Primulando, J.~Julio, P.~Uttayarat, JHEP \textbf{08}, 024 (2019).
\newblock \doi{10.1007/JHEP08(2019)024}

\bibitem{Ashanujjaman:2025scr}
S.~Ashanujjaman, P.S.B. Dev, J.~Huang, S.~Zhou, Phys. Rev. D \textbf{113}(5), L051704 (2026).
\newblock \doi{10.1103/mhds-pwkd}

\bibitem{PhysRevD.22.2227}
J.~Schechter, J.W.F. Valle, Phys. Rev. D \textbf{22}, 2227 (1980).
\newblock \doi{10.1103/PhysRevD.22.2227}.
\newblock \urlprefix\url{https://link.aps.org/doi/10.1103/PhysRevD.22.2227}

\bibitem{PhysRevD.25.2951}
J.~Schechter, J.W.F. Valle, Phys. Rev. D \textbf{25}, 2951 (1982).
\newblock \doi{10.1103/PhysRevD.25.2951}.
\newblock \urlprefix\url{https://link.aps.org/doi/10.1103/PhysRevD.25.2951}

\bibitem{Babu:2016gpg}
K.S. Babu, I.~Gogoladze, S.~Khan, Phys. Rev. D \textbf{95}(9), 095013 (2017).
\newblock \doi{10.1103/PhysRevD.95.095013}

\bibitem{Du:2018eaw}
Y.~Du, A.~Dunbrack, M.J. Ramsey-Musolf, J.H. Yu, JHEP \textbf{01}, 101 (2019).
\newblock \doi{10.1007/JHEP01(2019)101}

\bibitem{Akeroyd:2012ms}
A.G. Akeroyd, S.~Moretti, Phys. Rev. D \textbf{86}, 035015 (2012).
\newblock \doi{10.1103/PhysRevD.86.035015}

\bibitem{Akeroyd:2012rg}
A.G. Akeroyd, S.~Moretti, PoS \textbf{CHARGED2012}, 035 (2012).
\newblock \doi{10.22323/1.156.0035}

\bibitem{Akeroyd:2011ir}
A.G. Akeroyd, S.~Moretti, Phys. Rev. D \textbf{84}, 035028 (2011).
\newblock \doi{10.1103/PhysRevD.84.035028}

\bibitem{Haba:2016zbu}
N.~Haba, H.~Ishida, N.~Okada, Y.~Yamaguchi, Eur. Phys. J. C \textbf{76}(6), 333 (2016).
\newblock \doi{10.1140/epjc/s10052-016-4180-z}

\bibitem{Akeroyd:2010je}
A.G. Akeroyd, C.W. Chiang, Phys. Rev. D \textbf{81}, 115007 (2010).
\newblock \doi{10.1103/PhysRevD.81.115007}

\bibitem{Ashanujjaman:2021txz}
S.~Ashanujjaman, K.~Ghosh, JHEP \textbf{03}, 195 (2022).
\newblock \doi{10.1007/JHEP03(2022)195}

\bibitem{Aoki:2012yt}
M.~Aoki, S.~Kanemura, M.~Kikuchi, K.~Yagyu, Phys. Lett. B \textbf{714}, 279 (2012).
\newblock \doi{10.1016/j.physletb.2012.07.016}

\bibitem{Das:2016bir}
D.~Das, A.~Santamaria, Phys. Rev. D \textbf{94}(1), 015015 (2016).
\newblock \doi{10.1103/PhysRevD.94.015015}

\bibitem{Ashanujjaman:2024lnr}
S.~Ashanujjaman, S.~Banik, G.~Coloretti, A.~Crivellin, S.P. Maharathy, B.~Mellado, JHEP \textbf{04}, 003 (2025).
\newblock \doi{10.1007/JHEP04(2025)003}

\bibitem{Samarakoon:2023crt}
B.~Samarakoon, T.M. Figy, Phys. Rev. D \textbf{109}(7), 075015 (2024).
\newblock \doi{10.1103/PhysRevD.109.075015}

\bibitem{Ducu:2024xxf}
O.A. Ducu, A.E. Dumitriu, A.~Jinaru, R.~Kukla, E.~Monnier, G.~Moultaka, A.~Tudorache, H.~Xu, JHEP \textbf{06}, 020 (2025).
\newblock \doi{10.1007/JHEP06(2025)020}

\bibitem{Christensen:2008py}
N.D. Christensen, C.~Duhr, Comput. Phys. Commun. \textbf{180}, 1614 (2009).
\newblock \doi{10.1016/j.cpc.2009.02.018}

\bibitem{Alloul:2013bka}
A.~Alloul, N.D. Christensen, C.~Degrande, C.~Duhr, B.~Fuks, Comput. Phys. Commun. \textbf{185}, 2250 (2014).
\newblock \doi{10.1016/j.cpc.2014.04.012}

\bibitem{Hahn:1998yk}
T.~Hahn, M.~Perez-Victoria, Comput. Phys. Commun. \textbf{118}, 153 (1999).
\newblock \doi{10.1016/S0010-4655(98)00173-8}

\bibitem{Denner:1991kt}
A.~Denner, Fortsch. Phys. \textbf{41}, 307 (1993).
\newblock \doi{10.1002/prop.2190410402}

\bibitem{Hahn:2016ktb}
T.~Hahn, Comput. Phys. Commun. \textbf{207}, 341 (2016).
\newblock \doi{10.1016/j.cpc.2016.05.012}

\bibitem{Hahn:2004fe}
T.~Hahn, Comput. Phys. Commun. \textbf{168}, 78 (2005).
\newblock \doi{10.1016/j.cpc.2005.01.010}

\bibitem{Hahn:2000kx}
T.~Hahn, Comput. Phys. Commun. \textbf{140}, 418 (2001).
\newblock \doi{10.1016/S0010-4655(01)00290-9}

\bibitem{Nogueira:1991ex}
P.~Nogueira, J. Comput. Phys. \textbf{105}, 279 (1993).
\newblock \doi{10.1006/jcph.1993.1074}

\bibitem{Vermaseren:2000nd}
J.A.M. Vermaseren.
\newblock {New features of FORM} (2000).
\newblock ArXiv:math-ph/0010025

\bibitem{Cullen:2010jv}
G.~Cullen, M.~Koch-Janusz, T.~Reiter, Comput. Phys. Commun. \textbf{182}, 2368 (2011).
\newblock \doi{10.1016/j.cpc.2011.06.007}

\bibitem{Reiter:2009ts}
T.~Reiter, Comput. Phys. Commun. \textbf{181}, 1301 (2010).
\newblock \doi{10.1016/j.cpc.2010.01.012}

\bibitem{Mastrolia:2010nb}
P.~Mastrolia, G.~Ossola, T.~Reiter, F.~Tramontano, JHEP \textbf{08}, 080 (2010).
\newblock \doi{10.1007/JHEP08(2010)080}

\bibitem{Binoth:2008uq}
T.~Binoth, J.P. Guillet, G.~Heinrich, E.~Pilon, T.~Reiter, Comput. Phys. Commun. \textbf{180}, 2317 (2009).
\newblock \doi{10.1016/j.cpc.2009.06.024}

\bibitem{Cullen:2011kv}
G.~Cullen, J.P. Guillet, G.~Heinrich, T.~Kleinschmidt, E.~Pilon, T.~Reiter, M.~Rodgers, Comput. Phys. Commun. \textbf{182}, 2276 (2011).
\newblock \doi{10.1016/j.cpc.2011.05.015}

\bibitem{Guillet:2013msa}
J.P. Guillet, G.~Heinrich, J.F. von Soden-Fraunhofen, Comput. Phys. Commun. \textbf{185}, 1828 (2014).
\newblock \doi{10.1016/j.cpc.2014.03.009}

\bibitem{vanDeurzen:2013saa}
H.~van Deurzen, G.~Luisoni, P.~Mastrolia, E.~Mirabella, G.~Ossola, T.~Peraro, JHEP \textbf{03}, 115 (2014).
\newblock \doi{10.1007/JHEP03(2014)115}

\bibitem{Peraro:2014cba}
T.~Peraro, Comput. Phys. Commun. \textbf{185}, 2771 (2014).
\newblock \doi{10.1016/j.cpc.2014.06.017}

\bibitem{GoSam:2014iqq}
G.~Cullen, et~al., Eur. Phys. J. C \textbf{74}(8), 3001 (2014).
\newblock \doi{10.1140/epjc/s10052-014-3001-5}

\bibitem{Samarakoon:2025uyw}
B.~Samarakoon, Ph.D. thesis, Wichita State University (2025)

\bibitem{Kleiss:1985gy}
R.~Kleiss, W.J. Stirling, S.D. Ellis, Comput. Phys. Commun. \textbf{40}, 359 (1986).
\newblock \doi{10.1016/0010-4655(86)90119-0}

\bibitem{Platzer:2013esa}
S.~Plätzer,  (2013).
\newblock ArXiv:1308.2922 [hep-ph], DESY-13-145, MCNET-13-10

\bibitem{Campbell:2017hsr}
J.~Campbell, J.~Huston, F.~Krauss, \emph{The Black Book of Quantum Chromodynamics: A Primer for the LHC Era} (Oxford University Press, 2017).
\newblock \doi{10.1093/oso/9780199652747.001.0001}.
\newblock \urlprefix\url{https://doi.org/10.1093/oso/9780199652747.001.0001}

\bibitem{1934ZPhy...88..612W}
C.F.V. {Weizs{\"a}cker}, Zeitschrift fur Physik \textbf{88}(9-10), 612 (1934).
\newblock \doi{10.1007/BF01333110}

\bibitem{PhysRev.45.729}
E.J. Williams, Phys. Rev. \textbf{45}, 729 (1934).
\newblock \doi{10.1103/PhysRev.45.729}.
\newblock \urlprefix\url{https://link.aps.org/doi/10.1103/PhysRev.45.729}

\bibitem{Landau:1934zj}
L.D. Landau, E.M. Lifschitz, Phys. Z. Sowjetunion \textbf{6}, 244 (1934).
\newblock \doi{10.1016/B978-0-08-010586-4.50021-3}

\bibitem{Garosi:2023bvq}
F.~Garosi, D.~Marzocca, S.~Trifinopoulos, JHEP \textbf{09}, 107 (2023).
\newblock \doi{10.1007/JHEP09(2023)107}

\bibitem{Han:2020uid}
T.~Han, Y.~Ma, K.~Xie, Phys. Rev. D \textbf{103}(3), L031301 (2021).
\newblock \doi{10.1103/PhysRevD.103.L031301}

\bibitem{Han:2021kes}
T.~Han, Y.~Ma, K.~Xie, JHEP \textbf{02}, 154 (2022).
\newblock \doi{10.1007/JHEP02(2022)154}

\bibitem{Peskin:1995ev}
M.E. Peskin, D.V. Schroeder,  (Addison-Wesley, Reading, USA, 1995)

\bibitem{Budnev:1975poe}
V.M. Budnev, I.F. Ginzburg, G.V. Meledin, V.G. Serbo, Phys. Rept. \textbf{15}, 181 (1975).
\newblock \doi{10.1016/0370-1573(75)90009-5}

\bibitem{Frixione:1993yw}
S.~Frixione, M.L. Mangano, P.~Nason, G.~Ridolfi, Phys. Lett. B \textbf{319}, 339 (1993).
\newblock \doi{10.1016/0370-2693(93)90823-Z}

\bibitem{Haber:2013mia}
H.E. Haber, in \emph{{Proceedings of the 1st Toyama International Workshop on Higgs as a Probe of New Physics 2013}} (2013).
\newblock ArXiv:1401.0152

\bibitem{Plehn:2009nd}
T.~Plehn, Lect. Notes Phys. \textbf{844}, 1 (2012).
\newblock \doi{10.1007/978-3-642-24040-9}

\end{thebibliography}

\end{document}